\documentclass[12pt,a4paper]{article}
\usepackage{etex} 
\usepackage{silence}
\usepackage{jheppub,textcomp}
\usepackage[dvipsnames]{xcolor}
\usepackage{latexsym,amscd,amsmath,amssymb,amsfonts,xspace,mathrsfs,mathtools,amsthm,epsfig}
\usepackage[utf8]{inputenc}
\usepackage{bbm} 
\usepackage{bm}   
\usepackage{graphicx} 
\usepackage{tabu} 
\usepackage[matrix,arrow]{xy}
\usepackage{stmaryrd}
\SetSymbolFont{stmry}{bold}{U}{stmry}{m}{n}
\usepackage{array,booktabs} 
\usepackage{slashed}
\usepackage{subcaption}
\usepackage{tensor}
\usepackage[size=footnotesize]{caption,subcaption}
\usepackage[textsize=tiny]{todonotes}
\usepackage{paralist}
\usepackage{youngtab}
\usepackage{pifont}
\usepackage{cellspace, hhline}
\allowdisplaybreaks

\usepackage{lineno}

\usepackage{verbatim} 
\usepackage{rotating}
\usepackage{bm}
\usepackage{cleveref}
\usepackage{slashed}
\usepackage[normalem]{ulem}

\usepackage{bbding} 
\DeclareRobustCommand{\shamrock}{%
  \raisebox{-0.1ex}{\includegraphics[height=1.8ex]{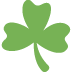}}%
}
 
\usepackage{tikz}
\usepackage{tikz-cd}
\usepackage{diagbox}
\usetikzlibrary{positioning}
\usetikzlibrary{calc}
\usetikzlibrary{decorations.pathreplacing,calligraphy}

\usepackage{xstring}
\usetikzlibrary{decorations.pathmorphing} 
\usetikzlibrary{decorations.markings} 
\usetikzlibrary{arrows, arrows.meta} 
\usetikzlibrary{shapes} 
\usetikzlibrary{matrix} 
\usetikzlibrary{positioning} 
\usepackage[english]{babel} 
\usepackage[autostyle]{csquotes}

\tikzstyle{brane}=[draw]
\tikzset{D7/.style={circle, draw=black, inner sep=0pt, fill=white, minimum size=3mm}}
\tikzset{hasse/.style={circle, fill,inner sep=2pt}}
\tikzset{flavour/.style={regular polygon,fill=white,regular polygon sides=4,inner sep=2.5pt, draw}}
\tikzset{gauge/.style={circle, draw,inner sep=2.5pt}}
\tikzset{gaugeb/.style={circle, draw,fill=black,inner sep=2.5pt}}
\tikzset{gauger/.style={circle, draw,fill=cyan,inner sep=2.5pt}}
\tikzset{gaugeg/.style={circle, draw,fill=red,inner sep=2.5pt}}
\tikzset{bd/.style={circle, draw=black, inner sep=0pt, fill=black, minimum size=2mm}}
\tikzset{wd/.style={circle, draw=black, inner sep=0pt, fill=white, minimum size=2mm}}
\tikzset{SUd/.style={circle, draw=black, inner sep=0pt, fill=yellow, minimum size=2mm}}
\tikzset{Dynkin/.style={circle, draw=black, inner sep=0pt, fill=white, minimum size=2mm}}
\tikzstyle{ligne}=[draw, thick] 
\tikzset{doublearrow/.style={ draw=black!75, color=black!75, thick, double distance=3pt, }}

\tikzset{->-/.style={decoration={
  markings,
  mark=at position #1 with {\arrow{>}}},postaction={decorate}}}
\tikzset{->>-/.style={decoration={
  markings,
  mark= between positions #1-0.05 and #1+0.05 step 0.1 with {\arrow{>}}
  },postaction={decorate}}}
\tikzset{->>>-/.style={decoration={
  markings,
  mark= between positions #1-0.1 and #1+0.1 step 0.1 with {\arrow{>}}
  },postaction={decorate}}}
  \tikzset{->>>>-/.style={decoration={
  markings,
  mark= between positions #1-0.2 and #1+0.2 step 0.1 with {\arrow{>}}
  },postaction={decorate}}}
    \tikzset{->>>>>-/.style={decoration={
  markings,
  mark= between positions #1-0.3 and #1+0.2 step 0.1 with {\arrow{>}}
  },postaction={decorate}}}
  \tikzset{->>>>>>-/.style={decoration={
  markings,
  mark= between positions #1-0.1 and #1+0.5 step 0.1 with {\arrow{>}}
  },postaction={decorate}}}

\newcommand{\SU}{\text{SU}}
\newcommand{\SO}{\text{SO}}

\newcommand{\eg}{\textit{e.g.}\ }

\numberwithin{equation}{section}

\newcommand{\bmat}[1]{\begin{bmatrix} #1 \end{bmatrix}}
\newcommand{\be}{\begin{equation}} 
\newcommand{\ee}{\end{equation}}
\newcommand{\bea}{\begin{equation} \begin{aligned}}
\newcommand{\eea}{\end{aligned} \end{equation}}

\newcommand{\bit}{\begin{itemize}} 
\newcommand{\eit}{\end{itemize}}

\newcommand{\bM}{\mathbb{M}}

\newcommand{\Z}{\mathbb{Z}}
\newcommand{\C}{\mathbb{C}}
\newcommand{\R}{\mathbb{R}}

\newcommand{\G}{\mathbb{G}}

\renewcommand{\t}{\tilde }
\renewcommand{\d}{\partial }

\renewcommand{\b}{\bar }

\newcommand{\half}{{1\over 2}}

\newcommand{\CA}{\mathcal{A}}

\newcommand{\CH}{\mathcal{H}}
\newcommand{\CI}{\mathcal{I}}

\newcommand{\CK}{\mathcal{K}}

\newcommand{\CM}{\mathcal{M}}
\newcommand{\CN}{\mathcal{N}}
\newcommand{\CO}{\mathcal{O}}

\newcommand{\CR}{\mathcal{R}}

\newcommand{\CU}{\mathcal{U}}
\newcommand{\CV}{\mathcal{V}}
\newcommand{\CW}{\mathcal{W}}

\newcommand{\h}{\widehat}

\DeclareMathOperator{\Tr}{Tr}

\newcommand{\ov}{\over}

\newcommand{\fM}{\mathfrak{M}}

\newcommand{\KK}{D_{S^1}}

\newcommand{\IW}{\boldsymbol{\CI}}
\DeclarePairedDelimiter{\ket}{\lvert}{\rangle}%
\DeclarePairedDelimiterX\braket[2]{\langle}{\rangle}{#1\delimsize\vert\mathopen{}#2}%
 
\usepackage{multirow}

\title{ Galois Covers of Calabi--Yau Quivers \\ and BPS State Counting
}

\author{Johannes Aspman${}^1$, Cyril Closset${}^2$, Elias Furrer${}^{2,3}$, Jan Manschot\textsuperscript{\shamrock} \\
\vspace{5pt}
{${}^1\ $\it Department of Computer Science, Czech Technical University in Prague, \\ Karlovo nam. 13, Prague 2, Czech Republic \\ \vspace{5pt}
${}^2\ $ \it School of Mathematics, University of Birmingham,\\
\it Watson Building, Edgbaston, Birmingham B15 2TT, United Kingdom \\ \vspace{5pt}
${}^3\ $ \it NHETC and Department of Physics and Astronomy, Rutgers University, \\ 
126 Frelinghuysen Rd., Piscataway NJ 08855, USA \\ \vspace{5pt}
\textsuperscript{\shamrock} \it School of Mathematics and Hamilton Mathematical Institute, Trinity College, \\
Dublin 2, Ireland 
\vspace{20pt} }}
   
\abstract{BPS quivers are central to our understanding of BPS states in 4d $\mathcal{N}=2$ supersymmetric field theories and of D-branes at Calabi--Yau threefold singularities. The two subjects are deeply interrelated through geometric engineering in Type II string theory, where a CY$_3$ quiver, also known as a 5d BPS quiver, describes fractional branes at a threefold singularity ${\bf X}$. We study the Galois cover $\tilde Q\rightarrow Q$ of any BPS quiver $Q$ by a finite abelian group $\mathbb{G}$, leading to a covering quiver $\t Q$. The Galois cover is determined by a $\mathbb{G}$-grading of the arrows of the quiver $Q$, which can be understood as an orbifolding procedure. In particular, if $Q$ is a CY$_3$ quiver for ${\bf X}$, then the Galois cover $\tilde Q$ is the CY$_3$ quiver for the orbifold singularity ${\bf X}/\G$. We explore such Galois covering procedures in the language of supersymmetric quiver quantum mechanics, in terms of fixed loci under $\G$ actions on moduli spaces of quiver representations, and in terms of homomorphisms between the Kontsevich--Soibelman algebras of $Q$ and $\tilde Q$. Our main result is an explicit covering formula for the BPS invariants of 4d $\mathcal{N}=2$ field theories, wherein the rational BPS invariant $\bar{\Omega}^Q(\gamma)$ of $Q$ is expressed as a sum of BPS invariants of $\tilde Q$. We derive this formula in various special cases, which include the case when $\gamma$ is a primitive charge vector, the case of general charge vectors for quivers without loops, and the case of CY$_3$ quivers for some simple geometries such as the conifold or local del Pezzo surfaces. The general formula is presented as a conjecture that can be verified in many examples.

\vspace{30pt}\noindent\today}

\preprint{}

\begin{document}
\maketitle

\section{Introduction}

The study of BPS states in 4d $\CN=2$ supersymmetric quantum field theories (SQFT) is a rich subject at the heart of physical mathematics. The determination of the full BPS spectrum of a given $\CN=2$ SQFT, at any given point of its Coulomb branch (CB) of vacua, remains an unsolved problem in general, although many important special cases are fully understood; seminal works include~\cite{Seiberg:1994rs, Ferrari:1996sv, Klemm:1996bj, Gaiotto:2012rg}. In many instances, the BPS states are elegantly encoded in a BPS quiver ~\cite{Douglas:1996sw, Denef:2002ru, Alim:2011kw, Cecotti:2012gh, Chuang:2013wt, Closset:2019juk, Beaujard:2020sgs, Mozgovoy:2020has, DelMonte:2021ytz}. The latter is essentially a gauged supersymmetric quantum mechanics (SQM) that describes the worldline of the BPS particles --- the BPS states being the supersymmetric ground states of the SQM \cite{Denef:2002ru, Denef:2007vg, Manschot:2010qz, Manschot:2013sya}. Mathematically, the BPS states are often identified with the cohomology of moduli spaces of semi-stable quiver representations \cite{Reineke_2003, Joyce:2004tk, Kontsevich:2008fj, Joyce:2008pc, Meinhardt2019}.

In this work, we explore new relationships between BPS states of distinct theories using the algebraic concept of {\it Galois cover} first introduced in the physics literature by Cecotti and Del Zotto~\cite{Cecotti:2015qha}. A Galois covering 
\be\label{Galois F intro}
F \; :\; \t Q \longrightarrow Q
\ee
of a quiver $Q$ by a quiver $\t Q$ is a relation determined by a finite abelian group $\G$ and a $\h\G$-grading of the arrows of $Q$. The covering quiver $\t Q$ has $|\G|$ times more vertices than $Q$, and it admits a free action of $\G$ such that 
\be
Q\cong \t Q/\G~.
\ee
BPS quivers of various theories can be identified as a Galois pair --- for example, the BPS quivers of the 4d $\CN=2$ $\SU(2)$ gauge theory with $N_f=0$ and $N_f=2$ hypermultiplets are related by a $\Z_2$ Galois cover~\cite{Gaiotto:2010okc, Cecotti:2015qha}. Similarly, the BPS quivers of the $E_0$ \& $E_6$ and $E_1$ \& $E_5$ 5d superconformal field theories (SCFTs) compactified on a circle form Galois covering pairs, while the BPS quiver of the 5d $\SU(N)_0$ theory on a circle can be viewed as a $\Z_N$ cover of the conifold quiver (that is, the Klebanov--Witten quiver~\cite{Klebanov:1998hh}) --- such covering relations have been instrumental to recent progress in the study of 5d SCFTs~\cite{Closset:2019juk, Longhi:2021qvz, DelMonte:2021ytz, Acharya:2024bnt, Collinucci:2025rrh, Dramburg:2025tlb}.

One can view Galois covering as an efficient device to construct rather complicated quivers from simple ones. Since BPS states are captured by the representation theory of a BPS quiver, it follows that there should exist a notion of Galois covering of BPS spectra, which is well established for the $N_f=0$ and $N_f=2$ 4d gauge theories mentioned above, as well as for various Argyres--Douglas theories~\cite{Gaiotto:2010okc, Cecotti:2015qha}. In this article, we revisit this question with a particular emphasis on quivers without oriented loops as well as on CY$_3$ quivers. The latter are the BPS quivers that describe D-branes with compact support on a (resolved) Calabi--Yau (CY) threefold singularity --- such quivers are known as fractional-brane quivers~\cite{Douglas:1996sw, Diaconescu:1997br, Klebanov:1998hh} or as 5d BPS quivers~\cite{Closset:2019juk}, depending on the context. The Galois covering then also corresponds to an orbifold action on the threefold singularity, as we will explain. The orbifold perspective was explored in recent works on CY$_3$ quivers~\cite{Bridgeland:2024saj, Collinucci:2025rrh, Dramburg:2025tlb}.

\medskip
\noindent
In the remainder of this introduction, we explain our approach and summarise our key results. Note that our discussion is physical and somewhat heuristic; we hope to discuss more rigorous proofs of some key statements in future work. 

\subsection{BPS quivers for 4d and 5d theories, and BPS states}

Let us briefly review BPS quivers for 4d $\CN=2$ SQFTs, and especially how one extracts BPS states from them. For our purpose, we distinguish between three types of 4d $\CN=2$ field theories, with their respective BPS quivers:
\begin{itemize}
\item Proper 4d $\CN=2$ SQFTs are well-defined field theories in four dimensions, which means that they have a UV completion as a four-dimensional SCFT or as an asymptotically-free gauge theory. The BPS quivers for these theories were systematically studied in~\cite{Alim:2011ae,Alim:2011kw, Cecotti:2012gh}. 

\item Kaluza-Klein (KK) 4d $\CN=2$ SQFTs are field theories which only admit a UV completion as higher-dimensional field theories. Here we focus on KK theories which correspond to 5d SCFTs compactified on a circle%
\footnote{One could also consider 6d theories on a torus, but very few BPS quivers are known in those cases. (See~{\it e.g.}~\protect\cite{Closset:2023pmc} for an important example.)}, in which case we consider so-called 5d BPS quivers~\cite{Closset:2019juk}.
 Threefold canonical singularities ${\bf X}$ --- for instance,  consider any toric Calabi--Yau threefold singularity --- are a rich source of 4d $\CN=2$ KK theories. Their 5d BPS quivers arise as so-called non-commutative crepant resolutions (NCCR) of ${\bf X}$, better known in string theory as fractional-brane quivers~\cite{Douglas:2000qw, Herzog:2003zc, Herzog:2005sy, Beaujard:2020sgs}. 

\item A third type of 4d $\CN=2$ field theories are supergravities. They arise, for instance, by compactifying Type II string theory on a compact Calabi--Yau threefold. In such cases, the BPS states include multi-centred black holes in space-time. Many aspects of their physics are captured by BPS quivers \cite{Denef:2002ru, Denef:2007vg, Manschot:2010qz, Bena:2012hf, Manschot:2012rx}, whose number of nodes and arrows are determined by the constituents of the bound state. The Coulomb branch formula \cite{Manschot:2012rx, Manschot:2014fua} for BPS invariants was originally developed using the gravitational perspective. While this work focuses on non-gravitational field theories, we will use that latter formula extensively in section \ref{sec:GaloisExamples}.

\end{itemize}

\medskip
\noindent
{\bf BPS states from BPS quivers.} Consider a 4d $\CN=2$ SQFT of rank $r$ (that is, $r$ is its Coulomb Branch dimension) and with $f$ the rank of its flavour symmetry group.%
\footnote{For KK theories, this always includes a Kaluza-Klein symmetry ${\rm U}(1)_{\rm KK}$, corresponding to five-dimensional momentum along the compactification circle.} We then have an extended Coulomb branch (CB) $\fM_{\rm ECB}$ of dimension $r+f$, consisting of VEVs for CB operators together with complex masses for the flavour symmetry. Any BPS excitation on $\fM_{\rm ECB}$  is determined by its electromagnetic and flavour charge
\be\label{Gamma in intro}
\gamma \in \Gamma \cong \Z^{2r+f}~.
\ee
The charge lattice $\Gamma$ admits an anti-symmetric pairing
\be
\label{eq:asymmpairing}
\Gamma \times \Gamma \rightarrow \Z \; : \; (\gamma, \gamma') \mapsto \langle \gamma, \gamma'\rangle~,
\ee
which is the standard Dirac pairing on electromagnetic charges --- note that the charge lattice includes flavour charges, which are the ones in the kernel of $\langle \gamma, -\rangle$, $\forall \gamma$. Finally, let us recall that the mass of a BPS particle at any given point $u\in \fM_{\rm ECB}$ is determined by the central charge function
\be
Z(u) \; : \; \Gamma \rightarrow \C \; : \; \gamma \mapsto Z_\gamma(u)~, 
\ee
which is known, in principle, in terms of the periods of the Seiberg--Witten (SW) geometry. In the case of 5d SCFTs defined by a singularity ${\bf X}$, the SW geometry is the family of local threefolds mirror to the quantum K\"ahler cone of ${\bf X}$ --- see {\it e.g.}~\cite{Chiang:1999tz, Closset:2021lhd}.

Mathematically, a quiver is simply a directed graph,  as we will review in detail in section~\ref{sec: 5d BPS quiver and Galois covers}.  Our BPS quivers are always quivers with superpotential,
\be
Q= (Q_0, Q_1, W)~,
\ee
where $Q_0$ is the set of vertices (also called nodes), $Q_1$ is the set of arrows between vertices, and $W$ is a linear combination of closed paths over $\C$ which generate quiver relations. 
The number of quiver nodes equals the dimension of the BPS charge lattice,
\be
|Q_0| = {\rm rank}(\Gamma) = 2r +f~.
\ee
The nodes correspond to a particular basis of $\Gamma$, called a quiver basis, which we denote by $\{\gamma_j\}$. Any charge $\gamma$ can then be expanded in terms of the quiver ranks $N_j$ as
\be\label{gamma to N intro}
\gamma= \sum_{j\in Q_0} N_j \gamma_j~.
\ee
At a given point on the extended CB, the BPS spectrum of the 4d $\CN=2$ SQFT admits a quiver description only if $N_j \geq 0$ ($\forall j$) for every BPS particle.%
\footnote{As opposed to anti-BPS particles. The full set of conditions for having a BPS quiver includes the existence of a finite number of hypermultiplets with central charges sitting in a strict half-plane and which form a basis for the full BPS spectrum~\protect\cite{Alim:2011kw}.} The net number of arrows from node $j$ to node $k$ is given in terms of the anti-symmetric pairing (\ref{eq:asymmpairing}) by:
\be
B_{jk} = \langle \gamma_j, \gamma_k\rangle~.
\ee
This $B$ is therefore the antisymmetrised incidence matrix of the quiver.

\medskip
\noindent
Assuming a quiver description exists, one reconstructs the BPS states with fixed   charge vector~\eqref{gamma to N intro} as follows:
\begin{enumerate}
    \item Consider the gauged $\CN=4$ SQM (also known as 1d GLSM) with unitary gauge group
    \be
G_\gamma = \prod_{j\in Q_0} {\rm U}(N_j)~,
    \ee
    with bifundamental chiral multiplets $\Phi_a$ associated to every arrow $a\in Q_1$. The abstract superpotential $W$ becomes a gauge-invariant superpotential in the chiral multiplets. Crucially, we turn on Fayet--Iliopoulos (FI) parameters $\zeta_j$ for each non-trivial ${\rm U}(N_j)$ gauge group, with the constraint 
    \be
    \sum_{j\in Q_0} \zeta_j N_j =0~. 
    \ee
    The FI parameters are fully determined by the central charges $Z_{\gamma_j}(u)$. Essentially, the BPS states with charge $\gamma$ are the supersymmetric ground states of this SQM.
    
    \item To compute these BPS states,  consider a rescaling of the $\zeta_i$'s that brings us to the Higgs phase of the $\CN=4$ SQM. In the absence of a superpotential, for simplicity, the Higgs branch of the SQM is the K\"ahler quotient 
    \be
       \CM_\gamma \cong M_\gamma  \sslash_\zeta \, G_\gamma^0~, \qquad\qquad M_\gamma \cong \bigoplus_{(a:j\rightarrow k)\in Q_1} \C^{N_j N_k}~,
    \ee
   with $M_\gamma$ the vector space spanned by the VEVs of the chiral multiplets, and where $G_\gamma^0 \equiv G_\gamma/{\rm U}(1)$ is the effectively acting gauge group. Equivalently, one constructs $\CM_\gamma$ as a GIT quotient describing the moduli space of semi-stable quiver representations, with  $\zeta$ determining the stability condition. 

   \item The supersymmetric ground states $\ket{\Psi; \gamma}$ are then obtained as the cohomology of the moduli space,
   \be\label{cohMgamma intro}
\ket{\Psi; \gamma}\in H^\bullet( \CM_\gamma)~,
   \ee
   as usual in any SQM~\cite{Witten:1982im}. Whenever the moduli space is a smooth compact K\"ahler manifold, this is simply its de Rham cohomology; more generally, $\CM_\gamma$ will be quite singular and some more general notion should be used --- for instance, intersection cohomology \cite{Meinhardt2019}. From a physics perspective, it is important to keep in mind that we are looking for the supersymmetric ground states, which is what determines the correct cohomology in principle.%
   \footnote{For nice enough singularities in target space, we would impose some $L^2$-normalisability condition of the wavefunction near the singular loci, and this should give us something akin to intersection cohomology.} In particular, the Witten index $\CI$ of the $\CN=4$ SQM is well-defined whenever the BPS spectrum is well-defined, and gives us the BPS invariant $\Omega(\gamma,\zeta)$ at charge $\gamma$. In the case of a smooth moduli space, it is simply given by the topological Euler characteristic:%
   \footnote{Up to some overall sign which is a matter of convention, and which we will discuss later.} 
   \be\label{Omega as chi intro}
 \IW[\text{$\CN=4$ SQM}] = \Omega(\gamma,\zeta) =(-1)^{\rm dim_\C(\CM_\gamma)} \chi(\CM_\gamma)~.
   \ee
   One can refine the Witten index by softly breaking the worldline supersymmetry from $\CN=4$ to $\CN=2$, with $y\neq 1$ the exponentiated soft mass, which gives us the refined BPS index $\Omega(\gamma,\zeta; y)$ that keeps track of the 4d spin of the massive BPS particles. For smooth moduli spaces, this computes the $\chi_{\bf y}$ genus (with ${\bf y}=-y^2$), namely:
     \be
 \CI[\text{$\CN=2$ SQM}] = \Omega(\gamma,\zeta; y) \propto \chi_{-y^2}(\CM_\gamma) = \sum_{p,q} (-1)^{p+q} y^{2q} h^{p,q}( \CM_\gamma)~,
   \ee
   with a $y$-dependent proportionality factor that we will recall in the main text. 
In particular, the 4d spin is determined by the Hodge numbers of the moduli space.%
\footnote{As we will review in the main text, this discussion assumes the no-exotic conjecture holds.}
\end{enumerate}

\noindent
Many tools have been developed to compute BPS indices. 
Some of the most efficient ones compute the Witten index of the SQM in a Coulomb phase \cite{Denef:2002ru, Denef:2007vg, Manschot:2011xc, Kim:2011sc, Manschot:2013sya, Hori:2014tda, Beaujard:2019pkn}, while the Higgs-phase approach \cite{Reineke_2003, Lee:2013yka, Duan:2020qjy, Longhi:2021qvz, DelMonte:2021ytz} amounts to computing some Euler characteristic of the target space as sketched above. It is this latter computation that has the simplest mathematical underpinning, since it amounts to studying moduli spaces of quiver representations. It is also in this geometric phase that we will be able to better understand Galois coverings.

\subsection{Galois covering formulas for rational BPS invariants}
We put forward a striking formula for the Galois covering of BPS indices, namely that the rational BPS invariant $\bar \Omega^Q(\gamma)$ for $Q$ --- defined in~\eqref{eq:DefRatInv} in the main text --- equals a sum of rational BPS invariants $\bar \Omega^{\tilde Q}(\tilde \gamma)$ of $\t Q$ for all the charges $\tilde \gamma$ with image $\gamma$ under the so-called push-down functor $F_\ast$ defined in~\eqref{eq:Flambdagamma}, divided by $|\G|$:
\be
\label{eq:CoverBPSInv0}
\bar \Omega^Q(\gamma,\zeta) ={1\ov |\G|} \sum_{\t\gamma\, |\,
  F_\ast(\t\gamma)=\gamma} \xi(\gamma,\tilde \gamma)\,\bar
\Omega^{\t Q}(\t\gamma,\tilde \zeta)~.
\ee
Here $\xi(\gamma, \t\gamma)\in \pm 1$ is a sign defined in~\eqref{eq:xigtg}. Since a given quiver $Q$ has infinitely many Galois covers, this relation also implies infinitely many relations between different covers of $Q$. 

We prove this formula in many cases using complementary approaches. In section~\ref{sec:gauging and ungauging}, we first arrive at~\eqref{eq:CoverBPSInv0} when $\gamma$ is a primitive charge --- that is, for $\bar\Omega^Q(\gamma)= \Omega^Q(\gamma)$ --- by viewing the Galois covering map as discrete gauging of $\t Q$ by $\G$. In section~\ref{subsec:circle action}, in favourable circumstances, we relate this formula to Atiyah--Bott--Kirwan localisation on the moduli space $\CM_\gamma$. 
 Finally, in section~\ref{sec:AlgHoms}, we derive the formula in many cases using a 
 homomorphism between the Kontsevitch--Soibelman (KS) Lie algebras of $\tilde Q$ and $Q$ --- the definition of the KS algebra is reviewed in section~\ref{sec:KSMonO} below. More precisely, the surjective homomorphism $\mathfrak{f}^{\G}_*$ \eqref{eq:fGlam} from the $\G$-invariant graded Lie algebra $\frak{g}^\G_{\tilde Q}$ to $\frak{g}_{Q}$ implies that the BPS invariants satisfy~\eqref{eq:CoverBPSInv0}, if this relation is satisfied for at least one choice of $\zeta$. By choosing $\zeta$ equal to attractor stability $\zeta_*$ \cite{Manschot:2013, Alexandrov:2018iao, Beaujard:2020sgs} or self-stability \cite{Bridgeland:2017}, this can be established for quivers without loops, as well as for specific CY$_3$ quivers with potentials for which attractor invariants are known.

We also introduce the notion of a {\it symmetric Galois cover} in section~\ref{subsec:symgalcov} as covers for which the gradings assigned to the arrows of $Q$ are invariant under $\G$. Such covers have the benefit that their BPS invariants $\Omega^{\tilde Q}(\tilde \gamma,\tilde \zeta)$ are unchanged upon a small deformation of $\tilde \zeta$ away from $F^\ast(\zeta)$. Moreover, we also establish for such covers a relation between non-commutative tori, which allows us to obtain a relation between refined BPS invariants $\Omega^{\tilde Q}(\tilde \gamma, \tilde \zeta;\tilde y)$ and $\Omega^Q(\gamma,\zeta;y)$. To this end, we apply the Coulomb Branch Formula \cite{Manschot:2011xc, Manschot:2012rx, Manschot:2013dua, Manschot:2014fua} and assign non-trivial single-centred indices $\Omega_{S}^{\t Q}$ to the vertices of $\tilde Q$ as in~\eqref{eq:assignOmS}. The invariants of both sides then satisfy
\be
\label{eq:OmQOmtQy}
\bar \Omega^Q(\gamma,\zeta;y)= \frac{y-y^{-1}}{y^{|\G|}-y^{-|\G|}} \sum_{\tilde \gamma\, |\,
  F_\ast(\tilde \gamma)=\gamma} \xi(\gamma,\tilde \gamma) \,\bar \Omega^{\tilde Q}(\tilde \gamma, \tilde \zeta;\tilde y)~,\qquad \tilde y=y^{|\G|}~.
\ee
Note that, due to the assignment of non-trivial $\Omega_{S}^{\tilde Q}$, the invariants $\bar \Omega_{S}^{\tilde Q}$ differ from the ``standard" ones, which correspond to single-centred invariants equal to 1~\cite{Manschot:2012rx}. 

Formulas \eqref{eq:CoverBPSInv0} and \eqref{eq:OmQOmtQy} have similarities with the abelianisation formula of~\cite{Manschot:2010qz}, which also expresses a BPS invariant in terms of a sum of other BPS invariants. The main difference between the two formulas is that the quiver $\tilde Q$ is fixed in the formulas above, whereas the abelianisation formula expresses $\Omega^Q(N,\zeta)$ as a sum $\sum_{Q'}$ of BPS invariants $\Omega_{Q'}(N',\zeta')$  of abelian representations $N'=(1,1,1,\dots)$, at the cost that the quivers $Q'$ have a larger (or equal) number of arrows and larger number of nodes than $Q$. While the formulas are quite different, there is also some overlap between the equivariant localisation discussion of section~\ref{sec:GaloisCoverFixedLoci} and the mathematical proof of the abelianisation formula~\cite{Reineke2009, Reineke_2012, Weist2013}.

\subsection{Future research directions}

We conclude this introduction by mentioning a few challenges for future work.

\begin{itemize}

\item In this paper, we focus on Galois covers for cyclic groups $\mathbb G=\mathbb Z_n$ acting freely on the nodes of $\t Q$. A first extension beyond this class is to consider Galois covers for which $\G$ is a non-Abelian finite group. This has been recently studied in the context of the generalised McKay correspondence for 5d BPS quivers~\cite{Collinucci:2025rrh, Dramburg:2025tlb}.

\item  A further avenue for future work would be to generalise covers for which $\mathbb G$ does not act freely. There are many important examples, including the BPS quivers for  4d $\CN=2$ $\SU(2)$ SQCD with $N_f=2$ and $N_f=3$ massless flavours. The latter has a $S_4$ symmetry with a fixed point, and so our result~\eqref{eq:CoverBPSInv0} does not, a priori, apply. From the mathematical point of view, the Galois functor has been generalised in various directions, including lifting the requirement of the free action~\cite{cibils2006skew,keller2005triangulated,asashiba2011generalization,asashiba2017generalization}. In section~\ref{sec:gauging and ungauging}, we explain how $\mathbb G$ can be thought of as a discrete symmetry of the SQM, and the quotient by $\mathbb G$ as a discrete gauging. If a node in $\t Q$ is fixed under $\mathbb G$, it appears as if it were an orbit under $\mathbb G$, that is, as if it were obtained by gauging. Thus, the generalisation of quotients by non-freely acting symmetries would be a simultaneous gauging and `ungauging' in the SQM.

\item Galois covers between $\CN=2$ theories of rank one are closely related to isogenies of their Seiberg--Witten curve. Isogenies are particular quotients of an elliptic surface and are generated by the Mordell--Weil group of the latter. A notable example is the isogeny between the SW curves of the pure $\SU(2)$ theory and the $\SU(2)$ theory with $N_f=2$, whose quivers form a Galois pair. 
For the four rank-one Galois covers described in Section~\ref{sec:galois cover}, the isogenies can be made fully explicit~\cite{Closset:2021lhd,Closset:2023pmc,Furrer:2024zzu}. There are however many more isogenies than Galois covers, which is again due to the symmetry not acting freely.

\item The KS algebra $\mathfrak{g}_Q$ is realised in $\CN=2$ quantum field theory on $\mathbb{R}^3\times S^1$ as the algebra of symplectomorphisms of a complexified torus \cite{Gaiotto:2010okc}. It will be interesting to connect the homomorphisms between graded Lie algebras associated to $Q$ and $\tilde Q$ of section~\ref{sec:AlgHoms} to that context. Closely related results were already obtained by Del Monte~\cite{DelMonte:2023vwv}, who constructed exact solutions to the TBA equations of~\cite{Gaiotto:2010okc} using precisely the fine-tuned stability conditions discussed in this paper.

\item Furthermore, the Schur and Macdonald indices of 4d $\CN=2$ SCFTs can be obtained as a trace of the KS monodromy $\mathbb{M}_Q$ \cite{Cordova:2015nma} and its refinement $\mathbb{M}_Q(y)$~\cite{Cecotti:2015lab, Andrews:2025tko, Kim:2025klh}. Since the homomorphisms between the Lie algebras relate the monodromies of $Q$ and $\t Q$, they imply in turn relations between Schur and Macdonald indices of the corresponding quantum field theories. This will be interesting to study more explicitly. 

\end{itemize}

\subsection*{Acknowledgements}
We thank Fabrizio Del Monte, Sergey Mozgovoy, Boris Pioline, Richard Thomas, and Graeme Wilkin  for useful discussions. We thank Horia Magureanu for collaboration in the early stage of this project. The work of CC is supported by a University Research Fellowship of the Royal Society.

\section{BPS quivers}
\label{sec: 5d BPS quiver and Galois covers}

In this section, we review relevant aspects of BPS quivers. We start by setting up our (standard) notation for quivers, viewed as abstract mathematical structures. We then review how BPS states and BPS invariants are extracted from BPS quivers, and briefly review aspects of the KS wall-crossing formula.

\subsection{Quivers and moduli spaces}
A quiver $Q=\{Q_0,Q_1,h,t\}$ consists of a collection of vertices (also called nodes) $Q_0$, a collection of oriented arrows $Q_1$ with a head
and a tail, and two maps $h: Q_1\to Q_0$ and $t: Q_1 \to Q_0$ which
assign the heads and the tails of arrows to vertices. We denote by $j\in Q_0$
the vertices of $Q$, and by $a\in Q_1$ the arrows. Hence for a given arrow $a:j\rightarrow j'$ in the quiver, we have $h(a)=j'$ and $t(a)=j$. 
 A quiver $Q$ is said to be connected if, for any decomposition $Q_0=Q_0' \cup
Q_0''$ with $Q_0',Q_0''\neq \emptyset$, there is an $a\in Q_1$ such
that $h(a)\in Q_0'$, $t(a)\in Q_0''$ or $t(a)\in Q_0'$, $h(a)\in
Q_0''$. Unless otherwise stated, we will always consider connected quivers.

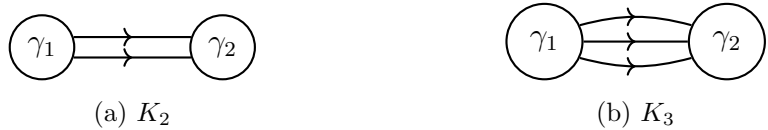
\begin{figure}[t]
\centering
\begin{subfigure}[b]{0.4\textwidth}
\centering
\begin{tikzpicture}[baseline=1mm, every node/.style={circle,draw},thick]
\node[] (1) []{$\gamma_1$};
\node[] (2) [right =1.5cm of 1]{$\gamma_2$};
\draw[->-=0.5] (1.18) to   (2.162); 
\draw[->-=0.5] (1.342) to   (2.198);
\end{tikzpicture}
\caption{$K_2$}
\label{QuiverK2}
\end{subfigure}
\hspace{.3cm}
\begin{subfigure}[b]{0.4\textwidth}
\centering
\begin{tikzpicture}[baseline=1mm, every node/.style={circle, draw, minimum size=1cm}, thick]
\node[] (1) at (0,-0.65){$\gamma_{1}$};
\node[] (2) at (2.4,-0.65){$\gamma_{2}$};
\draw[->-=0.5] (1.25)  to[bend left=15]  (2.155);
\draw[->-=0.5] (1.0)   to                (2.180);
\draw[->-=0.5] (1.335) to[bend right=15] (2.205);
\end{tikzpicture}
\caption{$K_3$}
\label{QuiverK3}
\end{subfigure}
\caption{The Kronecker quivers $K_n$ for $n=2,3$. The notation $\gamma_j$ is explained in the main text.}
\label{QuiverKn}
\end{figure}

A classic example of a connected quiver is the (generalised) Kronecker quiver $K_n$, {\it i.e.} the quiver with two vertices and $n$ arrows from vertex 1 to vertex 2 --- see figure~\ref{QuiverKn}.

\medskip
\noindent
{\bf Path algebra and superpotential.} Associated to a quiver $Q$ as defined above, we have the path algebra $\C Q$ generated by the arrows $a$ and by idempotent elements $e_j$ associated to the vertices, which consists of all paths obtained by concatenation. That is, we have the non-zero product $a b$ if $h(a)= t(b)$  while $e_j a e_k= a$ if $h(a)=j$ and $t(a)=k$.

If a subset of arrows of the quiver form an oriented path $p= a_{l_1} a_{l_2}\cdots a_{l_{|p|}}$ in $\C Q$, with $t(a_{l_s})= h(a_{l_{s-1}})$ and $l_0\equiv l_{|p|}$, one can introduce a non-trivial superpotential:
\be\label{def W on paths}
W = \sum_p C_p\,  \Tr\left(a_{s_1} a_{s_2}\cdots a_{s_{|p|}}\right)~,
\ee
where we sum over all closed paths with coefficients $C_p\in \C$; here the trace indicates that a closed path is defined modulo cyclic permutations. The superpotential gives us relations amongst paths, which are obtained by formal derivation along arrows, $\d_a W$, $\forall a\in Q_1$. This defines an ideal $(\d W)$ of the free associative algebra $\C Q$, and the path algebra of the quiver with superpotential $(Q,W)$ is defined as
\be\label{def AQ}
\CA_Q\equiv \C Q/(\d W)~.
\ee

\medskip
\noindent
{\bf Quiver representations.} A quiver representation $R=(\{V_j\}, \{X_a\})$ of $Q$ is a collection of vector spaces $V_j\cong \mathbb{C}^{N_j(R)}$ assigned to each $j\in Q_0$ together with a collection of linear maps%
\be\label{Xa def rep}
X_a: V_{t(a)}\to V_{h(a)}
\ee
assigned to each $a\in Q_1$. Note that we follow the physics convention for the composition of arrows, which means that the linear map~\eqref{Xa def rep} acts on vectors $v\in V_{{t(a)}}$ from the right, {\it i.e.}~$v\mapsto v X_a \in V_{h(a)}$ as a matrix multiplication. Then, any non-zero product $a b$ in the path algebra is represented by the matrix multiplication $X_{ab}=X_a X_b$. In particular, the superpotential~\eqref{def W on paths} is now represented by:
\be
\label{eq:Qpot}
W(R)=\sum_{p} C_{p}\,{\rm Tr}\left(X_{a_{s_1}}X_{a_{s_2}}\dots X_{a_{s_{|p|}}}\right)~,
\ee
A representation of $(Q,W)$ must satisfy the $F$-term relations:
\be
 \frac{\partial W(R)}{\partial X_a}=0,\qquad  a\in Q_1~.
\ee
Note that quiver representations are in one-to-one correspondence with right-$\CA_Q$ modules $V= \oplus_{j\in Q_0} V_j$, with the path algebra acting from the right as we just explained. 
A representation $R'=(\{W_j\},\{Y_a\})$ is a subrepresentation of $R$ if $W_j \subseteq
V_j$ for all $j$, and if for every $a\in Q_1$ we have the restriction
$X_a\vert_{W_j}=Y_a$. Let us also point out that quiver representations naturally form the objects of a category ${\bf Rep}(Q)$, where the morphisms $\varphi : R\rightarrow R'$ as linear maps $\varphi_j : V_j\rightarrow V_j'$ that commute with the arrows (that is $\varphi_j X_a' =X_a \varphi_k$ for all arrow $a:j\rightarrow k$); equivalently, this is the category of right-$\CA_Q$ modules.

\medskip
\noindent
{\bf Charge lattice and dimension vectors.} 
Any representation is partially characterised by its dimension vector, the vector of non-negative integers:
\be
N(R) = (N_1(R),\dots, N_{|Q_0|}(R))~.
\ee
To any quiver $Q$, we may assign the {\it charge lattice}:
\be\label{def Gamma Q}
\Gamma \cong \Z^{|Q_0|} = \Z\langle \gamma_1, \cdots, \gamma_{|Q_0|}\rangle~,
\ee
where we assign a basis vector $\gamma_j$ to each quiver node $j\in Q_0$. We then expand any vector of the charge lattice as
\be\label{gamma to N}
\gamma= \sum_{j\in Q_0} N_j \gamma_j~.
\ee 
If we assume that $N_j\geq 0$ for all $j$, we can interpret $N=(N_j)$ as the dimension vector of some representation $R$. Indeed, a natural question is to find all representations with a fixed dimension vector $N=(N_j)$. We will use the notation $\gamma$ and $N$ interchangeably to denote dimension vectors.

\medskip
\noindent
{\bf Stability conditions and moduli spaces of quiver representations.} 
 To construct proper moduli spaces of quiver representations, we need to introduce a notion of stability. Let us introduce the stability vector $\zeta\in \Gamma^\ast \otimes \mathbb{R}$, namely:
 \be
 \zeta=(\zeta_1,\dots ,\zeta_{|Q_0|})\in
\mathbb{R}^{|Q_0|}~.
\ee  
The slope $\mu(R,\zeta)$ of a representation $R$
with dimension vector $N(R)$ is defined as:
\be\label{def slope}
\mu(R,\zeta)=\frac{\sum_j \zeta_j\,  N_j(R)}{\sum_{j}N_j(R)}~.
\ee
A representation $R$ is stable (resp., semi-stable) with respect to $\zeta$ if, for every
subrepresentation $R'\subsetneq R$, we have $\mu(R',\zeta)<\mu(R,\zeta)$ (resp., $\mu(R',\zeta)\leq \mu(R,\zeta)$).

Consider a dimension vector $\gamma\equiv \sum_j N_j \gamma_j$ and the stability vector $\zeta$; without loss of generality, one can also assume that $\zeta(\gamma)\equiv \sum_{j\in Q_0} \zeta_j N_j=0$. The moduli space of semi-stable representations is well-studied~\cite{King:1994, Reineke_2003, Meinhardt2019} --- it is obtained as the GIT quotient:%
\footnote{We shall use both notations $\CM^Q(N, \zeta)$ and $\CM_\gamma^Q(\gamma)$  for~\protect\eqref{CMgamma GIT def}, and we may omit the superscript $Q$ or the stability vector $\zeta$ whenever they are clear from the context.}
\be\label{CMgamma GIT def}
\CM^Q_\gamma(\zeta) = {\rm Rep}_\gamma \sslash_\zeta  GL_\gamma^0 ~,
\ee
where we defined
\be
 {\rm Rep}_\gamma \equiv \bigoplus_{a \in Q_1} {\rm Hom}(V_{t(a)}, V_{h(a)})\Big/(\d W)~, \quad \qquad GL_\gamma^0 \equiv GL_\gamma/GL(1)~.
\ee
Namely, ${\rm Rep}_\gamma$ is the set of all quiver representations with fixed dimension vector~$\gamma$, while $GL_\gamma\equiv \prod_{i\in Q_0} GL(N_i)$ is the gauge group acting on the representations as:
\be\label{G action on V and X}
V_j \rightarrow  V_j  G_j~, \qquad X_a \rightarrow   G_{t(a)}^{-1} X_a G_{h(a)}~,
\ee
and $GL_\gamma^0$ is the effectively acting gauge group (with $GL(1)$ being the overall rescaling $G_j = \lambda \in \C^\ast$ which acts trivially on arrows). The quotient~\eqref{CMgamma GIT def} gives us the set of all gauge orbits of semi-stable representations. If $\gamma$ is a primitive vector and $Q$ has no oriented loops (and hence no superpotential), the moduli space $\CM^Q_\gamma(\zeta)$ is smooth and compact. More generally, $\CM^Q_\gamma(\zeta)$ may have singularities.

For any quiver without a superpotential, let us define the {\it virtual dimension} of the moduli space~\eqref{CMgamma GIT def} as:
\be
\label{eq:dimM}
\dim_\mathbb{C}\, \CM^Q_\gamma(\zeta)=1-(N,N)~,
\ee 
where $(N,N')$ is the Ringel--Tits form
\be
(N,N')=\sum_{j\in Q_0} N_j N_j'-\sum_{a\in Q_1} N_{t(a)}N'_{h(a)}~.
\ee 
The virtual dimension is simply the dimension of the vector space ${\rm Rep}_\gamma$ (with trivial $W$) minus the dimension of the gauge group. Depending on the stability vector, the moduli space is either empty (if no semi-stable representations exist) or its dimension is equal to the virtual dimension.

In the presence of a superpotential, the dimension of $\CM^Q(N,\zeta)$ depends both on the superpotential and on the stability parameters.  If the superpotential \eqref{eq:Qpot}
is generic, the expression \eqref{eq:dimM} for the dimension is expected to hold modulo 2~\cite{Denef:2007vg, Manschot:2012rx}. This fact will be useful for us. 
Let us also introduce the anti-symmetric form:
\be
\label{eq:antiNN}
\langle N,N' \rangle\equiv (N',N)-(N,N')~.
\ee

\subsection{BPS invariants from quiver quantum mechanics}

As already reviewed in the introduction, a {\it BPS quiver} for some 4d $\CN=2$ SQFT (including KK theories) is a quiver whose nodes $j\in Q_0$ correspond to `simple' objects or BPS states --- that is, stable BPS particles which are massive hypermultiplets --- of charges $\gamma_j\in \Gamma$, at a given point on the Coulomb branch. The charge lattice~\eqref{def Gamma Q} is then identified with the electromagnetic (and flavour) charge lattice introduced in~\eqref{Gamma in intro}. The Dirac--Schwinger--Zwanziger anti-symmetric inner product $\langle \gamma,\gamma'\rangle_{\rm DSZ}$ on $\Gamma$ equals the anti-symmetric product~\eqref{eq:antiNN} for dimension vectors $N$ and $N'$ if $\gamma$ ($\gamma'$) and $N$ ($N'$) are related as in~\eqref{gamma to N}. We thus suppress the subscript `DSZ' in the following, $\langle \gamma,\gamma'\rangle_{\rm DSZ}\equiv\langle \gamma,\gamma'\rangle = \langle N, N' \rangle$. 
The 4d $\CN=2$ SQFT also provides us with the central charge function
\be
Z(u) \; : \Gamma \to \mathbb{C}\; :\; \gamma \mapsto Z_\gamma(u)~,
\ee
at any point $u\in \fM_{\rm ECB}$ on the extended Coulomb branch --- as captured by the periods of the Seiberg--Witten geometry (equivalently, the IIB mirror geometry of IIA geometric engineering). At a {\it quiver point} $u_Q$, stable hypermultiplets with charges $\gamma_j$ have central charges $Z_{\gamma_j}(u_Q)$ that (approximately) align, and such that all BPS states can be obtained as bound states of the simple objects, with charges~\eqref{gamma to N}. The quiver description of the BPS spectrum is expected to be valid in some open set around a quiver point. 
The existence (or not) of quiver points (or loci) for a given SQFT is a hard question, but there are many known cases in which we can find BPS quivers by indirect arguments~\cite{Alim:2011ae}, not least in the case of 4d $\CN=2$ KK theories from three-fold singularities in Type IIA string theory (see {\it e.g.}~\cite{Closset:2019juk} and references therein). 

Given a BPS quiver $Q$, we may consider any fixed dimension vector $N$ --- that is, a charge $\gamma=\sum_{j\in Q_0} N_j \gamma_j$ --- and ask whether there exist stable BPS states with that chosen charge. A choice of $\gamma$ determines for us a gauged {\it supersymmetric quantum mechanics} (SQM), wherein we reinterpret each quiver node $j\in Q_0$ as a ${\rm U}(N_j)$ vector multiplet for 1d $\CN=4$ supersymmetry, and each arrow $a: j\rightarrow k$ as a 1d $\CN=4$ chiral multiplet in the fundamental of ${\rm U}(N_j)$ and the antifundamental of ${\rm U}(N_k)$.  We then have the 1d gauge group:
\be
G_\gamma \equiv \prod_{j\in Q_0} {\rm U}(N_j)~, \qquad G_\gamma^0\equiv G_\gamma/{\rm U}(1)~,
\ee
where $G_\gamma^0$ is the effectively acting gauge group. Note that the $GL_\gamma$ gauge group introduced above is the complexification of $G_\gamma$.  The stability parameters $\zeta_j$ are the Fayet-Iliopoulos (FI) parameters for the gauge groups ${\rm U}(N_j)$. 
 They are determined in terms of the central charge $Z=Z(u_Q)$ by:
\be
\label{eq:zetaZ}
\zeta_{j}=\lambda\, {\rm Im}\left(Z_{\gamma_j}\overline{ Z_\gamma}\right)~,
\ee 
where $\lambda \in \mathbb{R}^+$ can be chosen arbitrarily.

This quiver SQM is the 1d gauge-theory description of the worldline theory of a BPS particle --- such descriptions often arise from wrapped D-branes in Type II string theory. The actual BPS states are the supersymmetric ground states of this SQM. In the Higgs phase (that is, typically at large FI parameters), one can find the ground states semi-classically in two steps. First, we consider the classical Higgs branch of the $\CN=4$ SQM, which is the solution to the classical $D$-term and $F$-term equations moduli gauge transformations, and which gives us exactly the quiver moduli space~\eqref{CMgamma GIT def} --- the classical Higgs branch is naturally presented as a K\"ahler ({\it i.e.} symplectic) quotient by $G_\gamma$, which is famously equivalent to a GIT quotient by $GL_\gamma$ (by the Kempf--Ness theorem). Secondly, scaling $\lambda \rightarrow \infty$ in~\eqref{eq:zetaZ},  we should view the moduli space as the target space of a $\CN=4$ non-linear sigma model, and therefore the Hilbert space of supersymmetric ground states is obtained as the target space cohomology~\cite{Witten:1982im}:
\be\label{BPS as coho}
 \CH^Q_{\rm BPS}(\gamma, \zeta) \cong H^\bullet( \CM^Q_\gamma(\zeta))~,
\ee
as anticipated in~\eqref{cohMgamma intro}. More precisely, we can consider the de Rham cohomology if $\CM_\gamma^Q$ is smooth and compact, while more generally we might need to consider some appropriate cohomology related to finding the $L^2$-normalisable ground-state wave functions on the target --- following most of the physics literature, we will allow ourselves to remain glib on that important point.

\medskip
\noindent
{\bf BPS invariants as Witten indices.} The basic quantity of interest for us is the BPS invariant $\Omega^Q(\gamma; \zeta)$.  From the point of view of the four-dimensional field theory, it is defined as~\cite{Gaiotto:2010be}:
\be
\Omega(\gamma; \zeta)=-\frac{1}{2} {\rm Tr}_{\CH_\gamma}\left[ (2J_3)^2 (-1)^{2J_3}\right],
\ee
where $\CH_\gamma$ is the one-particle Hilbert space of states with electric-magnetic charge $\gamma$, and $J_3$ is the  generator of the $\SO(3)$ rotation group. In favourable circumstances, the BPS invariant is obtained physically as the Witten index of the quiver $\CN=4$ SQM at fixed $\gamma$, at least under favourable circumstances. For any quiver $\CN=4$ SQM at fixed dimension vector $\gamma$, we can define the Witten index
\be\label{def CI Witten}
\CI_\gamma^Q = \lim_{\beta \rightarrow \infty} \Tr\big((-1)^{\rm F} e^{-\beta H}\big)~,
\ee
where the trace is over the full SQM Hilbert space. This 1d Witten index can be subtle to compute whenever the SQM spectrum is not gapped --- for instance when the target space has non-compact directions and/or singularities. See~{\it e.g.}~\cite{Hori:2014tda,Lee:2017lfw} for a detailed discussion. This generally happens for a non-primitive dimension vector $\gamma$ (and/or for some particular choices of stability), wherein the SQM could encode contributions from multi-particle states (and/or marginally bound states at threshold). On the other hand, if the SQM has a smooth compact target space, we do have the equality
\be\label{Z equal Omega}
\CI_\gamma^Q = \Omega(\gamma; \zeta)~,
\ee
up to an overall sign which is a matter of convention. 
This interpretation of $\Omega$ as a 1d Witten index will be particularly important in this work, but it is equally important to distinguish between the two concepts in general.

More generally, we consider the refined BPS invariant $\Omega^Q(\gamma, y;  \zeta)$. In the same circumstances in which~\eqref{Z equal Omega} holds, the refined invariant can be computed as the refined $\CN=2^\ast$ Witten index obtained by softly breaking the $\CN=4$ supersymmetry to $\CN=2$ by an $R$-symmetry fugacity $y$ (with $y=1$ the $\CN=4$ limit). This index can be computed by supersymmetric localisation~\cite{Hori:2014tda}. 
There are various other approaches to  the evaluation of the BPS invariants of quivers, such as Reineke's formula for quivers without
superpotential \cite{Reineke_2003}. A powerful physical approach is the Manschot--Pioline--Sen (MPS) Coulomb
branch formula~\cite{Manschot:2010qz, Manschot:2011xc,
  Manschot:2012rx, Manschot:2013sya, Manschot:2014fua}, which
expresses the BPS invariants in terms of ``single centre invariants''; it is also directly connected to the supersymmetric localisation approach~\cite{Beaujard:2019pkn}. 
In certain cases related to local Calabi--Yau three-folds, these are
determined independently~\cite{Beaujard:2020sgs, Mozgovoy:2020has,
  Duan:2020qjy}. In this latter approach, it is important that the
FI parameters $\zeta_j$ are chosen to avoid being on a wall of marginal
stability. To ensure this, when necessary, a small random perturbation
$\varepsilon_{j\alpha}$ is included in the \texttt{CoulombHiggs} package \cite{CBPackage},
\be 
\label{eq:zetapert}
\zeta_{j\alpha}=\lambda\, {\rm Im}(Z(\gamma_j)\bar Z(\gamma))+\varepsilon_{j\alpha},
\ee 
with $|\varepsilon_{j\alpha}|\ll \lambda\,| {\rm Im}(Z(\gamma_j)\bar
Z(\gamma))|$. We will use this approach extensively in later sections. 

The supersymmetric localisation and MPS approaches just mentioned amount to computations of supersymmetric ground states in the Coulomb-branch approximation, taking $\lambda \sim 0$ in~\eqref{eq:zetaZ}, while the moduli space $\CM^Q_\gamma$ appears in the Higgs-branch approach, sending $\lambda \sim \infty$ in~\eqref{eq:zetaZ} --- that is, viewing $\CM^Q_\gamma$ as the target space of a 1d GLSM. The supersymmetric ground states extrapolate smoothly between the two approaches. For our purpose in this work, the target-space approach~\eqref{BPS as coho} provides the easier, geometric intuition.

\medskip
\noindent
{\bf BPS invariants as Euler characteristics.}  If a quiver description is available, $\Omega(\gamma; \zeta)$ can be related to the moduli space $\CM^Q_\gamma$ of quiver representations. This is most easily understood when the target space $\CM^Q_\gamma$ of the SQM is a smooth, compact manifold. Then, the unrefined invariant is simply the topological Euler characteristic up to an overall sign (in the standard conventions): 
\be\label{def OmegaQ}
\Omega^Q(\gamma; \zeta)=(-1)^{\dim_\mathbb{C}\left(\CM^Q_\gamma(\zeta)\right)} \,\chi\left(\CM_\gamma^Q(\zeta)\right)~. 
\ee 
Recall that the overall sign can be computed from the virtual dimension~\eqref{eq:dimM}. If $\CM^Q_\gamma$ is not smooth and compact, the geometric definition of $\Omega(\gamma; \zeta)$ requires the theory of Donaldson--Thomas (DT) invariants for moduli spaces of semi-stable objects~\cite{Kontsevich:2008fj, Joyce:2008pc, Meinhardt2019}.

The refined (or {\it motivic}) BPS index $\Omega^Q_{\rm r}$ is similarly defined in terms of the Poincar\'e polynomial of the moduli space:\footnote{More generally, one would like to consider the two-parameter Hodge--Deligne polynomial $E({\bf y}, t)= \sum_{p,q=0} {\bf y}^p (-1)^q t^{p-q}\, h^{p,q}$ keeping track of all the Hodge numbers $h^{p,q}(\CM_\gamma^Q)$.}
\be\label{P def of Omegay}
\Omega^Q_{\rm r}(\gamma, y; \zeta) =  (-y)^{-{\rm dim}_\C\,\CM_\gamma^Q} P_{\CM_\gamma^Q}(y)~, \qquad \quad\text{with}\;\;  P_\CM(y)\equiv \sum_{k=1}^{{\rm dim}_\R\,\CM}  y^k \, b_k(\CM)~.
\ee
For $\CM_\gamma$ a closed K\"ahler manifold, Poincar\'e duality implies that~\eqref{P def of Omegay}  is invariant under $y\rightarrow 1/y$. This index is not actually protected, in general. Instead, physicists generally consider the protected spin character~\cite{Gaiotto:2010be, Chuang:2013wt}, 
\be
\Omega^Q(\gamma, y; \zeta)=\frac{{\rm Tr}_{\CH_\gamma}\left[ (2J_3)(-1)^F (-y)^{2J_3+2I_3}\right]}{y-y^{-1}}~,
\ee
where $I_3$ is the generator of the $\SU(2)_R$ $R$-symmetry group. 
This corresponds to the $\chi_{\bf y}$-genus of the moduli space:
\be\label{chiygemus def of Omegay}
\Omega^Q(\gamma, y;\zeta) =  (-y)^{-{\rm dim}_\C\, \CM^Q_\gamma(\zeta)} \chi_{-y^2}(\CM^Q_\gamma(\zeta))~.
\ee
Here, we have set ${\bf y}=-y^2$ in the $\chi_{\bf y}$-genus of a compact complex manifold $\CM$, which is the graded Euler characteristic of the Dolbeault complex: 
\be
\chi_{{\bf y}}(\CM) \equiv \chi(\CM, \Lambda_{\bf y}^\bullet\Omega_{\CM}) = \sum_{p=0}^{{\rm dim}_\C \CM} {\bf y}^p \chi(\CM,\Omega^p) =  \sum_{p,q=0}^{{\rm dim}_\C \CM} {\bf y}^p (-1)^q\, h^{p,q}(\CM)~.
\ee
In many cases, the Hodge diamond of $\CM_\gamma^Q$ is diagonal, namely $h^{p,q}=0$ for $p\neq q$. Then, the refined index defined as in~\eqref{chiygemus def of Omegay} becomes equal to the motivic definition~\eqref{P def of Omegay}. 
The no-exotic conjecture~\cite{Gaiotto:2010be} states that this diagonal condition always holds for physical quivers describing 4d SQFTs --- see~{\it e.g.}~\cite{DelZotto:2014bga} for a detailed discussion --- such that $\Omega^Q_{\rm r}$ \eqref{P def of Omegay} and $\Omega^Q$ \eqref{chiygemus def of Omegay} coincide. While this clearly does not hold in general, for generic quivers with superpotential, this does hold for the explicit examples discussed below.

\medskip
\noindent
{\bf Rational BPS invariants.}  We also introduce the rational BPS invariant:
\be 
\label{eq:DefRatInv}
\bar \Omega^Q(\gamma; \zeta) \equiv \sum_{n| \gamma} \frac{\Omega^Q(\gamma/n; \zeta)}{n^2}~,
\ee 
where we sum over all positive integers $n$ that divide $\gamma$ in $\Gamma$.  
Similarly, the refined rational BPS invariant is given by:
\be \label{rational refined Omega}
\bar \Omega^Q(\gamma, y; \zeta)=\sum_{n|\gamma} \frac{y-y^{-1}}{n\,(y^n-y^{-n})}\Omega^Q(\gamma/n,y^n; \zeta)~.
\ee 
The rational invariants appear naturally in the mathematical literature as invariants for moduli stacks \cite{Joyce:2008pc}. In physics, these invariants play an important role for the effective description of BPS bound states using the Maxwell--Boltzmann statistics of distinguishable particles rather than Bose or Fermi statistics~\cite{Manschot:2010qz} --- essentially, the invariant~\eqref{eq:DefRatInv} counts both the stricly stable bound states captured by $\Omega^Q(\gamma)$ and the marginal bound states of $n$ identical particles of charge $\gamma/n$. Importantly, it is the rational invariants which appear most naturally in the celebrated  Kontsevich--Soibelman (KS) wall-crossing formula~\cite{Kontsevich:2008fj}, which we now review.

\subsection{The Kontsevich--Soibelman monodromy operator}
\label{sec:KSMonO}
BPS spectra and their wall-crossing properties are famously encoded in the KS monodromy operator~\cite{Kontsevich:2008fj, Gaiotto:2009hg, Gaiotto:2010okc}. We let $\frak{g}_Q$ be the $\Gamma$-graded Lie algebra, with elements $e_\gamma\in \frak{g}_Q$ for each charge $\gamma \in \Gamma$, satisfying the commutation relation,
\be 
\label{eq:gQcomm}
[e_{\gamma},e_{\gamma'}]=\left<\gamma,\gamma'\right> (-1)^{\left<\gamma,\gamma'\right>}e_{\gamma+\gamma'}~.
\ee 
We then define the formal quantities:
\be \label{def CKgamma}
\CK_\gamma=\exp\left( \sum_{n=1}^\infty \frac{1}{n^2} e_{n\gamma}\right)~.
\ee 
The BPS spectrum is encoded in the KS monodromy operator:
\be 
\label{MonodK}
\mathbb{M}_Q(\zeta) = \prod^\curvearrowright_{\gamma\in \Gamma_+} \CK_\gamma^{\Omega^Q(\gamma; \zeta)},
\ee 
where the product is over all BPS particles with central charges $Z_\gamma$ valued in a given half-plane of the complex plane (giving us a customary separation between particles and anti-particles). The ordering of the non-commutative product is determined by clockwise ordering of the phases of the central charges --- for BPS invariants, this corresponds to decreasing slopes~\eqref{def slope} for the stable representations in $\CM^Q_\gamma$.  The KS wall-crossing formula is the statement that $\mathbb{M}_Q\equiv \mathbb{M}_Q(\zeta)$ is actually independent of the choice of chamber in the space of stability conditions --- this gives us very intricate constraints on how the spectrum can change as we cross a wall of marginal stability.

It is also useful to reorder the product~\eqref{MonodK} in terms of the  (so-called) symplectomorphism $\CV_\gamma$ defined by
\be
\label{eq:Vgamma}
\CV_\gamma = \exp\left( \bar \Omega(\gamma)\,e_\gamma \right)~,
\ee 
written now in terms of the rational invariants $\bar \Omega$ defined in~\eqref{eq:DefRatInv}, so that:
\be
\label{MonodCV}
\mathbb{M}_Q=\prod^\curvearrowright_{\gamma\in \Gamma_+} \CV_\gamma= \prod^\curvearrowright_{\gamma\in \Gamma_+} \exp\left( \bar \Omega^Q(\gamma)\,e_\gamma \right)~.
\ee 
Note that, even if $\Omega(\gamma)$ vanishes for a specific $\gamma$, $\bar \Omega(\gamma)$ may be non-vanishing if $\gamma$ is non-primitive.

The monodromy operator provides us with a way to determine invariants at a degenerate stability condition or on a wall of marginal stability. In that case, we require that the charges $\gamma$ with aligned central charges appear in the same exponent:
\be 
\label{eq:sameslope}
\exp\left(\sum_{\gamma\,|\, {\rm Arg}(Z_\gamma)=\phi} \bar \Omega(\gamma)\,e_\gamma \right)~.
\ee 
We consider the coefficients of $e_\gamma$ so obtained  as the definition of the BPS invariants at such a non-generic stability condition. These can be determined using the Baker--Campbell--Hausdorff formula
\be 
\label{eq:BCH}
\begin{split}
&\exp\!\left(X\right)\,\exp\!\left(Y\right)=\\
&\quad \exp\left(X+Y+\tfrac{1}{2}[X,Y]+\tfrac{1}{12}[X,[X,Y]]+\tfrac{1}{12}[Y,[Y,X]]+\dots\right)~,
\end{split}
\ee 
where the dots represent higher commutators and $X$, $Y$ are any Lie algebra elements. We expect that the BPS invariants for such a degeneric stability parameters can be
alternatively derived using Harder--Narasimhan 
filtrations and wall-crossing for invariants of moduli stacks~\cite{Reineke_2003, Joyce:2004tk, Joyce:2008pc}. We leave this as an interesting challenge for future work.

\medskip
\noindent
{\bf Refined monodromy operator.} The quantum monodromy operator admits a refinement, denoted by $\mathbb{M}_Q(y)$, which depends on the refined BPS indices $\Omega(\gamma,y)$. To this end, we introduce:
\be 
\label{eq:DefU}
\CU_\gamma=\exp\left(\sum_{n=1}^\infty \frac{\Omega(\gamma,y^n)}{n}\frac{\hat e_{n\gamma}}{y^n-y^{-n}} \right)~, 
\ee 
with $\hat e_\gamma$ generators of the non-commutative quantum torus $\CR_{Q}$, which satisfy the relation,
\be 
\label{eq:QuTo}
\hat e_\gamma \hat e_{\gamma'}=(-y)^{\left<\gamma,\gamma' \right>}\hat e_{\gamma+\gamma'}~.
\ee 
In terms of the refined rational invariant \eqref{rational refined Omega}, we also introduce 
\be
\CW_\gamma=\exp\left( \bar \Omega(\gamma,y)\,\frac{\hat e_\gamma}{y-y^{-1}}\right).
\ee 
The quantum monodromy operator then reads
\be 
\label{eq:MQy}
\mathbb{M}_Q(y)=\prod^\curvearrowright_{\gamma\in \Gamma_+} \CU_\gamma=\prod^\curvearrowright_{\gamma\in \Gamma_+} \CW_\gamma.
\ee  
The unrefined case is reproduced in the limit $y\to 1$ with $e_\gamma=\lim_{y\to 1} \,  \hat e_\gamma/({y-y^{-1}})$.

\section{Galois covers of BPS quivers}
\label{sec:galois cover}

It has long been known that there exists BPS spectra which can be roughly interpreted as several copies of the BPS spectrum of another theory~\cite{Gaiotto:2010okc}, and that this fact can be explained in terms of {\it Galois covers} of BPS quivers~\cite{Cecotti:2015qha}. In this section, we review this concept and then discuss a number of illuminating examples of Galois covers relating the BPS quivers of distinct 4d $\CN=2$ theories, with a particular focus on 5d BPS quivers of 4d $\CN=2$ KK theories, which also describe D-branes on local CY threefolds~\cite{Closset:2019juk}.

\subsection{Galois covers of quivers with superpotential}
\label{sec:GCoverQuiver}

Consider a finite quiver $\t Q = \{ \t Q_0, \t Q_1, \t h, \t t\}$ with a superpotential $\t W$ which admits an action by a finite abelian group $\G$,
\be
\G\; :\;  (\t Q_0, \t Q_1)\rightarrow (\t Q_0, \t Q_1)\; : \; (j, a) \mapsto (g\cdot j, g\cdot a)~.
\ee
The action must leave the quiver invariant, namely:
\be
\t h(g\cdot a)=g\cdot\t h(a),\qquad \t t(g\cdot a)=g\cdot\t t(a)~, \qquad \text{for all}\; g\in \G~,
\ee
for every arrow $a\in \t Q$; it must also leave the superpotential $\t W$ invariant. The $\G$ action is said to be {\it admissible} if it is {\it free} on the nodes $j\in \t Q$. Without loss of generality, we can restrict the finite abelian group $\G$ to $\mathbb{Z}_n$, $n\in \mathbb{N}$.%
\footnote{We will also consider the limiting case of a $\mathbb{Z}$ action (sending $n\rightarrow \infty$), on a quiver with an infinite number of nodes, in section~\ref{sec:GaloisCoverFixedLoci} below.}

\medskip\noindent
{\bf Galois covering as a quotient.} 
To any quiver $\t Q$ with an admissible $\G$ action, we can associate a Galois cover~\cite{gabriel2006universal, bongartz1982covering,Cecotti:2015qha,cibils2006skew}:
\be\label{F Galois cover}
F\; : \; \t Q \longrightarrow  Q = \t Q/\G~.
\ee
We say that $\t Q$ is a $\G$ Galois cover of the `smaller' quiver $Q$. The nodes of $Q$ are indexed by the distinct orbits $\G j$ of nodes in $\t Q_0$. The set of arrows $(a : \G j\rightarrow \G k)$ in $Q_1$ is obtained from the set of arrows $(\t a : j \rightarrow k')$ in $\t Q_1$ for all $k'\in \G k$ (for an arbitrary reference node $j\in \G j$). The superpotential $W=\t W/\G$ is similarly obtained as a quotient.  It is important to note that $|\t Q_0|= |\G| |Q_0|$ and $|\t Q_1|=|\G|| Q_1|$.

\medskip\noindent
{\bf Galois covering of $Q$ from a grading assignment.}  Given any quiver $Q$ with a superpotential $W$, a Galois covering by $\G$  can be defined through a choice of {\it grading} of the arrows of $Q$:
\be\label{G grading arrows da}
d\; : \; Q_1 \rightarrow \h \G \; :\; a \mapsto d_a~,
\ee
where $\h \G$ is the group of characters of $\G$.%
\footnote{Since $\h\G\cong \G$ for a finite abelian group, we will be somewhat loose in our notation.} 
For $\G=\Z_n$, we choose $d_a \in \{0,1, \cdots, n-1\}$ for each arrow. In a quiver with superpotential, we also request that the superpotential~\eqref{def W on paths} has trivial grading:
\be\label{dW0 cond}
d(W)=0\qquad \Leftrightarrow \qquad  d(p)= \sum_{a\in p} d_a =0  \quad\; \forall\, p\in W~.
\ee
Then, the Galois cover $\t Q$ of $Q$ is defined as follows:
\begin{itemize}
    \item[$\t Q_0$:] The nodes of $\t Q$ are indexed by $(j, g)$, for $j\in Q_0$ and $g\in \G$. Fixing $\G=\Z_n$ and $\alpha\in \Z_n$, we simply denote the nodes by $j_\alpha \in \t Q_0$ with $\alpha =0, \cdots, n-1$. 
    \item[$\t Q_1$:] The set of arrows $(\t a : j_\alpha \rightarrow k_\beta)\in\t Q_1$ is given by the set of arrows $(a: j \rightarrow k)$ in $Q_1$ such that $\beta =\alpha +d_a$ (mod $n$). Note that the arrows $\t a\in \t Q_1$ can also be indexed by $(a, g)= a_\alpha$ for $a\in Q_1$, in which case the Galois covering head and tail maps of $\t Q$ are given by:
    \be
\label{eq:grading constraint arrows}
\tilde t(a_\alpha)=t(a)_\alpha,\qquad \tilde h(a_\alpha)=h(a)_\beta,\qquad
\beta=\alpha+ d_a \mod n~.
\ee
\item[$\t W$:] The superpotential for $\t Q$ is obtained from the superpotential~\eqref{def W on paths} for $Q$ according to: 
\be\label{tilde W}
\t W = \sum_p C_p \sum_{\alpha\in\mathbb Z_n}\,  \Tr\left((a_{s_1})_{\alpha} (a_{s_2})_{\alpha+ d_{a_{s_1}}}(a_{s_3})_{\alpha+d_{a_{s_1}}+d_{a_{s_2}}} \cdots (a_{s_{|p|}})_{\alpha - d_{a_{s_{|p|}}}}\right),
\ee
where we used the condition~\eqref{dW0 cond} which ensures that the loop in $Q$ lifts to a loop in $\t Q$.
\end{itemize}

 \noindent
 Once we have obtained a Galois covering $\t Q$ as described, it is clear that indeed $Q=\t Q/\G$, where the $\G=\Z_n$ action is the $\Z_n$ permutation of the nodes $j_\alpha$ and of the arrows $a_\alpha$ simultaneously.

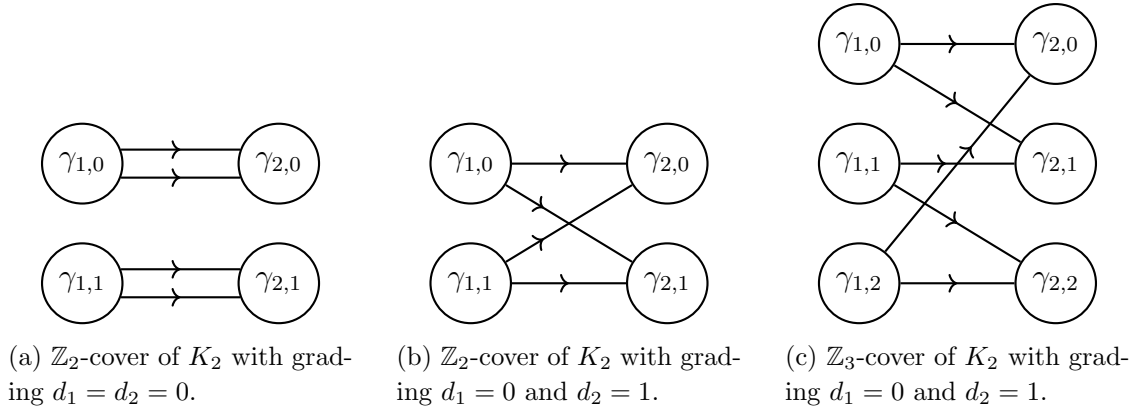
\begin{figure}
\centering
\begin{subfigure}[b]{0.3\textwidth}
\centering
 \begin{tikzpicture}[baseline=1mm,every node/.style={circle,draw},thick]
\node[] (1) []{$\gamma_{1,0}$};
\node[] (2) [right =1.5 of 1]{$\gamma_{2,0}$};
\node[] (3) [below =.5 of 1]{$\gamma_{1,1}$};
\node[] (4) [right =1.5 of 3]{$\gamma_{2,1}$};
\draw[->-=0.5] (1.20) to   (2.160); 
\draw[->-=0.5] (1.340) to   (2.200);
\draw[->-=0.5] (3.20) to   (4.160);
\draw[->-=0.5] (3.340) to   (4.200);
\end{tikzpicture}
\caption{$\mathbb{Z}_2$-cover of $K_2$ with grading $d_1=d_2=0$.}
\label{GQuiver1}
\end{subfigure}
\hspace{.3cm}
\begin{subfigure}[b]{0.3\textwidth}
\centering
 \begin{tikzpicture}[baseline=1mm,baseline=1mm,every node/.style={circle,draw},thick]
\node[] (1) []{$\gamma_{1,0}$};
\node[] (2) [right =1.5 of 1]{$\gamma_{2,0}$};
\node[] (3) [below =.5 of 1]{$\gamma_{1,1}$};
\node[] (4) [right =1.5 of 3]{$\gamma_{2,1}$};
\draw[->-=0.5] (1) to   (2);
\draw[->-=0.3] (1) to   (4);
\draw[->-=0.5] (3) to   (4);
\draw[->-=0.3] (3) to   (2);
\end{tikzpicture}
\caption{$\mathbb{Z}_2$-cover of $K_2$ with grading $d_1=0$ and $d_2=1$. }
\label{GQuiver2}
\end{subfigure}
\hspace{.3cm}
\begin{subfigure}[b]{0.3\textwidth}
\centering
\begin{tikzpicture}[baseline=1mm,baseline=1mm,every node/.style={circle,draw},thick]
\node[] (1) []{$\gamma_{1,0}$};
\node[] (2) [right =1.5 of 1]{$\gamma_{2,0}$};
\node[] (3) [below =0.5 of 1]{$\gamma_{1,1}$};
\node[] (4) [right =1.5 of 3]{$\gamma_{2,1}$};
\node[] (5) [below =0.5 of 3]{$\gamma_{1,2}$};
\node[] (6) [right =1.5 of 5]{$\gamma_{2,2}$};
\draw[->-=0.5] (1) to   (2);
\draw[->-=0.5] (1) to   (4);
\draw[->-=0.4] (3) to   (4);
\draw[->-=0.5] (3) to   (6);
\draw[->-=0.5] (5) to   (6);
\draw[->-=0.6] (5) to   (2);
\end{tikzpicture}
\caption{$\mathbb{Z}_3$-cover of $K_2$ with grading $d_1=0$ and $d_2=1$. }
\label{GQuiver3}
\end{subfigure}
\caption{Examples of Galois covers of the Kronecker quiver $Q=K_2$. Here, the nodes $j_\alpha \in \t Q_0$ are indexed by $\gamma_{j, \alpha}$. The arrow gradings $d_a$ are indicated in each case. \label{fig:K2 covers}
}
\end{figure}

\medskip\noindent
{\bf Simple example: Galois covers of $K_2$.} Some Galois covers of the Kronecker quiver $Q=K_2$ are shown in figure~\ref{fig:K2 covers}. Figure~\ref{GQuiver1} is a trivial example with $\G=\Z_2$ and trivial grading, which gives rise to a disconnected quiver $\t Q$.  
The second $\G=\mathbb{Z}_2$ Galois cover, displayed in figure~\ref{GQuiver2}, corresponds to grading $d_a=(0,1)$ for the two arrows of $K_2$. While the Kronecker quiver is the BPS quiver of 4d $\CN=2$ super-Yang-Mills with $\SU(2)$ gauge group, this $\Z_2$ cover is the BPS quiver of the $\SU(2)$ $\CN=2$ SQCD with $N_f=2$ --- this is the original example of a Galois cover in physics~\cite{Gaiotto:2010okc}, as we will review momentarily.  The third example, displayed in figure~\ref{GQuiver3},  is a $\G=\mathbb{Z}_3$ cover of $K_2$.


\subsection{Galois covers in 4d SQCD} 
\label{sec:4dSQCD}

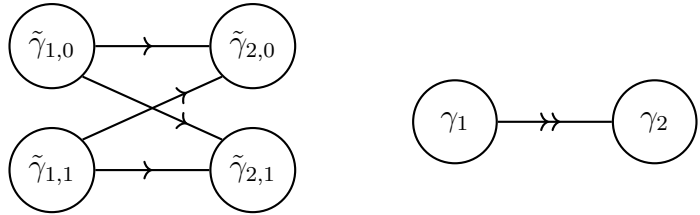
\begin{figure}[t]
\centering
\begin{tikzpicture}[baseline=1mm,baseline=1mm,every node/.style={circle,draw,minimum size=1.1cm},thick]
\node[] (1) []{$\t\gamma_{1,0}$};
\node[] (2) [right =1.5 of 1]{$\t\gamma_{2,0}$};
\node[] (3) [below =.5 of 1]{$\t\gamma_{1,1}$};
\node[] (4) [right =1.5 of 3]{$\t\gamma_{2,1}$};
\draw[->-=0.5] (1.0) to   (2.180); 
\draw[->-=0.75] (1.315) to   (4.135);
\draw[->-=0.5] (3.0) to   (4.180);
\draw[->-=0.75] (3.45) to   (2.225);
\node[] (5) [below right =0.2 and 4.5 of 1]{$\gamma_1$};
\node[] (6) [right =1.5 of 5]{$\gamma_2$};
\draw[->>-=0.5] (5.0) to   (6.180);
\end{tikzpicture}
\caption{Galois pair of the BPS quivers of the  4d $N_f=2$ (\textsc{Left}) and $N_f=0$ theories (\textsc{Right}). Note that we sometimes use the obvious notation of writing multiple arrows on top of each other.}
\label{fig:Nf0Nf2}
\end{figure}

The simplest examples of Galois covers of BPS quivers for 4d $\CN=2$ can be found in $\SU(2)$ SQCD with $N_f=0$ and $N_f=2$ massless hypermultiplets~\cite{Gaiotto:2010okc, Cecotti:2015qha}. A basis of BPS states for the pure, $N_f=0$ $\SU(2)$ theory is given by the monopole and dyon states,  which have magnetic-electric charges $\gamma_1 = (1,0)$ and $\gamma_2 = (-1,2)$. 
The BPS quiver is the $K_2$ quiver in figure~\ref{QuiverK2}, which we compare to the BPS quiver of the massless $N_f=2$ $\SU(2)$ theory in figure~\ref{fig:Nf0Nf2}.

The charges are  $\t\gamma_{1,0} = (1,0,-1,0)$,
$\t\gamma_{1,1} = (1,0,1,0)$, $\t\gamma_{2,0} =  (-1,1,0,-1)$, $ \t\gamma_{2,1} =
(-1,1,0,1)$. The last two entries are the $\mathfrak{so}(4)$ flavour charges. Notice that the magnetic-electric charges, which are the first two entries of $\t\gamma_{j,\alpha}$, do not depend on $\alpha$. Since the theory is massless, the central charges $Z_{\t\gamma_{j,\alpha}}$ also do not depend on $\alpha$.
The BPS quiver has an obvious $\mathbb G=\mathbb Z_2$ symmetry, exchanges the nodes $\t\gamma_{j,0}$ and $\t\gamma_{j,1}$ for both $j=1,2$. It is also clear that the arrows in $K_2$ are labelled by the gradings in $\mathbb Z_2$, which makes the $N_f=2$ quiver  $\t K_2$ a symmetric $\mathbb Z_2$-Galois cover of $K_2$.

The BPS spectrum can be determined explicitly from the representations of the Kronecker quiver, or, alternatively, by using the `mutation method' \cite{Alim:2011ae, Alim:2011kw}. For pure $\SU(2)$, the strong-coupling spectrum, with $\zeta_1<\zeta_2$, only contains the monopole and the dyon, while the weak-coupling spectrum, with $\zeta_1>\zeta_2$, also involves the states:
\bea    \label{BPS spectrum Nf=0}
   \text{pure } {\rm SU}(2) \colon \qquad  \quad \gamma_1 &+ k(\gamma_1 + \gamma_2)~, \\
      -\gamma_1 &+ (k+1)(\gamma_1 + \gamma_2)~, \qquad k \in \mathbb{Z}~,
\eea
together with the $W$-boson $ \gamma_1 +  \gamma_2$. 
The weak-coupling spectrum of dyons of the massless $N_f=2$, on the other hand, can be written as \cite{Alim:2011kw}:
\bea    \label{BPS spectrum massless Nf=2}
     N_f = 2 \colon \qquad    \t\gamma_{1,\beta} &+ k \textstyle\sum_{\alpha=0}^1 (\t\gamma_{1,\alpha}+\t\gamma_{2,\alpha})~, \\
       - \t\gamma_{1,\beta} &+ (k+1)\textstyle\sum_{\alpha=0}^1 (\t\gamma_{1,\alpha}+\t\gamma_{2,\alpha}) ~,  \qquad k \in \mathbb{Z}~,
\eea 
for  $\beta=0,1$, together with the quarks $\t\gamma_{1,\alpha}+\t\gamma_{2,\beta}$ for $\alpha,\beta=0,1$ and the W-boson $\sum_{\alpha=0}^1(\t\gamma_{1,\alpha}+\t\gamma_{2,\alpha})$.
The $\mathbb Z_2$ symmetry of the $N_f=2$ BPS quiver is therefore manifested by its BPS spectrum. Taking the $\mathbb Z_2$ quotient essentially recovers the spectrum of the pure ${\rm SU}(2)$ theory \eqref{BPS spectrum Nf=0}. 
This observation was given an interpretation in terms of the Kontsevich--Soibelman wall-crossing formulae for the two theories in \cite{Gaiotto:2010okc}, as a simple change of coordinates. We will discuss such homomorphisms induced by Galois covers in sections~\ref{sec:AlgHoms} and~\ref{sec:GaloisExamples}, where we will also explain the precise relations between the two BPS spectra --- see section~\ref{sec:Symm2Cover} in particular.

\subsection{Galois covers for \texorpdfstring{$\KK E_n$}{En} theories}\label{sec:Galois_covers_En}
For purely four-dimensional $\CN=2$ theories of rank one, the above Galois cover is the only known example. More interesting examples are found when considering 5d theories compactified on a circle, and when the Galois covering relates theories of different rank.

Two interesting Galois pairs can be obtained from the circle-compactification of the five-dimensional Seiberg $E_n$ theories~\cite{Seiberg:1996bd,Morrison:1996xf,Nekrasov:1996cz}. These 5d SCFTs are UV completions of $\SU(2)$ gauge theories with $N_f=n-1$ fundamental flavours. They are engineered in M-theory compactifications on non-compact Calabi--Yau threefolds $\bf X$, which are the canonical line bundles over del Pezzo surfaces $dP_n$ or the Hirzebruch surface $\mathbb F_0$. The circle compactification of the $E_n$ theory, which we denote by $\KK E_n$, is described in Type IIA string theory on the same $dP_n$ (or $\mathbb F_0$) singularity, while the local mirror description in Type IIB provides the Seiberg--Witten geometry~\cite{Katz:1997eq,Hori:2000ck, Hori:2000kt}. For $n=0,1,2,3$, the $E_n$ singularity in Type IIA is a toric Calabi--Yau manifold, while the theories for $n\geq 4$ are non-toric. 

Their BPS spectra are generally very difficult to determine. The BPS quivers for the theories associated with the toric geometries can be obtained using brane-tiling techniques, see \eg~\cite{Franco:2005rj,Hanany:2005ss,Hanany:2012hi,Cachazo:2001sg}. More generally, one can sometimes identify so-called {\it quiver points} $u_Q$, where the central charges of light BPS particles almost align. In those cases, the spectrum can conjecturally be determined as bound states of simple objects. 
 
\paragraph{The $\boldsymbol{\KK E_6 \leftrightarrow \KK E_0}$
Galois cover.} 
The first example of a Galois cover between two $\KK E_n$ theories is found between the $\KK E_6$ theory and the $\KK E_0$ theory.

The $E_0$ SCFT does not have any flavour symmetry, and there are therefore no relevant deformations or gauge-theory phases. It can be obtained in M-theory on local $dP_1$ by blowing down the exceptional curve, which gives the local $dP_0=\mathbb P^2$ geometry. The origin of the Kähler cone, where the $\mathbb P^2$ shrinks to zero size, is the orbifold singularity $\mathbb C^3/\mathbb Z_3$. The corresponding BPS quiver is the well-known $\mathbb C^3/\mathbb Z_3$ orbifold quiver---see \eg~\cite{Aspinwall:1993xz,Hori:2000ck,Douglas:2000qw,Aspinwall:2004jr} for a selection of the vast literature. 

\begin{figure}[t]
\centering
\begin{tikzpicture}[baseline=1mm,baseline=1mm,every node/.style={circle,draw,minimum size=1.1cm},thick]
\node[] (1) []{$\t\gamma_{1,1}$};
\node[] (4) [left =-0.01 of 1]{$\t\gamma_{1,0}$};
\node[] (7) [below right=0.17 and -0.235 of 4]{$\t\gamma_{1,2}$};
\node[] (5) [right =3 of 1]{$\t\gamma_{2,1}$};
\node[] (2) [left =-0.01 of 5]{$\t\gamma_{2,0}$};
\node[] (8) [below right=0.17 and -0.235 of 2]{$\t\gamma_{2,2}$};
\node[] (3) [below right =2.4 and 1.35 of 1]{$\t\gamma_{3,1}$};
\node[] (6) [left =-0.01 of 3]{$\t\gamma_{3,0}$};
\node[] (9) [below right=0.17 and -0.235 of 6]{$\t\gamma_{3,2}$};
\draw[->-=0.5] (1) to   (2);
\draw[->-=0.5] (8) to   (3);
\draw[->-=0.5] (6) to   (7);
\node[] at (7,-0.8) (1) []{$\gamma_{1}$};
\node[] (2) [right =1.5cm of 1]{$\gamma_{2}$};
\node[] (3) [below right=1.4cm and 0.525cm of 1]{$\gamma_{3}$};
\draw[->>>-=0.5] (1) to   (2);
\draw[->>>-=0.5] (2) to   (3);
\draw[->>>-=0.5] (3) to   (1);
\end{tikzpicture}
\caption{Galois pair of the BPS quivers of the 5d $\KK E_6$ (\textsc{Left}) and $\KK E_0$ (\textsc{Right}) theories. Note that the $E_6$ quiver is written in the block notation, which means that each arrow between from a block of $m_1$ nodes to a block of $m_2$ nodes means that there are $m_1 m_2$ arrows that connect each node in the first block to each node in the second block, with no arrows within a block --- hence this $E_6$ quiver contains 9 nodes and 18 arrows. 
\label{quiver E6 orbi}}
\end{figure}
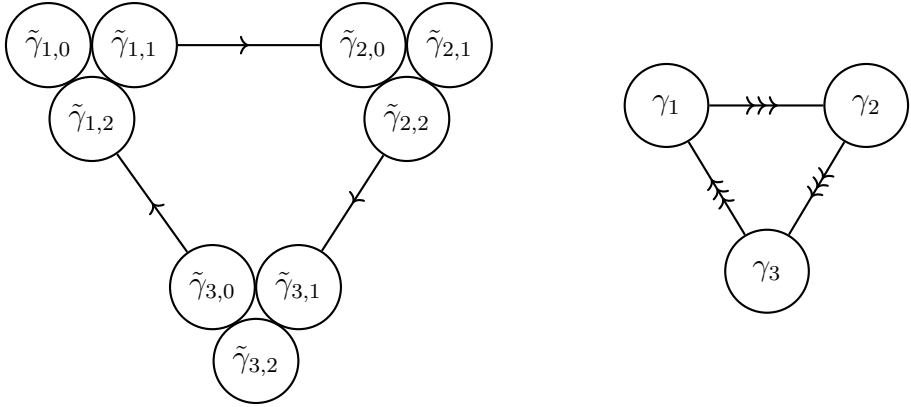

On the other hand, the $\KK E_6$ theory is engineered on local $dP_6$ and is non-toric. It can be viewed as a four-parameter non-toric family of deformations of the $\mathbb{C}^3/(\mathbb{Z}_3 \times \mathbb{Z}_3)$ orbifold (see \eg \cite{Wijnholt:2002qz}).
We are interested in the BPS quiver of the $\KK E_6$ theory known as the  $\mathbb{C}^3/(\mathbb{Z}_3\times \mathbb{Z}_3)$ orbifold quiver~\cite{Hanany:2012hi, Beaujard:2020sgs} displayed in figure~\ref{quiver E6 orbi}.

We claim that the BPS quiver $\t Q$ for the $\KK E_6$ theory forms a $\mathbb Z_3$ Galois cover of the quiver $Q$ of the $\KK E_0$ theory.
For simplicity, we use the `block' notation for the $\KK E_6$ quiver: each displayed arrow indicates a map between each node in a block to each node in the other block.

In each block of $\t Q$, there is a $\mathbb Z_3$ symmetry, which we make obvious in the notation. To check that figure~\ref{quiver E6 orbi} is indeed a Galois cover, the superpotentials need to be compatible with the cover. To show this, let us label the 9 arrows in $Q$ by $a_{ij}^a\colon i\to j$, where $a=1,2,3$ and $i,j=1,2,3$. The $\KK E_0$ superpotential can be recast in the compact form~\cite{Closset:2019juk}:
\bea
\label{eq:WE0}
 W_{E_0} = \sum_{a,b,c} \epsilon_{abc} \Tr \Bigl( a_{12}^a a_{23}^b a_{31}^c\Bigr)~, 
\eea
where the summation is over all $a,b,c\in\{1,2,3\}$, with the convention is $\epsilon_{123}=1$. To compare it with the cover, we choose the gradings $d_{a_{ij}^a}=a-1\mod 3$ of the arrows.

The superpotential $W_{E_6}$ of the quiver $\t Q$ for the $\KK E_6$ theory has been determined in~\cite[(3.1)]{Hanany:2012hi}. With our convention, the nodes of $\t Q$ are labelled as $j_\alpha$, with $j=1,2,3$ and $\alpha=0,1,2$, while the arrows are $(a_{ij}^a)_\alpha\colon i_\alpha \to j_{\alpha+d_a}$. The nodes of the quiver $\t Q$ in~\cite{Hanany:2012hi} are labelled by the integers $1,\dots, 9$, which we can bring to our convention of figure~\ref{quiver E6 orbi} by identifying $1_\alpha=1,7,4$, while $2_\alpha=5,8,2,$ and $3_\alpha=3,6,9$, each for $\alpha=0,1,2$. For those identifications, it is straightforward to check that it can be expressed as
\bea
\label{eq:WE6}
W_{E_6}
=\sum_{a,b,c}\epsilon_{abc}\sum_{\alpha\in\mathbb Z_3} \Tr \Bigl(
(a_{12}^a)_\alpha\,  (a_{23}^b)_{\alpha+a-1}\, (a_{31}^c)_{\alpha+a+b-2}
\Bigr)~.
\eea
This superpotential of the Galois cover is precisely of the form~\eqref{tilde W}: the sum runs over oriented paths in $Q$, with each arrow in $\t Q$ indexed by the gradings in $Q$. Moreover, the sum runs over all paths in $\t Q$ that collapse to paths present in the superpotential of $Q$ under the quotient map~\eqref{F Galois cover}. This shows that figure~\ref{quiver E6 orbi} indeed forms a Galois cover.

\paragraph{The $\boldsymbol{\KK E_1 \leftrightarrow \KK E_5}$ Galois cover.} 
The next example of a Galois cover of rank one theories is the pair of BPS quivers for the $\KK E_1$ and $\KK E_5$ theory. The $E_1$ theory is the UV completion of five-dimensional $\SU(2)_0$ gauge theory, and is geometrically engineered on the local Hirzebruch surface $\mathbb F_0\cong \mathbb P^1\times \mathbb P^1$; the $E_5$ theory similarly corresponds to 5d $SU(2)$ with $N_f=4$ and it is engineered with the local $dP_5$ surface --- see~\eg~\cite{Feng:2000mi,Feng:2001xr,Hanany:2012hi, Closset:2019juk, Beaujard:2020sgs,Bonelli:2020dcp,DelMonte:2021ytz,Bridgeland:2024saj} for a sample of the literature. Recall that there are generally infinitely many different BPS quivers related by quiver mutation (or 1d Seiberg duality). Here we discuss the two (so-called) toric phases for the $\mathbb{F}_0\cong \mathbb{P}^1\times \mathbb{P}^1$, dubbed Phases (a) and (b) in~\cite{Hanany:2012hi}:

\begin{itemize}
\item[\underline{Phase (a)}] We first study a particularly symmetric quiver with four nodes, which are exchanged under a  $\mathbb Z_4$ symmetry. 
We compare it to a 5d BPS quiver for the $\KK E_5$ theory, which has 8 nodes that organise into four blocks of two nodes~\cite{Wijnholt:2002qz,Hanany:2012hi} displayed in figure~\ref{Z4 symmetric quivers E5 E1}.

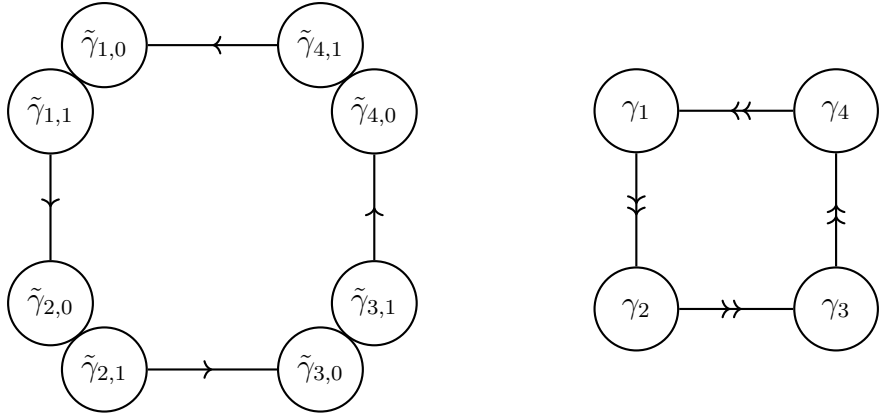
\begin{figure}[t]
\centering
\begin{tikzpicture}[baseline=1mm,baseline=1mm,every node/.style={circle,draw,minimum size=1.1cm},thick]
\node[] (1) []{$\t\gamma_{1,0}$};
\node[] (2) [below left=0.07 and -0.1 of 1]{$\t\gamma_{1,1}$};

\node[] (3) [below =1.4 of 2]{$\t\gamma_{2,0}$};
\node[] (4) [below right=0.07 and -0.1 of 3]{$\t\gamma_{2,1}$};

\node[] (5) [right =1.7 of 4]{$\t\gamma_{3,0}$};
\node[] (6) [above right=0.07 and -0.1 of 5]{$\t\gamma_{3,1}$};

\node[] (7) [above =1.4 of 6]{$\t\gamma_{4,0}$};
\node[] (8) [above left=0.07 and -0.1 of 7]{$\t\gamma_{4,1}$};

\draw[->-=0.5] (2) to   (3);
\draw[->-=0.5] (4) to   (5);
\draw[->-=0.5] (6) to   (7);
\draw[->-=0.5] (8) to   (1);

\node[] (1) at (7,-0.87) (1) []{$\gamma_{1}$};
\node[] (2) [below =1.5 of 1]{$\gamma_{2}$};
\node[] (3) [right =1.5 of 2]{$\gamma_{3}$};
\node[] (4) [above =1.5 of 3]{$\gamma_{4}$};
\draw[->>-=0.5] (1) to   (2);
\draw[->>-=0.5] (2) to   (3);
\draw[->>-=0.5] (3) to   (4);
\draw[->>-=0.5] (4) to   (1);
\end{tikzpicture}
\caption{Galois pair of the BPS quivers of the 5d $\KK E_5$ in the block notation (\textsc{Left}) and $\KK E_1$  (\textsc{Right}) theories, or local $dP_5$ and $\mathbb{F}_0$ respectively. The right-hand-side quiver corresponds to Phase~(a) of local $\mathbb{F}_0$.}
\label{Z4 symmetric quivers E5 E1}
\end{figure}
 
The quiver $\t Q$ for $dP_5$ again has an obvious $\mathbb Z_2$ symmetry. To show that it forms a $\mathbb Z_2$-Galois cover over the quiver $Q$ for the local $\mathbb F_0$ geometry, we thus only need to compare the superpotentials. Let us denote the nodes of $Q$ by $j=1,2,3,4$, with arrows $a_{ij}^a$, where we use again $a=1,2$ to enumerate the two arrows when $j=i+1$.

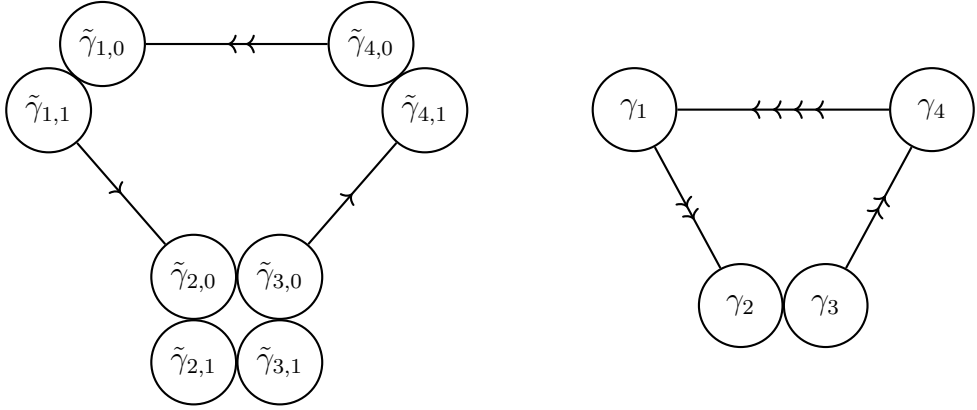
\begin{figure}[t]
\centering
\begin{tikzpicture}[baseline=1mm,baseline=1mm,every node/.style={circle,draw,minimum size=1.1cm},thick]
\node[] (1) []{$\t\gamma_{1,0}$};
\node[] (2) [below left=0.07 and -0.1 of 1]{$\t\gamma_{1,1}$};

\node[] (3) [below right =1.4 and 1.1 of 2]{$\t\gamma_{2,0}$};
\node[] (4) [below=-0.01 of 3]{$\t\gamma_{2,1}$};

\node[] (5) [right =-0.01 of 4]{$\t\gamma_{3,1}$};
\node[] (6) [above =-0.01 of 5]{$\t\gamma_{3,0}$};

\node[] (7) [above right =1.4 and 1.1 of 6]{$\t\gamma_{4,1}$};
\node[] (8) [above left=0.07 and -0.1 of 7]{$\t\gamma_{4,0}$};

\draw[->-=0.5] (2) to   (3);
\draw[->-=0.5] (6) to   (7);
\draw[->>-=0.5] (8) to   (1);

\node[] (1) at (7,-0.87) (1) []{$\gamma_{1}$};
\node[] (2) [below right =1.8 and 0.6 of 1]{$\gamma_{2}$};
\node[] (3) [right =-0.01 of 2]{$\gamma_{3}$};
\node[] (4) [above right =1.8 and 0.6 of 3]{$\gamma_{4}$};
\draw[->>-=0.55] (1) to   (2);
\draw[->>-=0.55] (3) to   (4);
\draw[->>>>-=0.55] (4) to   (1);
\end{tikzpicture}
\caption{Another Galois pair of BPS quivers of the 5d $\KK E_5$ (\textsc{Left}) and $\KK E_1$  (\textsc{Right}) theories, or local $dP_5$ and $\mathbb{F}_0$ respectively, both written in block notation. The right-hand-side quiver corresponds to Phase (b) of local $\mathbb{F}_0$; it can be obtained from the Phase~(a) quiver by a mutation on node $4$ of the latter (and with the relabelling of the nodes $3 \leftrightarrow 4$).
}
\label{fig:E1E5cover}
\end{figure}

The superpotential of the $\mathbb Z_4$ symmetric $\mathbb F_0$ quiver can be found for instance in~\cite[(17.1)]{Hanany:2012hi}, and it can be expressed as
\bea
\label{eq:WF0}
W_{\mathbb F_0}=\sum_{a,b,c,d}\epsilon_{ab}\epsilon_{cd} \Tr \Bigl( a^a_{12} a^{c}_{23} a_{34}^b a_{41}^d\Bigr)~,
\eea
where $a,b,c,d$ run from 1 to 2. For the $dP_5$ quiver, we again label the nodes by $j_\alpha$ with $j=1,2,3,4$ and $\alpha=0,1$, while the arrows are $(a_{ij}^a)_\alpha\colon i_\alpha \to j_{\alpha+d_a}$. We consider the symmetric gradings $d_{a_{ij}^a}=a-1$ for the arrows of $Q$. For the particular $dP_5$ quiver in figure~\ref{Z4 symmetric quivers E5 E1}, the superpotential has been worked out in~\cite[(6.1)]{Hanany:2012hi}. Using the map $j_\alpha=j+4\alpha$, where the right-hand side labels the nodes in the quiver~\cite[Figure 8]{Hanany:2012hi}, it can be expressed as: 
\bea
W_{dP_5}=\sum_{a,b,c,d}\epsilon_{ab}\epsilon_{cd}\sum_{\alpha\in\mathbb Z_2} \Tr \Bigl( (a^a_{12})_\alpha (a^{c}_{23})_{\alpha+a-1} (a_{34}^b)_{\alpha+a+c} (a_{41}^d)_{\alpha+d-1}\Bigr)~,
\eea
where we used the fact that $\alpha$ is a $\Z_2$-valued index, making the Galois covering structure manifest.

\item[\underline{Phase (b)}]
Performing a mutation on one node of the Phase (a) $\mathbb{F}_0$ quiver, one obtains the Phase (b) quiver shown in figure~\ref{fig:E1E5cover}, whose  superpotential~\cite[(17.13)]{Hanany:2012hi} we can write as
\bea
     W_{\mathbb F_0^{(b)}}=\sum_{a,b,c,d}\epsilon_{ab}\epsilon_{cd} \Tr \Bigl( a_{12}^a a_{24}^c A_{b,d}-a_{13}^a a_{34}^c A_{3-d, 3-b}     \Bigr)~, 
\eea
where $A_{b,d}$ is the matrix element of the matrix $A=\begin{psmallmatrix} a_{41}^3 & a_{41}^4 \\ a_{41}^2 & a_{41}^1\end{psmallmatrix}$. 
We consider the $\Z_2$-Galois cover defined through the gradings $d_{a_{ij}^a}=a-1 \mod 2$ . 
 The resulting quiver $\t Q$ is a known (pseudo-)$dP_5$ quiver~\cite{Feng:2000mi,Hanany:2012hi} displayed in figure~\ref{fig:E1E5cover}. 
One can check that (up to a relabelling of the nodes) the superpotential of the $\mathbb Z_2$ cover agrees with the superpotential~\cite[(6.16)]{Hanany:2012hi}:
\bea
\label{eq:WdP5}     W_{dP_5^{(d)}}=\sum_{a,b,c,d}\epsilon_{ab}\epsilon_{cd} \sum_{\alpha\in\mathbb Z_2}\Tr \Bigl(& (a_{12}^a)_{\alpha} (a_{24}^c)_{\alpha+a-1} (A_{b,d})_{\alpha+a+c} \\
     &-(a_{13}^a)_{\alpha} (a_{34}^c)_{\alpha+a-1} (A_{3-d, 3-b})_{\alpha+a+c}  \Bigr)~, 
\eea
Hence these two 5d BPS quivers also form a Galois pair.
\end{itemize}

\noindent
It is clear from this example that mutation and Galois covering essentially commute, in the sense that the Galois cover of the Phases (a) and (b) quivers for local $\mathbb{F}_0$ give us two distinct quivers for local $dP_5$ which are themselves related by `block mutation'. In general, a mutation at a node $j\in Q$ will correspond to a common mutation at all $|\G|$ nodes $(j,g)\in \t Q$. Interestingly, quiver mutation is only well-defined if the node $j$ has no self-loop ({\it i.e.} 1d adjoint chiral multiplet), which corresponds to the nodes $\{(j,g), g\in \G\}$ in $\t Q$ being a block of nodes with no arrows between them, so that they can be mutated independently within a block.

\subsection{Five-dimensional \texorpdfstring{$\SU(N)_0$}{SU(N)} examples}

The BPS quiver of the 5d SU$(2N)_0$ theory can be realised as a $\mathbb{Z}_N$ cover of the BPS quiver of the 5d $\SU(2)_0$ theory, also known as the $E_1$ theory. The latter is geometrically engineered with the local $\mathbb{F}_0\cong\mathbb{P}^1\times \mathbb{P}^1$ threefold geometry, and its quiver is displayed on the right-hand side of figure~\ref{Z4 symmetric quivers E5 E1} --- this is the so-called Phase (a) quiver~\cite{Hanany:2012hi}, which is related to the Phase (b) quiver shown in figure~\ref{fig:E1E5cover} by a mutation. Given the four nodes $j=1, \cdots, 4$, let us denote the arrows by:
\be
\h U_1 : 1\rightarrow 2~, \qquad
\h Z_1, \h Y_2 : 2\rightarrow 3~, \qquad
\h U_2 : 3\rightarrow 4~, \qquad
\h Y_0, \h Z_2 : 4\rightarrow 1~.
\ee
We can write the superpotential of the $\mathbb{F}_0$ quiver as:
\be\label{WF0a}
W_{\mathbb{F}_0}= \sum_{s=1}^2 \epsilon_{ab} \Tr\left(\h U_{s}^a \h Z_s \h U_{s+1}^b \h Y_{2s+2}\right)=  \epsilon_{ab}\Tr\left(\h U_{1}^a \h Z_1 \h U_{2}^b \h Y_{0}+\h U_{2}^a \h Z_2 \h U_{1}^b \h Y_{2} \right)~,
\ee
where the indices of $\h U$ and $\h Z$ are understood mod $2$ while the indices of $\h Y$ are given mod 4. 
 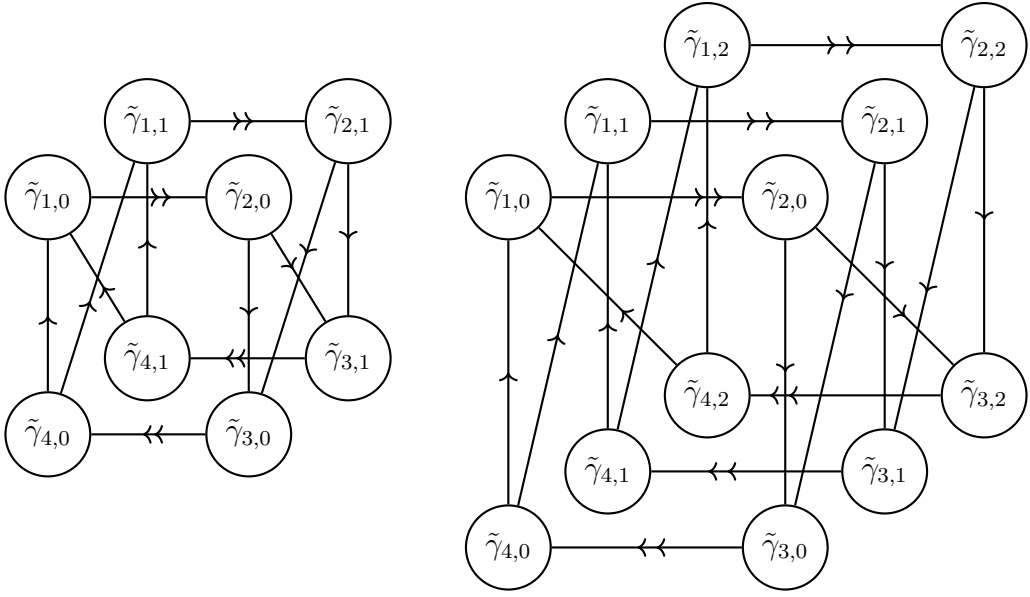
\begin{figure}[t]
\begin{center}
 \begin{tikzpicture}[baseline=1mm,baseline=1mm,every node/.style={circle,draw},thick]
\node[] (1) []{$\tilde\gamma_{1,0}$};
\node[] (2) [above right=0.2 and 0.5 of 1]{$\tilde\gamma_{1,1}$};
\node[] (3) [right =1.5 of 1]{$\tilde\gamma_{2,0}$};
\node[] (4) [above right=0.2 and 0.5 of 3]{$\tilde\gamma_{2,1}$};
\node[] (7) [below =2 of 1]{$\tilde\gamma_{4,0}$};
\node[] (8) [above right=0.2 and 0.5 of 7]{$\tilde\gamma_{4,1}$};
\node[] (5) [right =1.5 of 7]{$\tilde\gamma_{3,0}$};
\node[] (6) [above right=0.2 and 0.5 of 5]{$\tilde\gamma_{3,1}$};
\draw[->>-=0.65] (1) to   (3);
\draw[->>-=0.5] (2) to   (4);
\draw[->-=0.5] (3) to   (5);
\draw[->-=0.4] (3) to   (6);
\draw[->-=0.4] (4) to   (5);
\draw[->-=0.5] (4) to   (6);
\draw[->-=0.5] (7) to   (1);
\draw[->-=0.4] (7) to   (2);
\draw[->-=0.4] (8) to   (1);
\draw[->-=0.5] (8) to   (2);
\draw[->>-=0.5] (5) to   (7);
\draw[->>-=0.65] (6) to   (8);
\end{tikzpicture}
\qquad
 \begin{tikzpicture}[baseline=1mm,baseline=1mm,every node/.style={circle,draw},thick]
\node[] (1) []{$\tilde\gamma_{1,0}$};
\node[] (2) [above right=0.2 and 0.5 of 1]{$\tilde\gamma_{1,1}$};
\node[] (3) [above right=0.2 and 0.5 of 2]{$\tilde\gamma_{1,2}$};
\node[] (4) [right =2.5 of 1]{$\tilde\gamma_{2,0}$};
\node[] (5) [above right=0.2 and 0.5 of 4]{$\tilde\gamma_{2,1}$};
\node[] (6) [above right=0.2 and 0.5 of 5]{$\tilde\gamma_{2,2}$};
\node[] (10) [below =3.5 of 1]{$\tilde\gamma_{4,0}$};
\node[] (11) [above right=0.2 and 0.5 of 10]{$\tilde\gamma_{4,1}$};
\node[] (12) [above right=0.2 and 0.5 of 11]{$\tilde\gamma_{4,2}$};
\node[] (7) [right =2.5 of 10]{$\tilde\gamma_{3,0}$};
\node[] (8) [above right=0.2 and 0.5 of 7]{$\tilde\gamma_{3,1}$};
\node[] (9) [above right=0.2 and 0.5 of 8]{$\tilde\gamma_{3,2}$};
\draw[->>-=0.85] (1) to   (4);
\draw[->>-=0.6] (2) to   (5);
\draw[->>-=0.5] (3) to   (6);
\draw[->-=0.65] (4) to   (9);
\draw[->-=0.5] (4) to   (7);
\draw[->-=0.4] (5) to   (8);
\draw[->-=0.4] (5) to   (7);
\draw[->-=0.6] (6) to   (8);
\draw[->-=0.5] (6) to   (9);
\draw[->>-=0.5] (7) to   (10);
\draw[->>-=0.65] (8) to   (11);
\draw[->>-=0.85] (9) to   (12);
\draw[->-=0.5] (10) to   (1);
\draw[->-=0.5] (10) to   (2);
\draw[->-=0.4] (11) to   (2);
\draw[->-=0.5] (11) to   (3);
\draw[->-=0.4] (12) to   (1);
\draw[->-=0.5] (12) to   (3);
\end{tikzpicture}
\caption{\textsc{Left:} $\mathbb{Z}_2$-cover of the $\mathbb{F}_0$ quiver, giving us the quiver for 5d $\SU(4)_0$ SYM. \textsc{Right:} $\mathbb{Z}_3$-cover of the $\mathbb{F}_0$ quiver, giving us the quiver for 5d $\SU(6)_0$ SYM.}
\label{su40quiver}
\end{center}
\end{figure}
The BPS quiver for 5d SU$(2N)_0$ is then obtained by assigning non-trivial $\Z_N$ gradings to two arrows:
\be
d(\h Z_2)= 1~, \qquad d(\h Y_2)=-1~.
\ee
and zero grading to all other arrows. The Galois covering quiver has nodes $(j, \alpha)$ for $\alpha=0, \cdots, N-1$, with the arrows:
\bea
&\h U_{1,\alpha}= U_{1+2\alpha}~, \quad
&&\h Z_{1,\alpha}= Z_{1+2\alpha}~, \quad
&&\h Y_{0,\alpha}= \t Y_{4+4\alpha}~, \\
&\h U_{2,\alpha}= U_{2+2\alpha}~, \quad
&&\h Z_{2,\alpha}= Z_{2+2\alpha}~, \quad
&&\h Y_{2,\alpha}= \t Y_{6+4\alpha}~, 
\eea
to match the notation of~\cite{Closset:2019juk}, so that the superpotential is given by:
\be\label{WSU2N}
W_{2N}=\sum_{l=1}^{2N}\epsilon_{ab}\Tr\left(U_l^a Z_l U_{l+1}^b \t Y_{2l+2}\right)~,
\ee
with the indices of $U$ and $Z$ understood to be mod $2N$ and the index of $\h Y$ understood mod $4N$. The Galois covering quivers $\t Q$ for $N=2$ and $N=3$ are shown in figure~\ref{su40quiver}, reproducing the known SU$(4)_0$ and SU$(6)_0$ 5d BPS quivers. 
One easily checks that the superpotentials are also related by the Galois cover as in~\eqref{tilde W}. Rewriting~\eqref{WSU2N} as
\bea\label{W 2N}
W_{2N}=\sum_{s=1}^{2}\epsilon_{ab}\sum_{\alpha\in\mathbb Z_N} \Tr\left(U_{s+2\alpha}^a Z_{s+2\alpha} U_{s+1+2\alpha}^b \t Y_{2s+2+4\alpha}\right)
\eea
makes it obvious that it is indeed obtained as a Galois cover of~\eqref{WF0a}.

More generally, the 5d $\SU(N)_0$ BPS quiver is a $\Z_N$ Galois cover of the conifold quiver, which is the BPS quiver for the 5d hypermultiplet --- see section~\ref{sec:ConP1P1} for an explicit discussion in the $N=2$ case.

\subsection{The local \texorpdfstring{$dP_3$}{dP3} quiver as a Galois cover}\label{sec:Galois_cover_dP3}

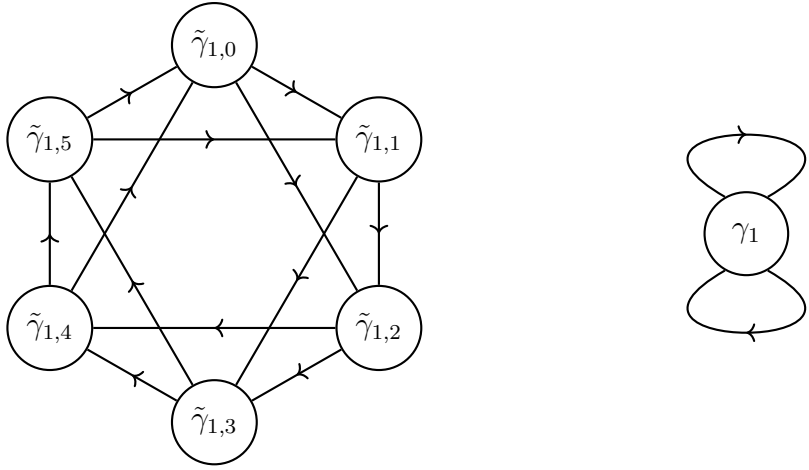
\begin{figure}[t]
\centering
\begin{tikzpicture}[baseline=1mm, every node/.style={circle,draw,minimum size=1.1cm}, thick]
  \foreach \i in {0,...,5} {
    \node (\i) at ({90 - 60*\i}:2.5cm) {$\t\gamma_{1,\i}$};
  }
  \foreach \i in {0,...,5} {
    \pgfmathtruncatemacro{\j}{mod(\i+1,6)}
    \draw[->-=0.5] (\i) -- (\j);
  }
  \foreach \i in {0,...,5} {
    \pgfmathtruncatemacro{\j}{mod(\i+2,6)}
    \draw[->-=0.5] (\i) to (\j);
  }
\node[] at (7,0) (1) []{$\gamma_1$};
\draw[->-=0.5] (1.120) to[out=150, in=30, looseness=10] (1.60);
\draw[->-=0.5] (1.300) to[out=330, in=210, looseness=10] (1.240);
\end{tikzpicture}
\caption{Galois pair of the BPS quiver $\t Q_{(6)}$ of $\KK E_3$ (\textsc{Left}) and a  single-node quiver $Q$ (\textsc{Right}). This $\Z_6$ Galois pair corresponds to the grading $d_a=(1,2)$ of the two arrows of $Q$.
\label{quivers dP3}}
\end{figure}

As another interesting example of a 5d BPS quiver which can be understood as a Galois cover, consider the $\KK E_3$ theory --- the circle compactification of the 5d SCFT corresponding to the 5d $SU(2)$ gauge theory with $N_f=2$ flavours. The corresponding BPS quiver, shown in figure~\ref{quivers dP3}, is a CY$_3$ quiver for the local $dP_3$ geometry. That particular quiver has an obvious $\Z_6$ symmetry permuting the 6 nodes, and indeed it can be understood as a $\G=\Z_6$ Galois cover of the one-node quiver $Q$ with two self-arrows shown on the right-hand side of figure~\ref{quivers dP3}. Let us denote by $a$ and $b$ the arrows of $Q$. The $\Z_6$ gradings $d_a=1$ and $d_b=2$ results in the local $dP_3$ quiver, as anticipated. Importantly, we also need to consider the superpotentials. We consider the following sextic superpotential for $Q$:
\be
W_Q = \Tr\left({1\ov 6} a^6 + {1\ov 3}b^3 -{1\ov 2} (ab)^2\right)~.
\ee
Then, denoting the arrows of the covering quiver by $a_\alpha : 1_\alpha\rightarrow 1_{\alpha+1}$ and $b_\alpha: 1_\alpha\rightarrow 1_{\alpha+2}$ with $\alpha\in \Z_6$, the superpotential of the $\Z_6$ Galois cover reads:
\bea
&W_{dP_3} &=& \;\;{ \Tr \Big(a_0 a_1 a_2 a_3 a_4 a_5 + b_0 b_2 b_4 + b_1 b_3 b_5} \\
&&&\qquad\quad\;{ - a_0 b_1 a_3 b_4 - a_1 b_2 a_4 b_5 - a_2 b_3 a_5 b_0\Big)~,}
\eea
which is exactly the $dP_3$ superpotential~\cite[equation (12.1)]{Hanany:2012hi}, thus proving that the $\KK E_3$ BPS quiver is indeed a $\Z_6$ Galois cover. Note that the one-node quiver $Q$ is not a CY$_3$ quiver --- we will explain in the next section that the Galois cover $\t Q$ of a CY$_3$ quiver $Q$ is itself a CY$_3$ quiver, but the converse is not necessarily true.

\begin{figure}[t]
\centering

\begin{tikzpicture}[baseline=1mm, every node/.style={circle,draw,minimum size=1.1cm}, thick]
  \foreach \i in {0,1,2} {
    \node (\i) at ({90 - 120*\i}:2cm) {$\tilde\gamma_{1,\i}$};
  }
  \draw[->-=0.5, bend left=15] (0) to (1);
  \draw[->-=0.5, bend left=15] (1) to (0);
  \draw[->-=0.5, bend left=15] (1) to (2);
  \draw[->-=0.5, bend left=15] (2) to (1);
  \draw[->-=0.5, bend left=15] (0) to (2);
  \draw[->-=0.5, bend left=15] (2) to (0);
  \node (A) at (5.5, 0.5) {$\tilde\gamma_{1,0}$};
  \node (B) at (7.5, 0.5) {$\tilde\gamma_{1,1}$};
 \draw[->-=0.5, bend left=20] (A) to (B);
  \draw[->-=0.5, bend left=20] (B) to (A);
  \draw[->] (A) edge[loop left]  (A);
  \draw[->] (B) edge[loop right] (B);
\end{tikzpicture}
\caption{The $\Z_3$ cover $\t Q_{(3)}$ (\textsc{Left}) and $\Z_2$ cover $\t Q_{(2)}$ (\textsc{Right}) of the one-node quiver $Q$ of figure~\protect\ref{quivers dP3}.
\label{quivers related to dP3}}
\end{figure}
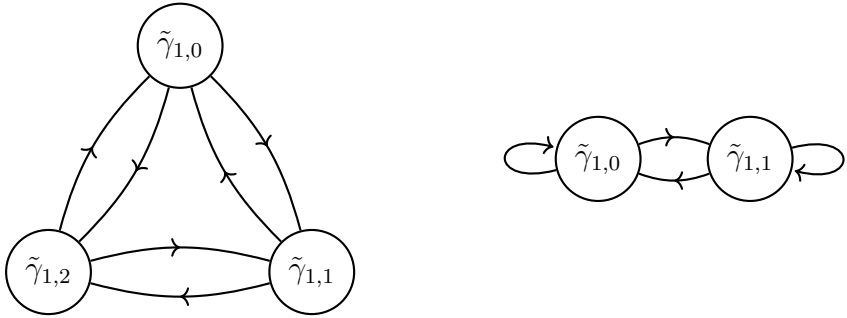

\medskip
\noindent
{\bf $\Z_2$ and $\Z_3$ covers of $Q$.} Since $\Z_6\cong \Z_2\times \Z_3$, there exists intermediate $\Z_3$ and $\Z_2$ Galois covers of the one-node quiver $Q$, denoted here by $\t Q_{(3)}$ and $\t Q_{(2)}$, respectively, such that
\be
\t Q_{(k)}/\Z_k = Q \quad (k=2,3,6),  \qquad \t Q_{(6)}/\Z_2 = \t Q_{(3)}~, \qquad  \t Q_{(6)}/\Z_3 = \t Q_{(2)}~,
\ee
where $\t Q_{(6)}$ denotes the $dP_3$ quiver. These intermediate Galois covers are displayed in figure~\ref{quivers related to dP3}. The superpotential of the $\Z_3$ cover reads
\bea
&W_{\t Q_{(3)}} &=& \;\;{ \Tr \Big(\half (a_0 a_1 a_2)^2 + b_0 b_2 b_1 } \\
&&&\qquad\quad\;{ - \half (a_0 b_1)^2 - \half (a_1 b_2)^2  - \half (a_2 b_1)^2\Big)~,}
\eea
where we have the arrows $a_\alpha: 1_\alpha\rightarrow 1_{\alpha+1}$ and  $b_\alpha: 1_\alpha\rightarrow 1_{\alpha+2}$ for $\alpha\in \Z_3$. Similarly, the superpotential for the $\Z_2$ cover is given by
\be
W_{\t Q_{(2)}}  = \Tr \Big({1\ov 3} (a_0 a_1)^3 + {1\ov 3} b_0^3+{1\ov 3}b_1^3 - a_0 b_1 a_1 b_0 \Big)~,
\ee
with $a_\alpha$ and $b_\alpha$ defined as before but with $\alpha\in \Z_2$. These two intermediate Galois covers were first considered in~\cite{Bridgeland:2024saj}.%
\footnote{We thank Fabrizio Del Monte for discussions on these examples.} None of the 3 quotients of the $dP_3$ quiver are themselves BPS quivers of known 4d $\CN=2$ theories, as far as we know.

\section{Graded representations and BPS states}

In this section, we explore the relationship between quiver representations of Galois pairs  $Q$ and $\t Q$. As a simple consequence of this discussion, we will see that Galois coverings of 5d BPS quivers correspond to orbifolds in M-theory. Finally, we will conjecture a precise formula relating BPS invariants of distinct BPS quivers related by a Galois covering.

\subsection{Induced actions on quiver representations}
Galois covers of quivers induce relationships between quiver representations --- more precisely, we have induced functors between the corresponding categories of quiver representations. 
There are two distinct Galois functors to consider, covariant and contravariant with respect to the Galois cover~\eqref{F Galois cover}, which correspond neatly to the two distinct definitions of Galois covering given in section~\ref{sec:GCoverQuiver}.

\subsubsection{Galois functors on quiver representations} \label{sec:Galois functors quiver rep}
Without any real loss of generality, we may fix $\G=\Z_n$.  For $\t Q$ any $\Z_n$ Galois cover of a quiver $Q$, we write the dimension vector of a representation $\t R$ of $\t Q$ as:
\be
\label{eq:tildeN}
\tilde N= (N_{j,\alpha})=(N_{1,0}, \dots, N_{1,n-1};N_{2,0}, \dots ,N_{2,n-1};\dots; N_{|Q_0|,0}, \dots ,N_{|Q_0|,n-1})~.
\ee
Given a Galois cover~\eqref{F Galois cover}, we have the following functors between categories of quiver representations. 

\medskip
\noindent {\bf Push-down functor.} 
 The push-down ({\it i.e.}~pushforward) functor
\be
F_\ast \; : \; {\bf Rep}(\t Q) \longrightarrow {\bf Rep}(Q)
\ee
maps a representation $\tilde R$ of $\tilde Q$ to a representation $R$ of $Q$ according to:
\be
V_j=\oplus_\alpha \tilde V_{j,\alpha}~,\qquad X_a=\oplus_\alpha \tilde X_{a,\alpha}~.
\ee
The entries of the dimension vector $N=F_\ast(\tilde N)$ are thus given by:
\be
N_j = F_\ast(\t N)_j =\sum_{\alpha=0}^{n-1} N_{j,\alpha}~.
\ee
Given the identification \eqref{gamma to N} between dimension vectors $N$ and $\gamma\in \Gamma$, we also use the notation $\tilde
\gamma=F^\ast(\gamma)$ and $\gamma=F_\ast(\tilde \gamma)$, with:
\be
\label{eq:Flambdagamma}
F_\ast\; : \;\t\Gamma \rightarrow \Gamma\; :\; \t\gamma \mapsto \gamma= F_\ast(\tilde \gamma)=\sum_{j\in Q_0} \left( \sum_{\alpha=0}^{n-1} N_{j,\alpha}\right) \gamma_{j}~.
\ee

\medskip
\noindent 
{\bf Pull-up functor.} 
The pull-up ({\it i.e.}~pull-back) functor
\be\label{pullup F}
F^\ast \; : \; {\bf Rep}(Q) \longrightarrow {\bf Rep}(\t Q)
\ee
 maps a representation $R=(\{V_j\},\{X_a\})$ of $Q$ to a
representation $\tilde R=(\{\tilde V_{j,\alpha}\},\{\tilde X_{a,\beta}\})$ of $\tilde Q$ according to:
\be
\tilde V_{j,\alpha}=V_j~,\qquad \tilde X_{a,\beta}=X_a,\qquad \text{for
  all }\alpha,\beta=0,\dots,n-1~.
\ee
The entries of the dimension vector $\tilde N=F^\ast(N)$ are thus given by $\tilde N_{j,\alpha}=N_j$ for all $\alpha\in \Z_n$.

\medskip
\noindent
The superpotentials $W$ and $\tilde W$ of the Galois pair $(Q, \t Q)$ satisfy
\be
\tilde W(F^\ast R)=n\,  W(R)~,\qquad 
 W(F_\ast \tilde R)=\tilde W(\tilde R)~.
\ee
One easily checks that the induced functors satisfy the relations:
\be
F_\ast F^\ast R= \bigoplus_{g\in \G} R = n R~,\qquad\quad
F^\ast F_\ast \t R=\bigoplus_{g\in \G} \t R^g~,
\ee
where $\t R^g$ denotes the $\G$-transformed representation of the quiver $\t Q$ under the action of $g\in \G$.

\subsubsection{Graded representations and equivariant modules}
To further explore how Galois covers relate quiver representations, it is convenient to use the standard isomorphism between quiver representations and path-algebra modules:
\be
{\bf Rep}(Q) \cong  {\bf mod}\text{-}\CA_{Q}~.
\ee
Here ${\bf mod}\text{-}\CA_{Q}$ denotes the category of right-$\CA_Q$ modules, where $\CA_Q$ is the path algebra~\eqref{def AQ}.

\medskip
\noindent {\bf Graded representations.} Given the grading~\eqref{G grading arrows da} on the arrows of $Q$, it is useful to define the notion of a {\it graded representation} $(R)$, which is a representation $R=(V, X)$ consisting of graded vector spaces for each node, and of linear maps that respect the grading:
\be
V_j = \bigoplus_{g\in \G} V_{j, g}~, \qquad \qquad X_a|_{V_{t(a), g}}
\, : \, V_{t(a), g} \rightarrow V_{h(a), g+d_a}~.
\ee
Equivalently, this construction defines a $\G$-equivariant $\CA_Q$-module. Note that $\G$ acts on the path algebra according to the grading. For $\G=\Z_n$, the group action is generated by:
\be\label{G action on ej a}
\G\; : \; \qquad e_j \rightarrow e_j~, \qquad a \rightarrow e^{{2\pi i \ov n}d_a} \,a~,
\ee
on idempotents and on simple arrows, respectively --- the grading on any path is additive.%
\footnote{More generally, for any abelian group $\G$, we have $a\rightarrow \chi_{d_a}(g) \,a$ for any $g\in \G$, where $\chi_{d_a} \in \h \G = {\rm Hom}(\G, {\rm U}(1))\cong \G$ is a character of $\G $.} 

\medskip
\noindent {\bf Relations amongst module categories.} 
The definition of the Galois cover $\t Q$ in terms of gradings of the arrows of $Q$ can be usefully recast in terms of the path algebras. The key result is that the path algebra of $\t Q$ is Morita-equivalent to the skew group algebra of $\CA_Q$ with $\G$~\cite{REITEN1985224, Demonet_2010} --- that is, their categories of modules are isomorphic:
\be
{\bf mod}\text{-}\CA_{\t Q} \cong {\bf mod}\text{-}\left(\CA_Q \ast \G\right)~.
\ee
The skew group algebra construction parallels the grading construction of section~\ref{sec:GCoverQuiver}. In particular, the algebra $ \CA_Q \ast \G$ is generated as a vector space by the elements $(p, g)$ for $p\in \CA_Q$ and $g\in \G$. The idempotent elements $e_j \in \CA_Q$ give us the $|\t Q_0|= |\G| |Q_0|$  idempotent elements $(e_j, g)\in \CA_Q\ast \G$, which is equivalent to the fact that we have the nodes $(j, g)\in \t Q_0$ in the Galois covering quiver. Similarly, the arrows $(a:j \rightarrow k)\in Q_1$ are the length-one paths in $\CA_Q$, and we correspondingly find the unit-length paths $(a, g) \in \CA_Q\ast \G$ --- this corresponds to the arrows 
\be
(a, g)\, :\, (j, g)\rightarrow (k, g+d_a)~, 
\ee
in $\t Q_1$, with $d_a$ the $\G$-grading of the arrow.
\footnote{Here for $\G=\Z_n$ or, more generally, using the abelian notation for the product `$+$' in $\G$.}  
More generally, a standard result on skew group algebras relates the above construction to $\G$-equivariant $\CA_Q$-modules:
\be\label{modAQtilde equals Gequiv modAQ}
{\bf mod}\text{-}\CA_{\t Q} = {\bf mod}^\G\text{-}\CA_{Q}~.
\ee
Namely, the category of quiver representations of $\t Q$ is isomorphic to the category of $\G$-graded representations of $Q$. Note that ${\bf mod}\text{-}\CA_{Q}$ embeds into the category~\eqref{modAQtilde equals Gequiv modAQ} by the pull-up functor~\eqref{pullup F}, whose image consists of the $\G$-invariant quiver representations of $\t Q$. 

\medskip
\noindent
{\bf Relations between centres.}  While the path algebra $\CA_Q$ is non-commutative, it also contains a non-trivial centre, $Z(\CA_Q)$. In particular, the grading induces a $\G$-action on the centre,
\be\label{G action on center of AQ}
\G \; : \; Z(\CA_Q)\rightarrow Z(\CA_Q)~, 
\ee
which can be determined from~\eqref{G action on ej a}.  
It follows from~\eqref{modAQtilde equals Gequiv modAQ} that the centre of $\CA_{\t Q}$ consists of the $\G$-invariant elements of $\CA_{Q}$:
\be\label{rel between centers}
Z(\CA_{\t Q})= Z(\CA_Q)^\G~.
\ee
This is particularly interesting when we consider 5d BPS quivers, as we now explain.

\subsection{Galois covers of 5d BPS quivers as orbifolds}\label{subsec:Galois and orbifold}

Consider a quiver $Q$ which is the non-commutative crepant resolution (NCCR) of a Calabi--Yau threefold singularity~\cite{bergh2002non}, as is the case for 5d BPS quivers of 5d SCFTs geometrically engineered in M-theory~\cite{Closset:2019juk} --- more precisely, the NCCR is the path algebra $\CA_Q$.

The Galois covering of $Q$ by $\G$ has a very simple geometric interpretation in this case:  $\t Q$ is a 5d BPS quiver corresponding to an orbifold ${\bf Y}={\bf X}/\G$  of the singularity ${\bf X}$  by $\G$. Indeed, by definition of a NCCR, the ring defining the affine variety ${\bf X}$ is the centre of the non-commutative algebra $\CA_Q$: 
\be
{\bf X}= {\rm Spec}\, Z(\CA_Q)~.
\ee
The $\G$-action~\eqref{G action on center of AQ} on  $Z(\CA_Q)\cong \CO_{\bf X}$ determines a discrete action on the affine variety ${\bf X}$ itself. 
It then directly follows from~\eqref{rel between centers} that:
\be
{\bf Y}= {\rm Spec}\,Z(\CA_{\t Q})= {\rm Spec}\, Z(\CA_{Q})^\G = {\bf X}/\G~.
\ee
One can check that $\t Q$ provides us with a NCCR of the orbifold singularity ${\bf Y}$, because the properties defining a NCCR are preserved under Galois covering. We should also note that, whenever the canonical singularity ${\bf X}$ admits a smooth crepant resolution $\pi: \t {\bf X}\rightarrow {\bf X}$, we should have a derived equivalence between $\CA_Q$-modules and compactly-supported coherent sheaves~\cite{bergh2002non}:
\be
D\left({\bf mod}\text{-}\CA_{Q}\right) \cong D_c(\t {\bf X})~.
\ee
Then the relation~\eqref{modAQtilde equals Gequiv modAQ} becomes:
\be\label{McKay rel}
D_c(\t {\bf Y}) \cong D_c^\G(\t {\bf X})~,
\ee
where $\t{\bf Y}$ is a crepant resolution of ${\bf Y}={\bf X}/\G$ and the right-hand side denotes the $\G$-equivariant derived category of $\t {\bf X}$~\cite{bridgeland1999mukai}.

This Galois-covering construction and the relations~\eqref{modAQtilde equals Gequiv modAQ} or~\eqref{McKay rel} are best known to physicists as the `generalised McKay correspondence' or as the Douglas--Moore construction of quivers for orbifolds~\cite{Douglas:1996sw, Kachru:1998ys} --- especially in the case of orbifolds of flat space, ${\bf X}= \C^3$. This very same construction was recently discussed in closely related contexts in~\cite{Bridgeland:2024saj, Collinucci:2025rrh, Dramburg:2025tlb}.%
\footnote{As also discussed in those references, the generalised McKay correspondence goes beyond Galois covers. This is because the orbifold group $\G$ need not be abelian and its `quantum' action on $\t Q$ need not be free. This lead to a number of generalisations beyond what we discuss in this paper, which we hope to come back to in future work.}

\subsection{Gauging and ungauging, and relations between BPS invariants}
\label{sec:gauging and ungauging}
We are interested in determining relations between the BPS invariants of the BPS quiver $Q$ and of its Galois cover $\tilde Q$. At broad strokes, such relations are easily understood by thinking of $\G$ as a symmetry of $\t Q$, with the Galois covering map~\eqref{F Galois cover} corresponding to a gauging of this discrete symmetry.

\medskip
\noindent 
{\bf Gauging and ungauging in $\CN=4$ SQM.} To the Galois covering quiver $\t Q$, we associate a `meta' $\CN=2$ supersymmetric quantum mechanics (MSQM) obtained by summing over all dimension vectors $\t N$:
\be\label{def MSQM}
{\rm MSQM}[\t Q] = \bigoplus_{\t\gamma\in \t\Gamma_+} {\rm SQM}_{\t\gamma}[\t Q]~.
\ee
Here we use the notation $\t\gamma \cong (\t N)$ for the dimension vectors of the quiver $\t Q$, corresponding to $\t N_{j,g}\geq 0$ for all $(j, g)\in \t Q_0$. The Galois covering group $\G$ acts as a discrete symmetry of this MSQM by permuting the charge vectors:
\be\label{G0 act on MSQM}
\G^{(0)} \, :\, {\rm MSQM}[\t Q]\rightarrow {\rm MSQM}[\t Q] \; :\; \t N_{j, g} \mapsto \t N_{j,g+h}~, \qquad \text{for any }\, h\in \G~.
\ee
In particular, each specific ${\rm SQM}_{\t\gamma}$ with a fixed charge $\t\gamma$ sits in a $\G$-orbit $[\t\gamma]$ of SQMs of cardinality $|[\t\gamma]|$ and with stabiliser ${\rm Stab}(\t\gamma)\subseteq \G$.

The MSQM~\eqref{def MSQM} only admits the discrete symmetry $\G$ if the stability parameters $\t\zeta_{j, g}$ are chosen to be $\G$-covariant inside each orbit $[\t\gamma]$. This amounts to choosing the FI parameters $\t\zeta_{j ,g}=\zeta_j$ for all $g\in \G$, where $\zeta_j$ will be identified with the FI parameters of $Q=\t Q/\G$ --- in other words, $\t\zeta= F^\ast(\zeta)$. Equivalently, we identify the central charges as:
\be\label{alligned central charges}
\t Z(\tilde \gamma_{j,g})=Z(\gamma_j)~, \qquad  \text{for all }\, g\in \G~,
\ee
which then determines the FI parameters as in~\eqref{eq:zetaZ}. It is important to note that this is a fine-tuned locus in the larger space of stability conditions for $\t Q$. In particular, since central charges align according to~\eqref{alligned central charges}, we are generally on walls of marginal stability for $\t Q$. Such fine-tuned loci are essentially the collimation chambers studied in the 5d BPS quiver context in~\cite{Closset:2019juk, DelMonte:2021ytz, DelMonte:2022kxh}; in the context of CY$_3$ quivers, these $\G$-invariant stability conditions were recently established in~\cite[theorem 3.1]{Bridgeland:2024saj}, where they were also recast in the language of Bridgeland stability conditions for triangulated categories.%
\footnote{In~\protect\cite{Bridgeland:2024saj}, it was found that the $\G$-invariant stability conditions of $\t Q$ correspond to the $\h\G$-invariant stability conditions of the quiver $Q$, when the two are related by a generalised McKay correspondence as discussed in subsection~\protect\ref{subsec:Galois and orbifold}. For our Galois cover with $\G$ abelian, the group of characters $\h \G\cong \G$ acts trivially on the stability parameters of $Q$ by assumption (it acts on the arrows but not on the nodes), so our discussion simplifies accordingly.}

Given the MSQM~\eqref{def MSQM} with a discrete symmetry $\G$, we can consider gauging the symmetry to obtain the MSQM for $Q=\t Q/\G$. We can certainly consider this gauging at the level of the BPS Hilbert space --- that is, the supersymmetric ground states of~\eqref{def MSQM}:
\be\label{tQ BPS H}
\CH^{(\t Q)}_{\rm BPS}\equiv \bigoplus_{\t\gamma\in \t\Gamma_+} \CH_{\t\gamma}^{(\t Q)}~,
\ee
on which $\G$ acts as $\t\gamma \rightarrow g\cdot \t \gamma$. The symmetry~\eqref{G0 act on MSQM} is an ordinary symmetry of the quantum mechanics --- that is, a zero-form symmetry $\G^{(0)}$ in modern terminology~\cite{Gaiotto:2014kfa}. Gauging $\G^{(0)}$ amounts to projecting over the $\G$-invariant states, which are simply the orbits $[\t \gamma]$. We then have:
\be
\CH_{\rm BPS}^{(Q)} \cong \CH_{\rm BPS}^{(\t Q)}\big/ \G^{(0)}~.
\ee
Working at the level of the (formal) Witten indices~\eqref{def CI Witten} written as:
\be
\CI_{\rm BPS}^{Q} \equiv \Tr_{\CH_{\rm BPS}^{(Q)}} \left((-1)^{\rm F}\right)= \sum_{\gamma\in \Gamma_+} \CI_\gamma^{Q}~,\qquad \CI_{\gamma}^{Q} \equiv  \Tr_{\CH_{\gamma}^{(Q)}}\left((-1)^{\rm F}\right)~,
\ee
the gauging of the one-form symmetry corresponds to the insertion of the $\G^{(0)}$ symmetry operators as follows:
\be
\CI_{\rm BPS}^{Q}={1\ov |\G|}\sum_{g\in \G} \Tr_{\CH_{\rm BPS}^{(\t Q)}} \left( g(-1)^{\rm F}\right)~.
\ee
More generally, the $\G$-equivariant states are indexed by the characters of $\G^{(0)}$ as:
\be
\ket{[\t\gamma]; \chi} \equiv {1\ov\sqrt{|\G|}} \sum_{g\in \G} \chi(g)\, \ket{g\cdot \t\gamma}~,\qquad\quad
\text{with }  \; \chi\in \h \G ={\rm Hom}(\t \G, {\rm U}(1))~.
\ee
The invariant states correspond to $\chi=1$. Like for any discrete symmetry in quantum mechanics, the group characters index superselection sectors and $\h\G\cong \G^{(-1)}$ is the `quantum' $(-1)$-form symmetry of the gauged theory. When gauging with the $\chi$-twist, we simply project onto the corresponding superselection sector --- the insertion of the character corresponds to the insertion of a 1d SPT phase for $\G^{(0)}$, in modern language. At the level of the Witten index, we have:
\be
\CI_{\rm BPS}^{Q,\, \chi} \equiv {1\ov |\G|} \sum_{g\in \G} \chi(g) \Tr_{\CH_{\rm BPS}^{(\t Q)}} \left( g(-1)^{\rm F}\right)~.
\ee
The original Hilbert space~\eqref{tQ BPS H} is recovered by summing over all characters --- this is the `ungauging' of the $\G$ symmetry. From this perspective, the grading action~\eqref{G grading arrows da} on $Q$ can be interpreted in terms of the $\G^{(-1)}$ action which permutes the distinct sectors of $\G$-equivariant $\CA_Q$-modules, and the ungauging procedure is the physical realisation of the algebraic relation~\eqref{modAQtilde equals Gequiv modAQ}.

\medskip
\noindent
{\bf Relations between 1d Witten indices.} Given a Galois pair $(Q, \t Q)$, the physical picture of $\G$ gauging and ungauging suggests simple relations between BPS states and, hence, between invariants of the two distinct quivers. First, consider decomposing~\eqref{tQ BPS H} using the pull-down functor on dimension vectors:
\be
\CH^{(\t Q)}_{\rm BPS}\equiv \bigoplus_{\gamma\in \Gamma_+} \CH_{\gamma}^{(\t Q)}~,\qquad \quad \text{with }\;  \CH_{\gamma}^{(\t Q)}\equiv \bigoplus_{\t\gamma\in \t\Gamma_+ \, |\, F_\ast(\t\gamma)=\gamma} \CH_{\t\gamma}^{(\t Q)}~.
\ee
In each $\gamma$-sector, we  then have
\be\label{rel between H BPS}
\CH_{\gamma}^{(Q)} \cong  \CH_{\gamma}^{(\t Q)}\big/\G^{(0)}~, 
\ee
and we are now dealing with finite-dimensional Hilbert spaces. We can then compute the Witten index of the $\gamma$-sector in the 1d gauged theory as
\be
\CI_{\gamma}^{Q} = {1\ov |\G|} \sum_{g\in \G}   \Tr_{\CH_{\gamma}^{(\t Q)}} \left( g(-1)^{\rm F}\right)
= {1\ov |\G|}\sum_{\t\gamma \, |\, F_\ast(\t\gamma)=\gamma}  \sum_{g\in {\rm Stab}(\t\gamma)} \CI^{\t Q}_{\t\gamma}~,
\ee
where we used the fact that the insertion of $g$ yields zero unless it is in the stabiliser of $\t\gamma$ since $\braket{\t\gamma}{g \cdot\t \gamma}=0$ if $g\cdot \gamma \neq \gamma$. We then find:
\be\label{rel between ZQs}
 \CI_{\gamma}^{Q} ={1\ov |\G|}\sum_{\t\gamma \, |\, F_\ast(\t\gamma)=\gamma}   |{\rm Stab}(\t\gamma)|\, \CI^{\t Q}_{\t\gamma}
 =\sum_{[\t\gamma] \, |\, F_\ast(\t\gamma)=\gamma}    \CI_{[\t\gamma]}^{\t Q}~.
\ee
In the last equality, we used the $\G^{(0)}$ symmetry which implies that the Witten index $\CI_{[\t\gamma]}^{\t Q}\equiv \CI_{\t\gamma}^{\t Q}$ must be the same for every $\t\gamma\in [\t\gamma]$, together with the orbit-stabiliser theorem. In other words, the gauging~\eqref{rel between H BPS} gives us BPS states of $Q$ which are indexed by the $\G^{(0)}$ orbits of BPS states of the quiver $\t Q$, as expected. 
We therefore derived a precise relationship between the Witten indices of the Galois pair, in principle. The difficulty in using such a formula, however, is that we generally do not have a good way to compute the 1d  $\CN=4$ Witten index~\eqref{def CI Witten} of the quiver SQM$_\gamma$, or even to properly define it, whenever the dimension vector $N$ is not primitive --- see~{\it e.g.}~\cite{Lee:2016dbm} for a detailed discussion. From the point of view of the 4d $\CN=2$ Coulomb-branch physics, this is related to the possible presence of bound states at threshold, and from the perspective of moduli spaces of quiver representations, this is closely related to the fact that the 1d target spaces should generally be understood as moduli stacks.

\medskip\noindent
{\bf Relations between rational BPS invariants.} 
To obtain a more useful relation, we need to recast the above discussion in terms of standard BPS invariants, which we can do whenever~\eqref{Z equal Omega} holds true. More generally, we propose the following explicit formula relating the {\it rational BPS invariants} of Galois pairs:
\be
\label{eq:CoverBPSInv}
\bar \Omega^Q(\gamma; \zeta) ={1\ov |\G|} \sum_{\t\gamma\, |\,
  F_\ast(\t\gamma)=\gamma}  {\xi(\gamma,\tilde \gamma)}\;\bar
\Omega^{\t Q}(\t\gamma; F^\ast(\zeta))~,
\ee
where $\bar \Omega$ was defined in~\eqref{eq:DefRatInv}. Here the stability parameters $\t\zeta= F^\ast(\zeta)$ are as discussed above~\eqref{alligned central charges}, and $\xi(\gamma,\tilde \gamma)$ is a sign that tracks the relative complex dimension of the moduli spaces $\CM^Q_\gamma(\zeta)$ and $\CM_{\t\gamma}^{\t Q}(\t\zeta)$, namely:
\be 
\label{eq:xigtg}
\xi(\gamma,\tilde \gamma) = (-1)^{{\dim}_{\mathbb{C}}(\CM^{\t Q}_{\t
    \gamma})-{\dim}_{\mathbb{C}}(\CM^Q_\gamma)}\in\{\pm 1\}~.
\ee
Recall that, for a quiver without superpotential, the virtual dimension of the
moduli space of charge $\gamma$ is given by~\eqref{eq:dimM}, and
that this formula remains true mod 2 for generic superpotential and FI
parameters, hence~\eqref{eq:xigtg} is reliably computed using~\eqref{eq:dimM} in all cases.

The relative sign~\eqref{eq:xigtg} appears in~\eqref{eq:CoverBPSInv} simply because of the conventional sign in the definition~\eqref{def OmegaQ}. The fact that the rational invariant $\bar\Omega$ appears instead of $\Omega$ is related to the non-trivial stabilisers. In the special case when $\gamma$ is primitive, then all $\t\gamma$ involved are primitive and have trivial stabiliser%
\footnote{This is easily seen by contradiction. If some $\t\gamma$ such that $F_\ast(\t\gamma)=\gamma$ is not primitive, it means that it is divisible by some integer $n>1$, and then so is $\gamma$. Moreover, if $\t\gamma$ had a non-trivial stabiliser $H\subseteq \G$, then $\gamma$ would be divisible by $|H|$.} 
and therefore the formula~\eqref{eq:CoverBPSInv} follows from~\eqref{rel between ZQs} with $\Omega$ instead of $\bar\Omega$. 
The appearance of $\bar\Omega$, in general, is suggested by a number of mathematical and physical considerations~\cite{Joyce:2008pc, Kontsevich:2008fj, Kontsevich:2010px, Manschot:2010qz, reineke2011cohomology, Meinhardt2019}; in particular, one should likely formalise our MSQM construction in terms of the cohomological Hall algebra (CoHA), in which context $\bar\Omega$ appears more naturally.  In this paper, we present~\eqref{eq:CoverBPSInv} as a {\it conjecture} for which we can present strong evidence. Besides the physics motivation in terms of $\G$ gauging explained above, we will give two more partial derivations of the formula in special cases  in sections~\ref{sec:GaloisCoverFixedLoci} and~\ref{sec:AlgHoms} below. We will also check it explicitly in many examples.

\medskip
\noindent
{\bf Simple examples.} The relation \eqref{eq:CoverBPSInv} is obviously true if the cover $\tilde Q$ is a disconnected sum of $|\G|$ copies of the original quiver, such as in figure~\ref{GQuiver1}. As a simple but non-trivial example, consider the $\Z_2$ cover of the Kronecker quiver $Q=K_2$ shown in figure~\ref{GQuiver2}.  
For $Q$, let us look at the examples $N=(1,1)$, $(2,1)$ and $(2,2)$, with stability parameters $\zeta_{1}>\zeta_2$. It is well-known that the invariants for these cases are
\begin{equation}
\label{eq:OmNf0}
    \Omega^Q((1,1);\zeta)=-2,\quad \Omega^Q((2,1);\zeta)=1,\quad \Omega^Q((2,2);\zeta)=0. 
\end{equation}
Since $(2,2)$ is non-primitive, we further have
\begin{equation}\label{checkN22}
    \bar\Omega^Q((2,2);\zeta)=\Omega^Q((2,2);\zeta)+\frac{1}{4}\Omega^Q((1,1);\zeta)=-\frac{1}{2}. 
\end{equation}
For the $\Z_2$ Galois cover $\tilde Q$ of Figure~\ref{GQuiver2} the relevant non-zero contributions for the pull-up of $N=(1,1)$ are from the four permutations of $\tilde N=(1,0;1,0)$ with $\tilde\zeta=F^\ast(\zeta)$ and
\begin{equation}
    \Omega^{\tilde Q}((1,0;1,0),\tilde\zeta)=1.
\end{equation}
The sign $\xi((1,1);(1,0;1,0))=-1$ and we thus see that
\begin{equation}
    \bar{\Omega}^Q((1,1);\zeta)=-\frac{1}{2}\sum_{\t\gamma\, |\,
  F_\ast(\t\gamma)=\gamma}  \bar
\Omega^{\t Q}(\t\gamma; F^\ast(\zeta))~,
\end{equation}
as predicted by \eqref{eq:CoverBPSInv}. Similarly, the non-zero contributions for the pull-ups of $N=(2,1)$ and $N=(2,2)$ are given by $\tilde N=(1,1;0,1)$ with permutation and $\tilde N=(1,1;1,1)$, respectively. The sign $\xi(N,\tilde N)=+1$ in both cases and we find that
\begin{equation}
    \Omega^{\tilde Q}((1,1;1,0);\tilde \zeta)=1,\qquad \Omega^{\tilde Q}((1,1;1,1),\tilde\zeta)=-2. 
\end{equation}
This again gives agreement with \eqref{eq:CoverBPSInv}. In particular, we see that for the non-primitive vector $N=(2,2)$ it is really necessary to consider the rational invariants $\bar\Omega$ rather than the integer ones, as is evident from the fact that $\Omega^Q((2,2);\zeta)=0$, while  $\Omega^{\tilde Q}((1,1;1,1),\tilde\zeta)=-2$. For $\bar\Omega$ we instead find perfect agreement, since
\begin{equation}
\label{eq:bOmtQ04}
    \bar\Omega^{\t Q}((1,1;1,1);\tilde\zeta)=-2+\frac{1}{4}\times 4=-1. 
\end{equation}
We return to this example with a more extensive discussion in Section \ref{sec:Symm2Cover}.

\subsection{Refined covering relation for symmetric Galois covers}\label{subsec:symgalcov}
 One of the difficulties encountered when relating BPS invariants of a Galois cover as in~\eqref{eq:CoverBPSInv} is that the stability condition $\t \zeta=F^*(\zeta)$ is non-generic and in many cases sits on a wall of marginal stability for $\t Q$. Such cases arise, in particular, for a charge $\t \gamma$ such that  $F_*(\t \gamma)=\gamma\in \Gamma$ is non-primitive, since two constituent charges $\t \gamma', \t \gamma''\in \t \Gamma$ with $F_*(\t \gamma')=F_*(\t \gamma'')$ have the same slope $\mu$ \eqref{def slope} for $\t \zeta=F^*(\zeta)$.

It is therefore useful to single out Galois covers for which the BPS invariants $\Omega(\t\gamma, F^*(\zeta))$ are independent of a small generic perturbation away from $F^*(\zeta)$.
To this end, we introduce the notion of a {\it symmetric} $\G$ Galois cover as follows. A symmetric Galois cover is a cover for which the set $\{d_a\}_{i,j}$ of gradings $d_a\in \h\G$ of the set of arrows $a\in Q_1$ with equal tail $i$ and head $j$ is $\G$ symmetric. In other words, the gradings are symmetrically distributed among the arrows. A necessary condition for a Galois cover to be $\G$ symmetric is that the number of arrows $(a: i\to j)$ between each pair of nodes $i,j \in Q_0$  is a multiple of $|\G|$. Moreover, the inner products for $Q$ and $\t Q$ of a symmetric Galois cover are related by
\be
\label{eq:symmCover<>}
|\G|\langle \tilde N, \tilde N' \rangle_{\t Q} = \langle F_\ast(\tilde
N),F_\ast(\tilde N')\rangle_Q~,\quad \text{for all } \tilde N, \tilde N'~.
\ee
Then, for all $\tilde N$, $\tilde N'$ such that $F_\ast(\tilde N)=F_\ast(\tilde N')$, the anti-symmetric product vanishes:
\be
\label{eq:<>=0}
F_\ast(\tilde
N)=F_\ast(\tilde N')\quad \Rightarrow \quad  \langle \tilde N,\tilde N'\rangle=0.
\ee
The vanishing of these inner products implies that the BPS invariant $\Omega(\tilde \gamma, \t\zeta)$ is independent for a small, generic perturbation away from $\t \zeta=F^*(\zeta)$. This aids the evaluation of the BPS invariants and ensures that these are integers. 

For such covers, we propose a generalisation of the covering relation \eqref{eq:CoverBPSInv} to refined rational invariants. To this end, we assign to the vertices in a block of the quiver $\tilde Q$ a non-trivial single-centred invariant $\Omega_S^{\tilde Q}$, which is an input in the Coulomb branch formula \cite{Manschot:2011xc, Manschot:2012rx, Manschot:2013dua, Manschot:2014fua}. In that context, the single-centred invariant $\Omega_S^{\tilde Q}$ represents the number of quantum states of a single black hole centre. We will explain in later sections that the general form of the motivic KS monodromy $\bM(y)$, leads us, for $\G=\Z_n$, to the assignment
\be
\label{eq:assignOmS}
\Omega_S^{\tilde Q}(\tilde \gamma_{j,\alpha},\tilde y)=\tilde y^{(2\alpha+1-|\G|)/|\G|},\qquad \alpha=0,\dots,|\G|-1.
\ee
Due to the $\mathbb{G}$-symmetry, we can of course permute the $\alpha$ on the right-hand side. This assignment thus mixes the space-time rotation symmetry and the $\mathbb{G}$-symmetry of the quiver. We then propose the following refinement of~\eqref{eq:CoverBPSInv}:
\be
\label{eq:OmQOmtQy2}
\bar \Omega^Q(\gamma,\zeta;y)= \frac{y-y^{-1}}{y^{|\G|}-y^{-|\G|}} \sum_{\tilde \gamma\, |\,
  F_\ast(\tilde \gamma)=\gamma} \xi(\gamma,\tilde \gamma) \,\bar \Omega^{\tilde Q}(\tilde \gamma, \tilde \zeta;\tilde y)~,\qquad \tilde y^{1/|\G|}=y~.
\ee
The phase of $y$ and $\t y$ is determined by requiring that the $\tilde y,y\to 1$ limits agree. We will discuss examples where this formula holds in section~\ref{sec:GaloisExamples}. 

The $K_2$ cover of figure~\ref{GQuiver2} discussed above is an example of a symmetric $\Z_2$ cover, while the $\Z_3$ cover of figure~\ref{GQuiver3} is not symmetric. Sections \ref{sec:Symm2Cover} to \ref{sec:E0E6} discuss more examples of symmetric $\Z_n$ covers and the evaluation of their BPS invariants, while section \ref{sec:nonsymGC} considers non-symmetric examples.

\section{Galois coverings and fixed loci of circle actions}
\label{sec:GaloisCoverFixedLoci}
In this section, we further explain how the Galois covering quiver $\t Q$ can be understood in terms of fixed loci under a $\G=\Z_n \subset \C^\ast$ action on moduli spaces of quiver representations of $Q$. We will also explain that it pays to further uplift the $\Z_n$ covering quiver $\t Q$ to a $\Z$ covering $\h Q$.

\subsection{Circle action from the arrow grading and its fixed loci}\label{subsec:circle action}

For a fixed $\G=\Z_n$, consider the `base' quiver with superpotential $Q=(Q_0, Q_1, W)$, with the $\G$-grading~\eqref{G grading arrows da} assigning some $d_a\in \Z$ for each arrow. In the following, it will be important to specify integers $d_a$, not only integers modulo $n$. In particular, the grading of the superpotential should be $d(W)=0$. The grading allows us to define a $\C^\ast$ action on quiver representations as
\be\label{def omega action}
\G_\C \; : \; R = (\{V_j\}, \{X_a\}) \mapsto g(\lambda)\cdot R=  (\{V_j\}, \{\lambda^{d_a} X_a\})~, \qquad \lambda\in \C^\ast~. 
\ee
This action commutes with the gauge group, and therefore $\G_\C$ naturally acts on the $GL_\gamma^0$ orbits $[R]$:
\be\label{GC on R orbits}
\G_\C \; :  \; [R]\mapsto g(\lambda)\cdot [R]= [g(\lambda)\cdot R]~,
\ee 
therefore it acts on the moduli space of semi-stable orbits $\CM^Q_\gamma$ defined in~\eqref{CMgamma GIT def}. We are interested in the fixed loci of that action. A point in $\CM^Q_\gamma$ will be invariant under~\eqref{GC on R orbits} if and only if the $\C^\ast$ action on $R$ is gauge equivalent to $R$, leaving the orbit invariant. This means that we must have 
\be\label{GCvsGLN}
\lambda^{d_a} X_a =G_i^{-1} X_a G_j~, \qquad \forall \, (a:i\rightarrow j)\in Q_1~,
\ee
where $G_i$ denotes the $GL(N_i)$ action on $V_i$ as in~\eqref{G action on V and X}; this is a non-trivial condition for every arrow $a\in Q_1$ which is represented non-trivially by $R$.  In other words, we need an embedding
\be\label{def sigma cochar}
\sigma\; :\; \G_\C\rightarrow GL_{\gamma}^0\; : \; \lambda \rightarrow \{G_i(\lambda)\}/GL(1)~.
\ee
The equations~\eqref{GCvsGLN} define this embedding up to the stabiliser of $R$. Focussing on $R$ a stable representation for now, its stabiliser is trivial in $GL_\gamma^0$, and therefore~\eqref{def sigma cochar} defines a cocharacter $\sigma$ of the complexified gauge group. Any such cocharacter is defined by a decomposition of the vector spaces $V_i$ into weight spaces, as:
\be\label{Vi decompose w} 
V_i \cong \bigoplus_{w\in \Z}  V_i^{(w)}~, \qquad\qquad \chi_w \; : \; V_i^{(w)}\rightarrow V_i^{(w)}\; : \; v \mapsto \lambda^{w} v~.
\ee
Here $\G_\C$ acts on each weight space as a character $\chi_w(\lambda)= \lambda^{w}$ of $\G_\C\cong \C^\ast$; these characters are indexed by the integers $w$. For each $V_i \cong \C^{N_i}$, the weight-space decomposition is equivalent to some partition of the rank,
\be\label{Ni w decomp}
N_i =\sum_{w\in \Z} N_{i, w}~, 
\ee
with $V_i^{(w)}\cong \C^{N_{i,w}}$. Let us denote by
\be\label{Xwiwj}
(X_a)_{w_j}^{w_i}\; : \; V_i^{(w_i)}\rightarrow V_j^{(w_j)}
\ee
the map $X_a$ between two given weight spaces. The fixed-point condition~\eqref{GCvsGLN} is then equivalent to
\be\label{grading condition on Xwiwj}
w_j = w_i + d_a~, \qquad \forall \, (a:i\rightarrow j)\in Q_1~,
\ee
for each arrow such that $X_a\neq 0$. We can therefore interpret the fixed loci as moduli spaces of quiver representations of an infinite Galois covering quiver of $Q$ by $\Z$, 
\be\
\h F \; :\; \h Q\longrightarrow Q~,
\ee
with the nodes indexed by $(i,w)$.  Each fixed locus of the circle action on the moduli space $\CM_\gamma^Q$ corresponds to a single charge vector of $\h Q$ denoted by
\be
\h\gamma = \sum_{(i,w)\in \h Q_0} N_{i,w} \h\gamma_{i,w}~,
\ee
up to the $\Z$-valued gauge transformation that shifts all ranks simultaneously ($N_{i, w}\rightarrow N_{i,w+1}$, $\forall i$), and the non-zero arrows in $\h Q$ correspond to those in~\eqref{Xwiwj} satisfying~\eqref{grading condition on Xwiwj}.  For each $\h\gamma$, we have the gauge group
\be
 G_{\h \gamma}= \prod_{(i,w)\in \h Q_0} {\rm U}(N_{i,w})~,
\ee
which encodes the gauge freedom one retains after the weight decomposition~\eqref{Vi decompose w}. The pull-down functor $\h F_\ast$ on quiver representations then induces an embedding of the moduli spaces of $\h Q$ inside the ones of $Q$. For any dimension vector $\h\gamma$ such that $F_\ast(\h\gamma)=\gamma$, we have
\be
\h F_\ast\; :\; \CM_{\h \gamma}^{\h Q} \mapsto \h F_\ast(\CM_{\h\gamma}^{\h Q})\cong \CM_{\h\gamma}\subseteq \CM_\gamma^Q~.
\ee
The above argument establishes this embedding if all orbits $[R]\in \CM_\gamma^Q$ are $\zeta$-stable. 
 If $R$ is strictly semi-stable --- {\it i.e.} polystable, assuming that $[R]\in \CM_\gamma^Q$ --- we have a decomposition into marginally bound constituents and a non-trivial stabiliser:
 \be
 R\cong \oplus_l R_l^{\oplus m_l}~,\qquad {\rm Stab}(R)= \prod_l GL(m_l)\Big/GL(1)~.
 \ee
 We can still run the above argument for each $R_l$ independently, and we conclude that polystable orbits of representations of $Q$ fixed by $\C^\ast$ simply give us polystable orbits of representations of $\h Q$.

We therefore conclude that fixed loci of the circle action~\eqref{GC on R orbits} on semi-stable representations are encoded in the $\Z$-covering quiver $\h Q$. Note that, while $\h Q$ has an infinite number of nodes, only a finite number are populated by any representation $\h R$ corresponding to fixed-points $[R]$ in $\CM_\gamma^Q$, for any finite $\gamma$.

\medskip
\noindent
{\bf $\G$ action versus $\C^\ast$ action.} To relate the above discussion to the $\G=\Z_n$ covering $\t Q$ of $Q$, we should restrict the circle action~\eqref{GC on R orbits} to the multiplicative $\G$ action on $\CM_\gamma^Q$ given by: 
\be\label{GZn on R orbits}
\G  \; :  \; [R]\mapsto g(\lambda_n)\cdot [R]= [g(\lambda_n)\cdot R]~,\qquad \lambda_n \equiv e^{2\pi i \ov n}~.
\ee 
The fixed loci under this $\Z_n$ action are generally larger than under the circle action, but it is clear that the latter are a subset of the former:
\be
(\CM^Q_\gamma )^{\C^\ast}\subseteq (\CM^Q_\gamma )^{\G}~.
\ee
It is also clear that the fixed loci under $\Z_n$ can be found using the same logic as above. The weight-space decomposition~\eqref{Vi decompose w} can be restricted to the $\G= \Z_n$ characters in ${\rm Hom}(\Z_n, \C^\ast)\cong \Z_n$, which we write as: 
\be\label{Vialpha weights}
V_i \cong \bigoplus_{\alpha\in \Z_n} \t V_{i, \alpha_i}~, \qquad\qquad N_i= \sum_{\alpha_i\in \Z_n} N_{i, \alpha_i}~,
\ee
with $V_{i,\alpha_i}\cong \C^{N_{i,\alpha_i}}$. 
The fixed-point condition~\eqref{grading condition on Xwiwj} on arrows now reads:
\be
\alpha_j = \alpha_i + d_a \; ({\rm mod } \, n)~, \qquad \forall \, (a:i\rightarrow j)\in Q_1~.
\ee
Therefore, we find that the Galois cover $\G: \t Q\rightarrow Q$ and the representation vectors $\t\gamma$ such that $F_\ast(\t\gamma)=\gamma$  precisely encode the fixed loci in the moduli spaces $\CM_\gamma^Q$  under the induced $\G$ action~\eqref{GZn on R orbits} --- any representation $\t R$ of $\t Q$ gives us a representation $R=F_\ast(\t R)$ of $Q$ that sits in a $\G$-invariant orbit $[R]$. For each $\t\gamma$, We have the corresponding SQM gauge group
\be
 G_{\t \gamma}= \prod_{(i,\alpha)\in \t Q_0} {\rm U}(N_{i,\alpha})~.
\ee 
There are many cases where we can argue that, actually, the fixed points under $\C^\ast$ and the $\Z_n$ action coincide --- for instance, this is the case in any quiver without superpotential with $d_a\in \{0, 1, \cdots, n-1\}$. More generally, the fixed loci under $\C^\ast$ are a strict subset of those under $\Z_n$. There is a $\Z/\Z_n$ Galois covering of $\t Q$ by $\h Q$ that corresponds to the decomposition of the vector spaces of any representation $\t R$ of $\t Q$ into weight spaces for $\h Q$:
\be\label{Qhat to Qtilde V}
\t V_{i, \alpha} = \bigoplus_{l\in \Z} V_i^{(w+l n)}~, \qquad \forall (i,\alpha)=i _\alpha \in \t Q_0~.
\ee
A necessary condition for having $(\CM^Q_\gamma)^{\C^\ast} = (\CM^Q_\gamma)^{\G}$  is that the decomposition~\eqref{Qhat to Qtilde V} is trivial ({\it i.e.} with a single summand $\t V_{i,\alpha} = V_i^{(w)}$ for each node $(i,\alpha)\in \t Q_0$).

 \medskip
 \noindent
 {\bf $\Z_n$ orbits and stacky fixed points.} It is important to emphasise that the cocharacter $\sigma$ in~\eqref{def sigma cochar} is only defined up to a simultaneous rescaling of the weights, $w_i \rightarrow w_i + 1$, for all $i\in Q_0$, and similarly when restricted to the multiplicative action by $\G$. In other words, the $\Z_n$ transformation
  \be\label{G0 action}
 \G^{(0)}\; : \; \t \gamma= (N_{i, \alpha})\mapsto \lambda_n \cdot \t\gamma = (N_{i, \alpha+1 \,({\rm mod} \, n)}) 
 \ee
 on the gauge ranks of the $\t Q$ quiver descends to a gauge transformation $\sigma|_{\lambda=\lambda_n}$ in the $\CN=4$ SQM for the quiver $Q$. This means that two $\t Q$ moduli spaces for two distinct dimension vectors related by~\eqref{G0 action} are not just isomorphic (this much is clear from the $\G$ symmetry of $\t Q$), they also descend to the same fixed locus in $\CM_\gamma^Q$. 
 
As in section~\ref{sec:gauging and ungauging}, let us denote by $[\t\gamma]$ the orbit of $\t\gamma$ under $\G^{(0)}$, and by $\CM^{\t Q}_{[\t\gamma]}\cong \CM_{\gamma}^{\t Q}$ the corresponding moduli space.
  In general, for a fixed dimension vector $\t\gamma$ of $\t Q$, the action~\eqref{G0 action} might have a non-trivial stabiliser ${\rm Stab}(\t\gamma)\subseteq  \G^{(0)}$. The corresponding fixed points inside $\CM_\gamma^Q$ are then understood as stacky points. Physically, we simply have the finite gauge group ${\rm Stab}(\t\gamma)\subset G_\gamma^0$ which remains un-Higgsed on the classical Higgs branch. We then formally write down the fixed loci of $\CM^Q_\gamma$ under the $\G$ action as a sum over distinct $\G^{(0)}$ orbits $[\t\gamma]$:
\be\label{decompose fixed G locus}
   (\CM^Q_\gamma)^{\G} = \sum_{[\t\gamma]\; | \; F_\ast(\t\gamma)=\gamma}   {1\ov |{\rm Stab}(\t\gamma)|}  \CM^{\t Q}_{[\t \gamma]} ={1\ov n}   \sum_{\t\gamma\; | \; F_\ast(\t\gamma)=\gamma}  \CM^{\t Q}_{\t \gamma}~,
\ee
where the second equality follows from the orbit-stabiliser theorem. This should be compared to~\eqref{rel between ZQs}. 
 
  \medskip
 \noindent
 {\bf BPS index from Atiyah--Bott--Kirwan localisation.} Our main conjecture \eqref{eq:CoverBPSInv} is intimately related to the decomposition~\eqref{decompose fixed G locus} of the fixed locus. In favourable cases, we can see this explicitly. For instance, let us assume that $(\CM^Q_\gamma)^{\C^\ast}= (\CM^Q_\gamma)^{\G}$, and that all the moduli spaces involved are smooth. Then, viewing the moduli spaces $\CM^{\t Q}_{\t\gamma}$ as fixed loci of the circle action~\eqref{def omega action} inside $\CM_\gamma^Q$, one can recover the homology of $\CM_\gamma^Q$ from Atiyah--Bott--Kirwan localisation~\cite{kirwan1984cohomology} --- see also~\cite{wilkin2019equivariant} for a related discussion. Indeed, in this case, the topological Euler characteristic can be written as
 \be
\chi(\CM^Q_\gamma) = \sum_{[\t \gamma]} \chi(\CM^{\t Q}_{[\t\gamma]}) = {1\ov n} \sum_{\t\gamma} \chi(\CM^{\t Q}_{\t\gamma})~.
 \ee
It directly follows from~\eqref{def OmegaQ} that, for $\CM_\gamma^Q$ smooth, 
\be\label{Omega sum fixed points unrefined}
  \Omega^Q(\gamma;\zeta) ={1\ov n} \sum_{\t\gamma\, |\,
  F_\ast(\t\gamma)=\gamma} \xi(\gamma,\tilde \gamma)\,
\Omega^{\t Q}(\t\gamma;\tilde \zeta)~,
\ee 
with the sign~$\xi(\gamma,\tilde \gamma)$ defined in~\eqref{eq:xigtg}. More generally, it is natural to conjecture that the similar relation~\eqref{eq:CoverBPSInv} for $\b\Omega$ also follows from the decomposition~\eqref{decompose fixed G locus} when properly interpreted in terms of moduli stacks. We hope to return to this point in future work.

\subsection{Refined indices from fixed loci and their normal bundles}
The non-trivial distinction between fixed loci under $\C^\ast$ and under $\G$ is further illuminated by considering how fixed-point theorems give us new relations between refined BPS indices of $Q$ and of their $\Z$ Galois covers. Recall that the refined BPS index is obtained as the $\chi_{\bf y}$-genus~\eqref{chiygemus def of Omegay}, assuming the moduli space is smooth, and it can therefore be computed by taking advantage of the circle action. Indeed, the $\C^\ast$ action~\eqref{GC on R orbits} on the K\"ahler manifold $\CM_\gamma^Q$ is naturally Hamiltonian, with the moment map schematically given by:
\be
\mu_{\C^\ast} = \sum_{a\in Q_1}  d_a\,  |X_a|^2~.
\ee
This gives us a Morse-Bott function whose critical points are the fixed loci $\CM^{\t Q}_{\h \gamma}\subset \CM^Q_\gamma$. We then have an explicit localisation of the $\chi_{\bf y}$-genus as
\be
\chi_{- y^2}(\CM_\gamma^Q)= \sum_{\h \gamma\, |\, \h F_\ast(\h \gamma)=\gamma} y^{2\, {\rm rank}(N_{\h\gamma}^+)} \,\chi_{-y^2}(\CM_{\h\gamma}^{\h Q})~,
\ee
where the sum is over all fixed loci as indexed by dimension vectors of $\h Q$, and with $N_{\h\gamma}^+$ denoting the positive-weight part of the normal bundle to $\CM_{\h\gamma}^{\h Q}\subset \CM^Q_{\gamma}$. This formula is most easily derived in the language of Morse theory~\cite{kirwan1984cohomology}, where $2\, {\rm rank}(N_{\h\gamma}^+)$ is also the real Morse index at the fixed loci.%
\footnote{In our conventions,  ${\rm rank}(N_{\h\gamma}^+)$ gives the number of complex directions in $\CM_\gamma$ along which $\mu_{\C^\ast}$ increases; that is the Morse index if we choose $f=-\mu_{\C^\ast}$ to be the Morse function.} The rank of $N_{\h\gamma}^+$ can be easily derived from the weight decomposition, at least in terms of virtual dimensions, by counting representations of positive weights minus the residual gauge transformations in $GL_\gamma/GL_{\h\gamma}$ acting on them:
\be\label{rkNp formula}
{\rm rk}(N^+_{\h\gamma}) = \sum_{a\in Q_1} \sum_{\substack{w_{t(a)}, w_{h(a)} \\ w_{h(a)} < w_{t(a)} + d_a}} N_{t(a),w_{t(a)}} N_{h(a),w_{h(a)}} - \sum_{j \in Q_0} \sum_{w' > w} N_{j,w} N_{j,w'}~.
\ee
We therefore obtain a completely explicit formula relating the refined BPS indices of $Q$ and of its $\Z$ Galois covering $\h Q$, as follows:
\be\label{Omega ref sum fixed loci}
\Omega^Q(\gamma, y; \zeta) = \sum_{\h \gamma\, |\, \h F_\ast(\h \gamma)=\gamma} \, (-y)^{{\rm dim}_\C(\CM_{\h\gamma})-{\rm dim}_\C(\CM_{\gamma})}\,  y^{2\, {\rm rank}(N_{\h\gamma}^+)} \,\Omega^{\h Q}(\h \gamma, y; \h \zeta)~,
\ee
with $\h\zeta=\h F^\ast(\zeta)$. 
Note that this formula depends crucially on the weight decomposition~\eqref{Vi decompose w}-\eqref{Ni w decomp} and on the fact that the gradings and weights are $\Z$-valued. There is no such formula in terms of the $\G$ fixed loci only --- that is, even in cases where $(\CM^Q_\gamma)^{\C^\ast}=(\CM^Q_\gamma)^\G$ so that $\CM_{\h\gamma}^{\h Q}= \CM_{[\t\gamma]}^{\t Q}$, we still need to consider the $\Z$ covering quiver $\h Q$ to compute the Morse indices~\eqref{rkNp formula}. Note also that setting $y=1$ in~\eqref{Omega ref sum fixed loci} in cases when $\CM_{\h\gamma}^{\h Q}= \CM_{[\t\gamma]}^{\t Q}$ for every $\h\gamma$ reproduces the unrefined relation~\eqref{Omega sum fixed points unrefined}.

\subsection{Examples of fixed-loci decompositions}
Let us consider a few simple examples of the above discussion. We will focus on 2-node quivers for simplicity, but one can easily verify the formula~\eqref{Omega ref sum fixed loci} in more complicated examples. These and similar examples will also be discussed in section~\ref{sec:GaloisExamples}.

\begin{figure}
\centering
\begin{subfigure}[b]{0.4\textwidth}
\centering
\begin{tikzpicture}[baseline=1mm, every node/.style={circle, draw, minimum size=1cm}, thick]
\node[] (1) at (0,-0.65){$\gamma_{1}$};
\node[] (2) at (2.4,-0.65){$\gamma_{2}$};
\draw[->-=0.5] (1.25)  to[bend left=15]  (2.155);
\draw[->-=0.5] (1.0)   to                (2.180);
\draw[->-=0.5] (1.335) to[bend right=15] (2.205);
\end{tikzpicture}
\caption{$K_3$ quiver.}
\label{fig:K3 quiver}
\end{subfigure}
\hspace{.3cm}
\begin{subfigure}[b]{0.4\textwidth}
\centering
\begin{tikzpicture}[baseline=1mm, every node/.style={circle, draw, minimum size=0.9cm}, thick]
\node[] (1) []{$\gamma_{1,0}$};
\node[] (2) [right =1.5 of 1]{$\gamma_{2,0}$};
\node[] (3) [below =.5 of 1]{$\gamma_{1,1}$};
\node[] (4) [right =1.5 of 3]{$\gamma_{2,1}$};
\draw[->-=0.5] (1.15) to   (2.165);
\draw[->-=0.5] (1.345) to   (2.195);
\draw[->-=0.3] (1) to   (4);
\draw[->-=0.5] (3.15) to   (4.165);
\draw[->-=0.5] (3.345) to   (4.195);
\draw[->-=0.3] (3) to   (2);
\end{tikzpicture}
\caption{$\Z_2$-cover of $K_3$ with $d_a=(0,0,1)$.}
\label{K3 cover}
\end{subfigure}
\hspace{.3cm}
\caption{A $\Z_2$-cover of the $K_3$ quiver. \label{fig:K3 Z2 cover}}
\end{figure}
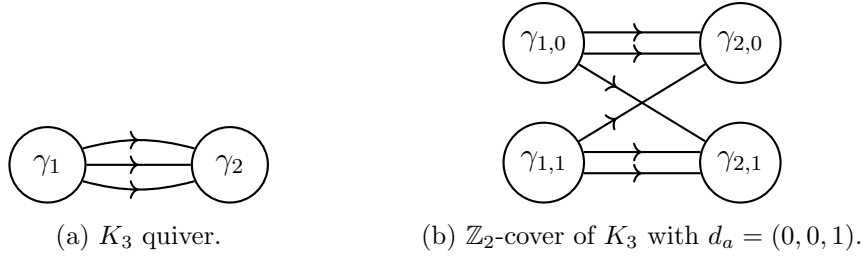

\newcommand{\quivpad}[1]{%
  \raisebox{0pt}[\dimexpr\height+4pt\relax][\dimexpr\depth+4pt\relax]{#1}}
\begin{table}[ht]
\centering
\renewcommand{\arraystretch}{1.2}
\begin{tabular}{|@{\hspace{5pt}}c@{\hspace{5pt}}|c|c|c|}
\hline
$\h Q$ & ${\rm dim}(\CM_{\h \gamma}^{\h Q})$ & ${\rm rank}(N^+_{\h \gamma})$ & $\Omega^{\h Q}(\h\gamma, y)$ \\
\hline\hline
\quivpad{\begin{tikzpicture}[baseline=1mm, every node/.style={circle, draw, minimum size=0.9cm}, thick]
\node (L0) at (0,0) {3};
\node (R0) at (2,0) {2};
\draw[->-=0.5] (L0.15) to (R0.165);
\draw[->-=0.5] (L0.345) to (R0.195); 
\end{tikzpicture}} & 0 & 6 & $1$ \\
\hline 
\quivpad{\begin{tikzpicture}[baseline=1mm, every node/.style={circle, draw, minimum size=0.9cm}, thick]
\node (L0) at (0,0) {2};
\node (L1) at (0,-1) {1};
\node (R1) at (2,-1) {2};
\draw[->-=0.5] (L0) to (R1);
\draw[->-=0.5] (L1.15) to (R1.165);
\draw[->-=0.5] (L1.345) to (R1.195);
\end{tikzpicture}} & 0 & 0 & $1$ \\
\hline
\quivpad{\begin{tikzpicture}[baseline=1mm, every node/.style={circle, draw, minimum size=0.9cm}, thick]
\node (L0) at (0,0) {2};
\node (L1) at (0,-1) {1};
\node (R0) at (2,0) {1};
\node (R1) at (2,-1) {1};
\draw[->-=0.5] (L0.15) to (R0.165);
\draw[->-=0.5] (L0.345) to (R0.195);
\draw[->-=0.5] (L0) to (R1);
\draw[->-=0.5] (L1.15) to (R1.165);
\draw[->-=0.5] (L1.345) to (R1.195);
\end{tikzpicture}} & 2 & 3 & $y^2+2+y^{-2}$ \\
\hline
\quivpad{\begin{tikzpicture}[baseline=1mm, every node/.style={circle, draw, minimum size=0.9cm}, thick]
\node (L0) at (0,0) {1};
\node (L1) at (0,-1) {2};
\node (R1) at (2,-1) {2};
\draw[->-=0.5] (L0) to (R1);
\draw[->-=0.5] (L1.15) to (R1.165);
\draw[->-=0.5] (L1.345) to (R1.195);
\end{tikzpicture}} & 2 & 2 & $y^2+1+y^{-2}$ \\
\hline
\quivpad{\begin{tikzpicture}[baseline=(current bounding box.center), every node/.style={circle, draw, minimum size=0.9cm}, thick]
\node (L0) at (0,0) {1};
\node (L1) at (0,-1) {2};
\node (R0) at (2,0) {1};
\node (R1) at (2,-1) {1};
\draw[->-=0.5] (L0.15) to (R0.165);
\draw[->-=0.5] (L0.345) to (R0.195);
\draw[->-=0.5] (L0) to (R1);
\draw[->-=0.5] (L1.15) to (R1.165);
\draw[->-=0.5] (L1.345) to (R1.195);
\end{tikzpicture}} & 1 & 6 & $0$ \\
\hline
\quivpad{\begin{tikzpicture}[baseline=(current bounding box.center), every node/.style={circle, draw, minimum size=0.9cm}, thick]
\node (L0) at (0,0) {1};
\node (L1) at (0,-1) {2};
\node (R1) at (2,-1) {1};
\node (R2) at (2,-2) {1};
\draw[->-=0.5] (L0) to (R1);
\draw[->-=0.5] (L1.15) to (R1.165);
\draw[->-=0.5] (L1.345) to (R1.195);
\draw[->-=0.5] (L1) to (R2);
\end{tikzpicture}} & 1 & $-1$ & $0$ \\
\hline
\quivpad{\begin{tikzpicture}[baseline=(current bounding box.center), every node/.style={circle, draw, minimum size=0.9cm}, thick]
\node (L0) at (0,0) {1};
\node (L1) at (0,-1) {1};
\node (L2) at (0,-2) {1};
\node (R1) at (2,-1) {1};
\node (R2) at (2,-2) {1};
\draw[->-=0.5] (L0) to (R1);
\draw[->-=0.5] (L1.15) to (R1.165);
\draw[->-=0.5] (L1.345) to (R1.195);
\draw[->-=0.5] (L1) to (R2);
\draw[->-=0.5] (L2.15) to (R2.165);
\draw[->-=0.5] (L2.345) to (R2.195);
\end{tikzpicture}} & 2 & 1 & $y^2+2+y^{-2}$ \\
\hline
\end{tabular}
\caption{Quiver $\h Q$ and fixed-loci data for the dimension vector $\gamma=(3,2)$ of the $K_3$ quiver with grading $d_a=(0,0,1)$. The ranks $N_{1,w}$ and $N_{2,w}$ are indicated inside the nodes, with the position $w\in \Z$ indicated from top to bottom --- for instance, the last quiver shown corresponds to $N_{1,0}=N_{1,1}=N_{1,2}=1$ and $N_{2,1}=N_{2,2}=1$, and all other ranks vanishing. The dimension of the moduli space indicated is the virtual dimension~\protect\eqref{eq:dimM}.}
\label{tab:quivers Qhat K3 expls}
\end{table}

\medskip
\noindent
{\bf $\Z_2$-cover of $K_3$.} Consider the $Q=K_3$ quiver of figure~\ref{fig:K3 quiver} with dimension vector $N=(1,1)$ and stability parameters $\zeta=(1,-1)$, which gives us the moduli space 
\be
\CM_{(1,1)}^Q=\mathbb{P}^2\cong \{[a_0, a_1, a_2]\}~,
\ee
where $(a_0,a_1, a_2)$ are the three arrows of the $K_3$ quiver, and hence the homogeneous coordinates of $\mathbb{P}^2$ in this example.  We have the $\C^\ast$ action:
\be\label{circle action on K3}
\C^\ast\; :\; (a_0, a_1, a_2)\rightarrow (a_0, a_1, \lambda a_2)~,
\ee
and the corresponding $\Z_2$ action for $\lambda=-1$. There are two fixed loci of the circle action, which are also the fixed loci of the $\Z_2$ action, at $a_2=0$ and at $a_0=a_1=0$. The corresponding moduli spaces of the $\t Q$ quiver are:
\be
\CM_{(1,0;1,0)}^{\t Q}= \mathbb{P}^1 \cong \{[a_0,a_1, 0]\}~, \qquad\quad
\CM_{(1,0;1,0)}^{\t Q}= \mathbb{P}^0 \cong \{[0,0, a_2]\}~,
\ee
respectively. The corresponding ranks $(N_{1,0},N_{1,1},N_{2,0},N_{2,0})$ given above sit in full orbits of the $\Z_2$ action on $\t Q$. The relation~\eqref{Omega sum fixed points unrefined} is then clearly satisfied, since $\Omega^{Q}_{\gamma=(1,1)}= 3$ while $\Omega^{\t Q}_{\gamma=(1,0;1,0)}=-2$ and $\Omega^{\t Q}_{\gamma=(1,0;0,1)}=1$. Moreover, uplifting the covering quiver $\t Q$ to the $\Z$ covering $\h Q$, we easily compute the rank of $N_{\h \gamma}^+$ for the two fixed loci:
\be
{\rm rank}(N_{(1,0;1,0)}^+) = 1~,\qquad {\rm rank}(N_{(1,0;1,0)}^+) = 0~,
\ee
hence the relation~\eqref{Omega ref sum fixed loci} is indeed satisfied:
\be
\Omega^{Q}((1,1), y) = y^2+1 +y^{-2}= (- y)^{-1} y^2 \times \left(- y - {1\ov y}\right) +(-y)^{-2} y^0 \times 1~.
\ee
One can easily check more involved examples. For instance, consider the ranks $N=(3,2)$, for which we have the refined BPS index
\be
\Omega^{Q}((3,2), y) = y^6+y^4+3 y^2+3+\frac{3}{y^2}+\frac{1}{y^4} +\frac{1}{y^6}~.
\ee
Note also that ${\rm dim}(\CM_{(3,2)}^Q)=6$. 
Using the same grading $d_a=(0,0,1)$ for the three arrows, we can consider the quivers corresponding to the circle action~\eqref{circle action on K3} on $\CM_{(3,2)}^Q$, which gives us the $\h Q$ ranks and moduli space data collected in table~\ref{tab:quivers Qhat K3 expls}. One again finds that~\eqref{Omega ref sum fixed loci} holds. Note that, in this example, the fixed loci under the circle action are strict subsets of the $\Z_2$ fixed loci.

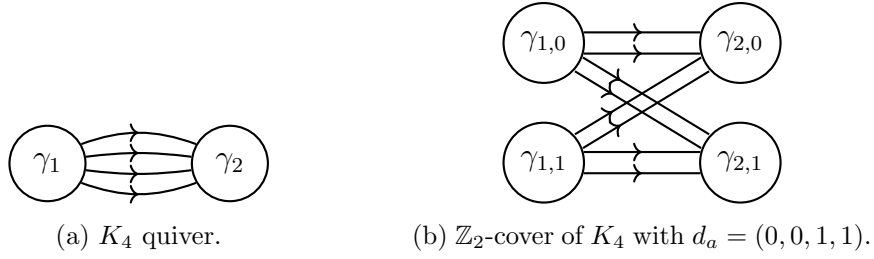
\begin{figure}
\centering
\begin{subfigure}[b]{0.4\textwidth}
\centering
\begin{tikzpicture}[baseline=1mm, every node/.style={circle, draw, minimum size=1cm}, thick]
\node[] (1) at (0,-0.65){$\gamma_{1}$};
\node[] (2) at (2.4,-0.65){$\gamma_{2}$};
\draw[->-=0.5] (1.30)  to[bend left=20]  (2.150);
\draw[->-=0.5] (1.10)  to[bend left=7]   (2.170);
\draw[->-=0.5] (1.350) to[bend right=7]  (2.190);
\draw[->-=0.5] (1.330) to[bend right=20] (2.210);
\end{tikzpicture}
\caption{$K_4$ quiver.}
\label{fig:K4 quiver}
\end{subfigure}
\hspace{.3cm}
\begin{subfigure}[b]{0.4\textwidth}
\centering
\begin{tikzpicture}[baseline=1mm, every node/.style={circle, draw, minimum size=0.9cm}, thick]
\node[] (1) []{$\gamma_{1,0}$};
\node[] (2) [right =1.5 of 1]{$\gamma_{2,0}$};
\node[] (3) [below =.5 of 1]{$\gamma_{1,1}$};
\node[] (4) [right =1.5 of 3]{$\gamma_{2,1}$};
\draw[->-=0.5] (1.15)  to (2.165);
\draw[->-=0.5] (1.345) to (2.195);
\begin{scope}[transform canvas={xshift=1.5pt, yshift=2.6pt}]
  \draw[->-=0.3] (1) to (4);
\end{scope}
\begin{scope}[transform canvas={xshift=-1.5pt, yshift=-2.6pt}]
  \draw[->-=0.3] (1) to (4);
\end{scope}
\draw[->-=0.5] (3.15)  to (4.165);
\draw[->-=0.5] (3.345) to (4.195);
\begin{scope}[transform canvas={xshift=-1.5pt, yshift=2.6pt}]
  \draw[->-=0.3] (3) to (2);
\end{scope}
\begin{scope}[transform canvas={xshift=1.5pt, yshift=-2.6pt}]
  \draw[->-=0.3] (3) to (2);
\end{scope}
\end{tikzpicture}
\caption{$\Z_2$-cover of $K_4$ with $d_a=(0,0,1,1)$.}
\label{fig:K4 cover}
\end{subfigure}
\hspace{.3cm}
\caption{A $\Z_2$-cover of the $K_4$ quiver. \label{fig:K4 Z2 cover}}
\end{figure}

\begin{table}[th]
\centering
\renewcommand{\arraystretch}{1.4}
\begin{tabular}{|@{\hspace{5pt}}c@{\hspace{5pt}}|c|c|c|}
\hline
$\h Q$ & ${\rm dim}(\CM_{\h \gamma}^{\h Q})$ & ${\rm rank}(N^+_{\h \gamma})$ & $\Omega^{\h Q}(\h\gamma, y)$ \\
\hline\hline
\quivpad{\begin{tikzpicture}[baseline=1mm, every node/.style={circle, draw, minimum size=0.9cm}, thick]
\node (L0) at (0,0) {2};
\node (R0) at (2,0) {1};
\draw[->-=0.5] (L0.15) to (R0.165);
\draw[->-=0.5] (L0.345) to (R0.195);
\end{tikzpicture}} & 0 & 4 & $1$ \\
\hline
\quivpad{\begin{tikzpicture}[baseline=1mm, every node/.style={circle, draw, minimum size=0.9cm}, thick]
\node (L0) at (0,0) {2};
\node (R1) at (2,-1) {1};
\begin{scope}[transform canvas={xshift=1.5pt, yshift=4.5pt}]
  \draw[->-=0.5] (L0) to (R1);
\end{scope}
\begin{scope}[transform canvas={xshift=-1.5pt, yshift=-1.5pt}]
  \draw[->-=0.5] (L0) to (R1);
\end{scope}
\end{tikzpicture}} & 0 & 0 & $1$ \\
\hline
\quivpad{\begin{tikzpicture}[baseline=1mm, every node/.style={circle, draw, minimum size=0.9cm}, thick]
\node (L0) at (0,0) {1};
\node (L1) at (0,-1) {1};
\node (R1) at (2,-1) {1};
\draw[->-=0.5] (L1.15) to (R1.165);
\draw[->-=0.5] (L1.345) to (R1.195);
\begin{scope}[transform canvas={xshift=1.5pt, yshift=4.5pt}]
  \draw[->-=0.5] (L0) to (R1);
\end{scope}
\begin{scope}[transform canvas={xshift=-1.5pt, yshift=-1.5pt}]
  \draw[->-=0.5] (L0) to (R1);
\end{scope}
\end{tikzpicture}} & 2 & 1 & $y^2+2 + y^{-2}$ \\
\hline
\end{tabular}
\caption{Quiver data for the $\h Q$ covers of $K_4$ with dimension vector $N=(2,1)$ and with grading $d_a=(0,0,1,1)$. The conventions are the same as in~\protect\ref{tab:quivers Qhat K3 expls}.}
\label{tab:k4_quivers}
\end{table}

\medskip
\noindent
{\bf $\Z_2$-cover of $K_4$.} As another simple example, take the $Q=K_4$ quiver of figure~\ref{fig:K4 quiver} with $\Z$ grading $d_a=(0,0,1,1)$ for the four arrows, hence with the circle action
\be\label{Cstar action K4}
 \C^\ast\; :\; (a_0, a_1, a_2, a_3)\rightarrow (a_0, a_1, \lambda a_2,\lambda a_3)~. 
\ee
The $\Z_2$ cover is shown in figure~\ref{fig:K4 cover}. Let us consider the dimension vector $N=(2,1)$. The corresponding moduli space is a Grassmannian manifold:
\be
\CM^Q_{(2,1)} \cong {\rm Gr}(2,4)\cong \left\{\bmat{a^1_0 &a^1_1 &a^1_2 &a^1_3 \\ a^2_0 &a^2_1 &a^2_2 &a^2_3}\right\}~,
\ee
which is parameterized by the maximal-rank $2\times 4$ matrix $a_{s}^k$ ($k=1,2$ and $s=0,1,2,3$) modulo $GL(2)$ transformations acting from the left. There are three fixed loci of the circle action~\eqref{Cstar action K4},  namely:
\be
\mathbb{P}^0\cong \left\{\bmat{a^1_0 &a^1_1 &0&0 \\ a^2_0 &a^2_1 &0 &0}\right\},\;
\mathbb{P}^0\cong  \left\{\bmat{0 &0 &a^1_2 &a^1_3 \\ 0 &0 &a^2_2 &a^2_3}\right\},\;
\mathbb{P}^1\times \mathbb{P}^1\cong \left\{\bmat{0 &0 &a^1_2 &a^1_3 \\ a^2_0 &a^2_1 &0 &0}\right\}.
\ee
The corresponding $\h Q$ quiver data is shown in table~\ref{tab:k4_quivers}, from which we can reproduce
\be
\Omega^Q((2,1), y)= y^4+y^2+2 +{1\ov y^2}+{1\ov y^4}
\ee
using the relation~\eqref{Omega ref sum fixed loci}.

\section{Galois Covers and homomorphisms of graded Lie algebras}
\label{sec:AlgHoms}

In this section, given a Galois pair of quivers $Q$ and $\t Q$, we discuss several homomorphisms between the Kontsevich--Soibelman graded Lie algebras $\frak{g}_{Q}$ and $\frak{g}_{\tilde Q}$, or between subalgebras of these algebras, which help us to establish Galois covering relations between BPS invariants. In subsection~\ref{sec:GenGalHom} we consider general Galois pairs, while in subsection~\ref{sec:SymGalHom} we consider additional homomorphisms that can be defined for symmetric Galois covers. In subsection~\ref{subsec:hom and BPS rels}, we explain how these homomorphisms imply non-trivial relations between BPS invariants of Galois pairs. 

\subsection{Homomorphism for general Galois covers}
\label{sec:GenGalHom}

The KS algebra $\frak{g}_{Q}$ for $Q$ is the $\Gamma$-graded Lie algebra with commutation relations~\eqref{eq:gQcomm}. Similarly, the KS algebra  $\frak{g}_{\tilde Q}$ is the $\tilde \Gamma$-graded Lie algebra with commutation relations:%
\be
\label{eq:tildecomm}
[\tilde e_{\tilde \gamma}, \tilde e_{\tilde\gamma'}]=\left<\tilde\gamma, \tilde\gamma'\right> (-1)^{\left<\tilde\gamma, \tilde\gamma'\right>}\, \tilde e_{\tilde\gamma+\tilde\gamma'}~.
\ee
Recall that the Galois group $\G$ acts on $\t\Gamma$ as:
\be
\G\; :\; \t\Gamma \rightarrow \t\Gamma \; : \; \t\gamma \mapsto  g\tilde \gamma=\sum_{(j, h) \in \tilde Q_0} \tilde N_{j,h} \tilde \gamma_{j,h+g}~, \qquad \forall g\in \G~.
\ee
One can easily show that, for $\gamma= F_\ast(\t\gamma)$ and $\gamma'= F_\ast(\t\gamma')$, we have the identities:
\be
\label{eq:gtgprops}
\langle g\tilde \gamma, g \tilde \gamma'\rangle=\langle \tilde \gamma, \tilde \gamma'\rangle~,\qquad \qquad \sum_{g\in
  \G} \langle \tilde \gamma, g \tilde \gamma'\rangle = \langle
\gamma,\gamma'\rangle~.
\ee

\medskip\noindent
{\bf $\G$-invariant subalgebra.} For our purpose, it is very useful to define the subalgebra
\be
\mathfrak{g}^\G_{\tilde Q} \subset \mathfrak{g}_{\tilde Q}
\ee
generated by the $\G$-invariant elements:
\be\label{eq:defeG}
\tilde e^\G_{\tilde \gamma}\equiv \sum_{g\in \G} \tilde e_{g\tilde \gamma}.
\ee
Note this definition only depends on the orbit $[\t\gamma]$ of $\t \gamma$, and that the right hand side is a sum of $|\G|$ elements, even if $\t \gamma$ has non-trivial stabiliser group.%
\footnote{Thus we should rather write this element as $\tilde e^\G_{[\tilde \gamma]}$, but prefer to write it as $\tilde e^\G_{\tilde \gamma}$ to avoid clutter.} 
To prove that the elements~\eqref{eq:defeG} indeed generate a subalgebra, we compute the commutator using~\eqref{eq:tildecomm}:
\bea\label{eq:tildec}
&\left[ \tilde e^\G_{\tilde \gamma} , \tilde e^\G_{\tilde \gamma'} \right]&=&\;\sum_{g\in \G} \sum_{g'\in \G} \langle g\tilde \gamma,g'\tilde
\gamma'\rangle (-1)^{\langle g\tilde \gamma,g'\tilde
\gamma'\rangle}\,\tilde e_{g\tilde \gamma+g'\tilde \gamma'}\\
&&=&\; \sum_{g \in \G}\sum_{g''\in \G} \langle g \tilde \gamma,g g''\tilde
\gamma'\rangle (-1)^{\langle g \tilde \gamma,g g''\tilde
\gamma'\rangle}\tilde e_{g(\tilde \gamma+g''\tilde \gamma')} \\
&&=&\; \sum_{g \in \G} \langle \tilde \gamma,g\tilde
\gamma'\rangle (-1)^{\langle \tilde \gamma,g\tilde
\gamma'\rangle}\,\tilde e^\G_{\tilde \gamma+g\tilde \gamma'}~.
\eea
Here we set $g'=g\,g''$ in the second line and we then use the first identity in~\eqref{eq:gtgprops}, so that the sum over $g$  simply returns the $\G$-invariant elements $\tilde e^\G_{\tilde
  \gamma+g''\tilde \gamma'}$.

\medskip\noindent
{\bf The KS algebra homomorphism $\mathfrak{f}_\ast^\G$.}  It is natural to try and define a graded Lie algebra homomorphism between $\frak{g}_{\t Q}$ and $\frak{g}_{Q}$ that would somehow correspond to the pull-down functor $F_\ast$ that acts on quiver representations. In the general case, the relevant homomorphism actually acts on the $\G$-invariant elements only. We define:
\be
\label{eq:fGlam}
\mathfrak{f}^\G_\ast\; : \; \mathfrak{g}^\G_{\tilde Q}\to \mathfrak{g}_{Q} \; :\; \tilde e^\G_{\tilde \gamma}\mapsto \mathfrak{f}^\G_\ast(\tilde e^\G_{\tilde \gamma}) =
\xi(\gamma,\tilde \gamma)\, e_{\gamma}~,
\ee
where $\gamma=F_\ast(\tilde \gamma)$ and the sign $\xi(\gamma,\tilde \gamma)$ was defined in~\eqref{eq:xigtg}. 
This is well-defined on $\mathfrak{g}^\G_{\tilde Q}$ since the parity of $\dim_{\mathbb{C}}(\CM^{\t Q}_{\t \gamma})$ is independent of $\t\zeta$ and $\xi$  is $\G$-invariant --- that is:
\be
\label{eq:xiG}
\xi(\gamma,\tilde \gamma)=\xi(\gamma,g\tilde \gamma)~, \qquad \forall g\in \G~.
\ee
It also follows from~\protect\eqref{eq:dimM} and~\protect\eqref{eq:antiNN} that $\xi(\gamma,\tilde \gamma)$ satisfies the cocycle relation:
\be
\label{eq:xicocy}
\xi(\gamma+\gamma',\tilde \gamma+\tilde \gamma')\, \xi(\gamma,\tilde
\gamma)\, \xi(\gamma',\tilde
\gamma')=(-1)^{\langle\gamma,\gamma'
  \rangle+\langle\tilde \gamma,\tilde \gamma' \rangle}~.
\ee
As an alternative to~\protect\eqref{eq:xigtg}, one may define $\xi$ iteratively through this cocycle condition after setting $\xi(\gamma_{j},\tilde \gamma_{j,g})=1$ for the vertices $j\in Q_0$ and $(j,g)\in \tilde Q_0$.%

To prove that~\eqref{eq:fGlam} is indeed a Lie algebra homomorphism, we first note that
\be
\label{eq:lhsegcomm}
\mathfrak{f}^\G_\ast\left(\left[ \tilde e^\G_{\tilde \gamma} , \tilde e^\G_{\tilde \gamma'} \right]\right) = \xi(\gamma,\tilde \gamma) \xi(\gamma',\tilde \gamma') \left[ e_{\gamma} , e_{\gamma'} \right].
\ee
On the other hand, $\mathfrak{f}^\G_\ast$ acting  on the third line of~\eqref{eq:tildec} gives us
\be
\sum_{g \in \G} \langle \tilde \gamma,g\tilde
\gamma'\rangle\, (-1)^{\langle \tilde \gamma,g\tilde
\gamma'\rangle} \,\xi(\gamma+\gamma', \t \gamma + g \t \gamma')\,  e_{\gamma+\gamma'}~.
\ee
Using the cocycle identity \eqref{eq:xicocy} and  the property~\eqref{eq:xiG}, this reduces to:
\be
\begin{split}
&\xi(\gamma,\tilde \gamma) \xi(\gamma',\tilde \gamma')\sum_{g \in \G} \langle \tilde \gamma,g\tilde
\gamma'\rangle (-1)^{\langle  \gamma,
  \gamma'\rangle}  e_{\gamma+\gamma'}\\
&=\xi(\gamma,\tilde \gamma)
\xi(\gamma',\tilde \gamma') \langle \gamma, 
\gamma'\rangle (-1)^{\langle  \gamma,
\gamma'\rangle}  e_{\gamma+\gamma'}~,
\end{split}
\ee
which is the $\mathfrak{g}_{Q}$ commutator for the right-hand side of~\eqref{eq:lhsegcomm}.

\subsection{Homomorphisms for a symmetric Galois cover}
\label{sec:SymGalHom}
There exists a richer structure of algebra homomorphisms if the Galois cover is symmetric, that is to say if the cover satisfies the constraint~\eqref{eq:symmCover<>}. In this case, it is not necessary to project $\mathfrak{g}_{\tilde Q}$ to the $\mathbb{G}$-invariant subalgebra $\mathfrak{g}^{\mathbb{G}}_{\tilde Q}$ considered above. Instead, we can directly apply the forgetful map $F_\ast: \tilde \Gamma \to \Gamma$ provided by the push-down functor on the charge lattice, as defined in \eqref{eq:Flambdagamma}. 
Let us introduce the $\Gamma$-graded Lie algebra $\mathfrak{\underline g}_{\tilde Q}$ generated by $\underline e_\gamma$, $\gamma\in \Gamma$, with the commutation relations
\be
\label{eq:underlinecommute}
\left[\underline e_{\gamma}, \underline e_{\gamma'}\right]=\left< \tilde \gamma, \tilde \gamma'\right>\, (-1)^{\left< \tilde \gamma, \tilde \gamma'\right>} \, \underline e_{\gamma+\gamma'}~,
\ee
where $\tilde \gamma,\tilde \gamma'\in \tilde \Gamma$ are charges such that $F_\ast(\tilde \gamma)=\gamma \in \Gamma$ and $F_\ast(\tilde \gamma')=\gamma' \in \Gamma$. The symmetric condition~\eqref{eq:symmCover<>} ensures that $\left< \tilde \gamma, \tilde \gamma'\right>$ is independent of the choice of $\tilde \gamma$, $\tilde \gamma'$. We then have the forgetful homomorphism:
\be
\label{eq:frakuh}
\mathfrak{{\underline  h}}\;:\; \mathfrak{g}_{\tilde Q}  \to \mathfrak{\underline g}_{\tilde Q} \; :\; \tilde e_{\tilde \gamma} \mapsto \mathfrak{{\underline  h}}(\tilde e_{\tilde \gamma})= {\underline e}_{\gamma}~,
\ee
with $F_\ast(\tilde \gamma)=\gamma$. 
Indeed, the commutation relation \eqref{eq:tildecomm} implies \eqref{eq:underlinecommute}. 
The algebra $\mathfrak{\underline g}_{\tilde Q}$ can be realised as a subalgebra of $\mathfrak{g}_{\tilde Q}$ in many ways. For example, for each $j\in Q_0$, $\underline e_{\gamma_j}$ can be identified with one of the elements $\tilde e_{\tilde \gamma_{j,g}}$, giving us $|Q_0|^{|\G|}$ distinct realizations. Note that the image of the $|\mathbb{G}|$-invariant element~\eqref{eq:defeG} is
\be
\mathfrak{\underline h}(\tilde e_{\tilde \gamma}^{\mathbb{G}})=|\G|\,\underline e_\gamma~.
\ee
It follows from (\ref{eq:underlinecommute}) that there is a bijective homomorphism $\mathfrak{\underline f}_*$ between the two $\Gamma$-graded algebras $\mathfrak{\underline g}_{\tilde Q}$ and $\mathfrak{g}_{Q}$, with
\be
\label{eq:frakfdownlam}
\mathfrak{\underline f}_*({\underline e}_{{\gamma}})=\frac{\xi(\gamma, \tilde
  {\gamma})}{|\G|}\, e_\gamma~,\qquad F_\ast(\tilde \gamma)=\gamma~,
\ee
with $\xi$ as discussed in the previous subsection. This is again independent of the choice of $\tilde \gamma\in \tilde \Gamma$ by the properties of a symmetric cover. 

\medskip
\noindent
{\bf Surjective homomorphism and motivic lift.}
The composition $\mathfrak{f}_*=\mathfrak{\underline f}_*\circ \mathfrak{\underline h}$ provides the surjective algebra homomorphism:
\be
\mathfrak{f}_*=\mathfrak{\underline f}_*\circ \mathfrak{\underline h}\;:\; \mathfrak{g}_{\tilde Q}\to \mathfrak{g}_{Q} \; :\; \tilde e_{\tilde \gamma}\mapsto \mathfrak{f}_*(\tilde e_{\tilde \gamma})=\frac{\xi(\gamma,\tilde
  \gamma)}{|\G|}\,e_{\gamma}~, \qquad \gamma=F_\ast(\tilde \gamma)~.
\ee
This is a homomorphism since it maps the commutation relation~\eqref{eq:tildecomm} to:  
\be
\frac{\xi(\gamma,\tilde \gamma) \xi(\gamma',\tilde
  \gamma')}{|\G|}\, [ e_{\gamma},  e_{\gamma'}]=
\frac{\xi(\gamma+\gamma',\tilde \gamma+\tilde \gamma')}{|\G|}
\langle \t\gamma,\t\gamma'\rangle\, (-1)^{\left<\tilde\gamma, \tilde\gamma'\right>}\,e_{\gamma+\gamma'}~,
\ee
which reproduces~\eqref{eq:gQcomm} upon using the symmetric property~\eqref{eq:symmCover<>} and the cocycle relation~\eqref{eq:xicocy}.  This relation for symmetric covers may be lifted to a relation between quantum tori. We let $\hat {e}_{\gamma}$ be the generators of the quantum torus $\CR_Q$ associated to $\mathfrak{g}_Q$ satisfying ~\eqref{eq:QuTo}, and $\hat {\underline e}_{\gamma}$ the generators of the quantum torus $\underline \CR_{\t Q}\subset \CR_{\t Q}$ associated to $\mathfrak{\underline g}_{\t Q}$, satisfying
\be
\left[ \hat {\underline e}_\gamma , \hat {\underline e}_{\gamma'}\right]=(-\t y)^{\langle \t \gamma,\t \gamma'\rangle}\, \hat {\underline e}_{\gamma+\gamma'}, \qquad F_*(\tilde \gamma)=\gamma,\,\,F_*(\tilde \gamma')=\gamma'.
\ee
The homomorphism $\mathfrak{\underline f}_*$ then acts on the quantum tori as follows:
\be 
\label{eq:frakQT}
\mathfrak{\underline f}_*:\qquad \hat { \underline e}_{\gamma}\to \xi(\gamma,\t\gamma)\, \hat {e}_{\gamma},\qquad \t y^{1/|\G|}=y~.
\ee 

\medskip
\noindent
{\bf Injective homomorphism.} Finally, symmetric covers also allow for the injective homomorphism:
\be
\label{eq:frakfuplam}
\frak{\underline f}^*:\mathfrak{g}_{Q}\to \mathfrak{\underline g}_{\tilde Q},\qquad \mathfrak{\underline f}^*( {e}_{\gamma})=\frac{\xi(\gamma, |\G|\tilde
  {\gamma})}{|\G|}\,  {\underline e}_{
  {|\G|\gamma}},\qquad F_\ast(\tilde \gamma)=\gamma~.
\ee
Note this maps only to elements ${\underline e}_{
  {\gamma}}$ with charges $\gamma$ divisible by $|\G|$, and is thus not surjective.

\subsection{Homomorphisms and covering relations for BPS invariants}\label{subsec:hom and BPS rels}
The homomorphisms described above will act on the KS monodromy operator $\mathbb{M}_{\tilde Q}$. For any Galois cover, the BPS invariants $\Omega(\tilde \gamma,\tilde \zeta)$ are $\G$-invariant if $\tilde \zeta=F^\ast(\zeta)$, namely:
\be
\label{eq:OmegaG}
\Omega(g\tilde \gamma,\tilde \zeta)=\Omega(\tilde \gamma,\tilde \zeta)~, \qquad \forall g\in \G~.
\ee
Even for a generic $\zeta$, the choice $\tilde \zeta=F^\ast(\zeta)$ is a highly non-generic stability condition in $\t Q$: many quiver representations will have the same slope even though their dimension vectors are not necessarily parallel. The prescription~\eqref{eq:sameslope} for such states sitting on the same ray implies that $\mathbb{M}_{\tilde Q}$ can be written in terms of the $\G$-invariant elements~\eqref{eq:defeG}, as we now explain. Firstly, assuming that $\zeta$ is a generic stability condition for $Q$, the monodromy operator of $\t Q$ for the fine-tuned stability condition can be written in terms of a product over primitive elements of $\Gamma_+$, as:
\be
\mathbb{M}_{\tilde Q}(\t\zeta=F^*(\zeta))=\prod^\curvearrowright_{\substack{ \gamma\in  \Gamma_+ \\ \gamma\,\,\text{primitive}}}\exp\left( \sum_{\ell=1}^\infty \sum_{\t\gamma \vert F_*(\t\gamma)=\ell \gamma} \bar \Omega(\t \gamma,\t\zeta)\,\t e_{\t \gamma} \right).
\ee
Here we put in the same exponent the generators $\t e_{\t\gamma}$ for the charges $\t\gamma$ that are sent to the same $\gamma$ of $Q$, allowing for the possibility that the latter might not be primitive. Then, using the orbit-stabiliser theorem, we can write the exponent as a sum over $\G$-orbits $[\t\gamma]$ as
\be
\exp\left( \sum_{\ell=1}^\infty \sum_{[\t\gamma] \vert F_*(\t\gamma)=\ell\gamma} \sum_{g\in \G} \frac{\bar \Omega(g\t \gamma,\t\zeta)}{|\text{Stab}(\t\gamma)|}\,\t e_{g\t \gamma} \right),
\ee
with $\text{Stab}(\t\gamma)\subset \G$ the stabilizer group of $\tilde \gamma$. Using~\eqref{eq:OmegaG}, we can express this in terms of the $\G$-invariant elements $\t e^{\G}_{\t\gamma}$:
\be
\exp\left( \sum_{\ell=1}^\infty \sum_{[\t\gamma] \vert F_*(\t\gamma)=\ell\gamma} \frac{\bar\Omega(\t \gamma,\t\zeta)}{|\text{Stab}(\t\gamma)|} \,\t e^\G_{\t \gamma} \right).
\ee
Finally, writing the sum over orbits $[\t\gamma]$ as a sum over distinct pre-images $\t\gamma$ of $\ell\gamma$ gives us
\be
\exp\left(\frac{1}{|\G|} \sum_{\ell=1}^\infty \sum_{\t\gamma \vert F_*(\t\gamma)=\ell\gamma} \bar\Omega(\t \gamma,\t\zeta) \,\t e^\G_{\t \gamma} \right)~.
\ee

\medskip
\noindent
{\bf Relating the monodromy operators.} 
Given the above, acting with the surjective homomorphism~\eqref{eq:fGlam} on $\mathbb{M}_{\tilde Q}$ and using the fact that the $e_{\ell \gamma}$'s commute, we find:
\be
\label{eq:expfrakf}
\mathfrak{f}^\G_*(\mathbb{M}_{\tilde Q})=\prod^\curvearrowright_{ \gamma\in  \Gamma_+}\exp\left(\frac{1}{|\G|} \sum_{\t\gamma \vert F_*(\t\gamma)=\gamma} \bar\Omega(\t \gamma,\t\zeta) \, e_{\gamma} \right).
\ee
Comparing with $\mathbb{M}_{Q}$ given in~\eqref{MonodCV}, we see that this is precisely of the desired form to derive the covering relation~\eqref{eq:CoverBPSInv} for rational BPS invariants. 
To complete the argument, we need to demonstrate that 
\be\label{cov rel for MQs}
\mathfrak{f}^\G_*(\mathbb{M}_{\tilde Q}(F^\ast(\zeta))) = \mathbb{M}_{Q}(\zeta)
\ee
for at least one choice $\zeta$. Since $\mathbb{M}_{\tilde Q}$ and $\mathbb{M}_{Q}$ are independent of the stability condition, this would establish~\eqref{eq:CoverBPSInv} for every $\zeta$. 
While proving~\eqref{cov rel for MQs} is an open question for Galois covers of arbitrary quivers, it can be established for large families of quivers $Q$ where `easy' stability conditions exist, including:
\begin{itemize}
\item \underline{Quivers without loops}: One can choose the FI parameters such that $\langle \gamma_i,\gamma_j\rangle\, (\zeta_i-\zeta_j)<0$ for all pairs $i,j\in Q_0$. As a result, the only non-vanishing BPS invariants are $\Omega(\gamma_j)$ with $\gamma_j$ the charge associated to $j\in Q_0$. Since the Galois cover $\t Q$ has precisely $|\G|$ nodes for each node of $Q$, the relation~\eqref{eq:CoverBPSInv} is satisfied and $\mathfrak{f}^\G_*$ indeed relates the KS monodromies $\mathbb{M}_{\tilde Q}$ and $\mathbb{M}_Q$ for this choice of $\zeta$. Then, by wall-crossing invariance, it holds for any choice $\zeta$.
\item \underline{Quivers for which empty attractor chambers can be established}: The attractor point $\zeta^\star$ \cite{Alexandrov:2018iao, Beaujard:2020sgs}, or self-stability \cite{Bridgeland:2017}, is a special choice of FI parameter $\zeta$ determined by the quiver $Q$ and dimension vector $N$,
\be
\zeta^\star_i=-\sum_{j\in Q_0} \langle \gamma_i,\gamma_j\rangle\,N_j~.
\ee
For Calabi--Yau quivers based on del Pezzo surfaces or $\KK E_n$ quivers, it can be shown that the expected dimension of the moduli space is negative for $\zeta^\star_i$, such that the corresponding BPS invariant vanishes \cite{Beaujard:2020sgs}. Since the wall-crossing away from $\zeta^\star_i$ is governed by the Lie algebra, the Lie algebra homomorphism again ensures the relation (\ref{eq:CoverBPSInv}). This argument does not hold for charges in the kernel of $\langle - , \gamma\rangle$ for all $\gamma$, whose BPS invariants are independent of $\zeta$. These charges include multiples of $\delta=\sum_{j\in Q_0} \gamma_j$, or D0-branes from the string theory perspective. For these charges, one can compare the attractor invariants determined in \cite{Mozgovoy:2020has} to verify (\ref{eq:CoverBPSInv}). 
\end{itemize}

\medskip\noindent
{\bf Symmetric Galois covers and monodromy operators.} 
Symmetric Galois covers are a special case, where we expect that further relations can be established using the homomorphisms of subsection~\ref{sec:SymGalHom}.  Recall that the BPS invariants of symmetric Galois covers $\t Q$ are invariant under small perturbation of $\tilde \zeta$, which facilitates their evaluation. Moreover, the homomorphism $\mathfrak{f}_*^\G$ factorizes, $\mathfrak{f}_*^\G=\mathfrak{\underline f_*}\circ \mathfrak{\underline h}$. Denoting by $\underline {\bM}_{\tilde Q}$ the KS monodromy operator in terms of the element $\underline e_{\gamma}$, we have the maps
\be
\begin{split}
&\mathfrak{\underline h}:\qquad {\bM}_{\tilde Q}\to \underline {\bM}_{\tilde Q}~,\\
&\mathfrak{\underline f_*}:\qquad \underline {\bM}_{\tilde Q}\to {\bM}_{Q}~.
\end{split}
\ee
For symmetric covers, we have proposed the relation \eqref{eq:OmQOmtQy2} for the refined BPS invariants. Similarly to the above discussion leading to~\eqref{cov rel for MQs}, this relation can be understood from a homomorphism mapping the refined monodromy operator ${\bM}_{\tilde Q}(\t y)$ to ${\bM}_{Q}(y)$. This map is induced by the homomorphism of non-commutative tori defined in~\eqref{eq:frakQT}. To see this, let us fix $\t \zeta=F^*(\zeta)$ and introduce the product $\tilde {\underline \CU}_\gamma$ over all charges $\t \gamma$ with $F_*(\t \gamma)=\gamma$ defined as
\be
\label{eq:deftU}
\tilde {\underline \CU}_\gamma=\exp\left( \sum_{\t\gamma | F_*(\t \gamma)=\gamma} \frac{\Omega(\t \gamma,\t y;\t \zeta)}{n} \frac{\hat {\underline e}_{n\gamma}}{\t y^n - \t y^{-n}}\right)~.
\ee
Combined with the assignment \eqref{eq:assignOmS} of non-trivial single-centred indices and the following relation for $\t y^{1/|\G|}=y$:
\be
\frac{\sum_{\alpha=0}^{|\G|-1} \tilde y^{(2\alpha+1-|\G|)/|\G|}}{ \tilde y- \tilde y^{-1}} = \frac{1}{y- y^{-1}}~,
\ee 
one verifies that, for $\gamma_j\in Q_0$,
\be
\mathfrak{f}_*(\tilde {\underline \CU}_{\gamma_j})=\CU_{\gamma_j}~,
\ee
since $\Omega^Q(\gamma_j,\zeta)=1$. This confirms the relation~\eqref{eq:OmQOmtQy2} for these simple dimension vectors, and indeed this was our original motivation to assign single-centred invariants as given in~\eqref{eq:assignOmS}. Using the same arguments as for the bullet points above, one can then argue that \eqref{eq:OmQOmtQy2} also holds for general charge vectors of those families of quivers.

Finally, we comment on the action of $\frak{\underline f}^*$. Since this map is not surjective, it does not provide a map from $\mathbb{M}_{Q}$ to ${\mathbb{\underline  M}}_{\tilde Q}$, in general, and therefore cannot be used as a general strategy to verify relations between BPS invariants. Nonetheless, it can be useful in specific instances where the support of the BPS invariants of $\tilde Q$ is such that all the relevant information is encoded in an appropriate subset of the KS algebra. We see in section~\ref{sec:Symm2Cover} below that the Galois pair of the 4d $\CN=2$ $SU(2)$ $N_f=0$ and $N_f=2$ BPS quivers gives one such example. 

\section{Examples of Galois covers and their BPS invariants}\label{sec:GaloisExamples}
In this last section, we study various specific examples of Galois covers and their BPS spectra, verifying our covering formulas explicitly. Sections~\ref{sec:Symm2Cover}--\ref{sec:E0E6} deal  with symmetric Galois covers, while subsection~\ref{sec:nonsymGC} discusses non-symmetric covers.

\subsection{Symmetric 2-cover of the Kronecker quiver \texorpdfstring{$K_2$}{K2}}
\label{sec:Symm2Cover}
Recall that the Kronecker quiver $K_d$ is the quiver with two vertices and $d$ arrows from vertex 1 to vertex 2. For a symmetric cover $\tilde Q$ of
order $n$, any vertex in one block of $\tilde Q$ is connected to each vertex in the other block by $d/n$ arrows. In the following, we will consider the 2-cover of $K_2$, the 3-cover of the $K_3$, and the 2- and 4-cover of $K_4$.

\medskip\noindent
{\bf The $K_2$ quiver: pure $SU(2)$ $\CN=2$ super-Yang--Mills and $N_f=2$ cover.} The Kronecker quiver shown in figure~\ref{fig:Nf0Nf2} is famously the BPS quiver $Q$ of the 4d $\CN=2$ $\SU(2)$ theory ($N_f=0$), as already discussed in section~\ref{sec:4dSQCD}. Recall that we label the electric-magnetic charge of this gauge theory as $\gamma=(m,e)$, with the standard Dirac pairing:
\be
\left<\gamma,\gamma' \right>=me'-em'~.
\ee
 In this basis for $\Gamma\cong \Z^2$, the charges of the two quiver nodes are 
 \be
 \gamma_1=(1,0)~, \qquad \gamma_2=(-1,2)~,
 \ee
 with $\gamma_1+\gamma_2=(0,2)$ the charge of the $W$-boson.%
 \footnote{The fact that this choice of $\{\gamma_1, \gamma_2\}$ does not form a primitive basis of $\Z^2$ is related to the existence of a one-form symmetry $\Z_2^{(1)}$ of the 4d gauge theory --- see~{\it e.g.}~\protect\cite{Closset:2023pmc} for a detailed discussion. \label{footnote:Z21form}}

 The symmetric 2-cover $\tilde Q$ (shown in figure~\ref{fig:Nf0Nf2} as well) is the BPS quiver of the 4d $SU(2)$ $N_f=2$ gauge
theory, as discussed in section \ref{sec:galois cover}.  The charge lattice of this 4d gauge theory is four-dimensional, with two flavour charges $f_1$, $f_2$ for the Cartan of the $\mathfrak{so}(4)$ symmetry and with the same electromagnetic charges as for $N_f=2$. In that basis, we write any charge as $\t\gamma=(m, e, f_1, f_2) \in \t\Gamma \cong \Z^4$, and the four quiver nodes correspond to the charges: 
\bea
&\tilde \gamma_{1,0}=(1,0,-1,0)~,\qquad && \tilde\gamma_{2,0}=(-1,1,0,-1)~,\\
&\tilde \gamma_{1,1}=(1,0,1,0), \qquad
&&\tilde \gamma_{2,1}=(-1,1,0,1)~.
\eea
The standard Dirac pairing, $\langle \t \gamma,\t \gamma'\rangle= me'-e m'$, reproduces the arrows of $\t Q$. 
The pull-down functor acts surjectively on the set
of vertices and maps the charges according to:
\be 
F_\ast(\tilde \gamma_{j,\alpha}) = \gamma_j~, \qquad \alpha=0,1~,\qquad \quad
F_\ast((m,e,f_1,f_2))=(m,2e)~. 
\ee 
We also have that $F^\ast((m,e))=2(m,e,0,0)$, but the relation $F_\ast(\t\gamma)=\gamma$ is the most important one for our purposes.

\begin{table} 
\center
\renewcommand{\arraystretch}{1.4}
\begin{tabular}{|c|c|r|}
\hline 
$N$ & $\gamma$ & $\Omega^Q(N, \zeta)$ \\
\hline \hline
$(1,1)$ & $(0,2)$ & $-2$  \\
$(k+1,k)$ & $(1,2k)$ & 1  \\
$(k,k+1)$ & $(-1,2(k+1))$ & 1  \\
\hline 
\end{tabular}
\caption{Non-vanishing BPS invariants $\Omega^Q(N,\zeta)$ for the $K_2$ Kronecker quiver with $\zeta_1>\zeta_2$. }
\label{BPSInvK2}
\end{table}

\begin{table} 
\center
\renewcommand{\arraystretch}{1.4}
\begin{tabular}{|c|c|r|r|}
\hline 
$\tilde N$ & $\tilde \gamma$ & $\Omega^{\t Q}(\tilde N,\tilde \zeta)$ & \# permutations\\
\hline \hline
$(1,0;1,0)$ & $(0,1,-1,-1)$ & 1 & 4 \\
$(1,1;1,1)$ & $(0,2,0,0)$ & $-2$ & 1\\
$(n,n;n\pm 1,n)$ & $(\mp 1,2n\pm 1,0,\mp 1)$ & 1 & 2 \\
$(n\pm 1,n;n,n)$ & $(\pm 1,2n,\mp 1,0)$ & 1 & 2 \\
\hline 
\end{tabular}
\caption{Non-vanishing BPS invariants $\Omega^{\t Q}(\tilde N,\tilde \zeta)$ for the symmetric $2$-cover
  $\tilde Q$ of $K_2$ with $\tilde \zeta=F^\ast(\zeta)$. }
\label{BPSInvQK2cover}
\end{table}

The BPS spectra for this Galois pair were already spelled out in section~\ref{sec:4dSQCD} and are displayed in tables~\ref{BPSInvK2} and~\ref{BPSInvQK2cover}. When specialised to this case, our general Galois covering formula~\eqref{eq:CoverBPSInv} for rational BPS invariant reads:
\be  
\label{CoverRelInv}  
\sum_{\tilde \gamma | F_\ast(\tilde \gamma)=\gamma}\bar \Omega^{\tilde Q}(\tilde \gamma)=2\,(-1)^{(m+e)e}
\,\bar \Omega^{K_2}(\gamma), \qquad \gamma=(m,2e)~,
\ee
with $(m,e)$ the electromagnetic charges of the $N_f=2$ theory, and
where the sign $(-1)^{(m+e)e}$ is the specialization of $\xi(\gamma,\tilde \gamma)$ defined in~\eqref{eq:xigtg}. The identity~\eqref{CoverRelInv} is easily checked for primitive dimension vectors, and the case of the non-primitive charge $\gamma=(0,4)$ ({\it i.e.} $N=(2,2)$) was already discussed below~\eqref{checkN22}.

\medskip
\noindent
{\bf Another covering relation.} The symmetric $\mathbb{Z}_2$ cover of $K_2$ is special in that the
corresponding BPS spectra satisfy a second relation, namely
\be
\label{RatRelInv2}
\sum_{\tilde \gamma | F_\ast(\tilde \gamma)=\gamma}\bar\Omega^{\tilde Q}(2\tilde \gamma)=\frac{1}{2}\,\bar \Omega^{K_2}(\gamma).
\ee 
For example, for $\gamma=(0,2)$, we have for the left hand side $-1$ by \eqref{eq:bOmtQ04}, which agrees with half of
$\bar \Omega(\gamma)=\Omega(\gamma)=-2$. This relation is different
from the general relation (\ref{eq:CoverBPSInv}) --- in fact, it does not seem to have a natural generalisation to other Galois covers discussed below --- but it can be understood in terms of the homomorphisms discussed in section~\ref{sec:SymGalHom}, as we now explain. 

\medskip
\noindent
{\bf Relations between monodromy operators.} The KS monodromy operator for $Q$ can be written down either in the strong-coupling ($\zeta_1<\zeta_2$) or and weak coupling ($\zeta_1>\zeta_2$) region of stability parameters. It reads:
\be 
\label{KSNf0}
\mathbb{M}_Q=\left\{ \begin{array}{rr} 
\CK_{-1,2} \CK_{1,0}~, & \quad \zeta_1<\zeta_2~,\\
(\CK_{1,0} \CK_{1,2} \CK_{1,4}\cdots) \CK_{0,2}^{-2}(\cdots \CK_{-1,6} \CK_{-1,4} \CK_{-1,2})~, &\quad \zeta_1>\zeta_2~, 
\end{array} \right.
\ee 
in terms of the operators $\CK_\gamma= \CK_{m,e}$ defined in~\eqref{def CKgamma}. 
On the other hand, the KS monodromy operator for the $N_f=2$ BPS quiver can be most easily written for the subalgebra $\frak{\underline g}_{\tilde Q}$ obtained from  $\frak{g}_{\tilde Q}$ by ``forgetting'' the flavor charges, using the homomorphism $\mathfrak{\underline h}$ defined in~\eqref{eq:frakuh} --- this is because any two elements $e_{\t\gamma}$ distinguished only by their flavour charges commute. More generally, the homomorphism~\eqref{eq:frakuh} will be similarly useful whenever the Galois pairs correspond to 4d $\CN=2$ SQFTs with the same rank and with the covering corresponding to adding in flavour charges. The KS product $\mathbb{\underline M}_{\tilde Q}$ for the $N_f=2$ theory then reads:%
\footnote{In these subsections on symmetric Galois covers, in particular subsections~\protect\ref{sec:Symm2Cover} and~\protect\ref{sec:Sym3Cover}, we will only consider the $\Gamma$-graded Lie algebra $\mathfrak{\underline g}_{\tilde Q}$ and not the $\tilde \Gamma$-graded Lie algebra $\mathfrak{g}_{\tilde Q}$. To simplify notation and keep it closer to $Q$ versus $\t Q$, we will denote the elements of $\mathfrak{\underline g}_{\tilde Q}$ by $\tilde e_{\gamma}$ instead of $\underline e_{\gamma}$, and similarly use $\tilde \CK_{\gamma}$ for the corresponding Lie group elements. To stress that $\mathfrak{\underline g}_{\tilde Q}$ is only a subalgebra of $\mathfrak{g}_{\tilde Q}$, we continue to use the underline notation for the monodromy $\mathbb{\underline M}_{\tilde Q}$ and homomorphisms $\mathfrak{\underline f}^*$, $\mathfrak{\underline f}_*$.
}
\be \label{KSNf2}
\mathbb{\underline M}_{\tilde Q}=\left\{\begin{array}{rr}
\tilde \CK^2_{-1,1} \tilde \CK_{1,0}^2~,& \quad \zeta_{1,\alpha} <\zeta_{2,\alpha}~, \\
(\tilde \CK_{1,0}^2\tilde \CK_{1,1}^2\tilde \CK_{1,2}^2\cdots)\tilde \CK_{0,1}^4\tilde \CK_{0,2}^{-2}(\cdots \tilde \CK_{-1,3}^2\tilde \CK_{-1,2}^2\tilde \CK_{-1,1}^2)~,& \quad \zeta_{1,\alpha} >\zeta_{2,\alpha}~, 
\end{array}
\right. 
\ee 
 at strong and weak coupling, respectively. 
Let us now see how the formalism of section~\ref{sec:AlgHoms} applies to this Galois pair. As we will show, the two covering relations~\eqref{CoverRelInv} and~\eqref{RatRelInv2} are consequences of the identities
\be\label{rels between M and M SU2expl}
\mathfrak{\underline f}_*(\mathbb{\underline M}_{\tilde Q})=\mathbb{M}_{Q}
\qquad\qquad \text{and}\qquad\qquad
\frak{\underline f}^*(\mathbb{M}_{Q}) =  \mathbb{\underline M}_{\tilde Q}~,
\ee
respectively, which are written in terms of homomorphisms relating the KS algebras $\frak{g}_Q$ and $\frak{\underline g}_{\tilde Q}$. 

\medskip
\noindent
{\bf Homomorphism induced by the push-down functor.}
 Consider the homomorphism~\eqref{eq:frakfdownlam}, namely:
\be 
\frak{\underline f}_*\;:\; \frak{\underline g}_{\tilde Q}\to \frak{g}_{Q}\; :\; \qquad 2\,\tilde e_{m,e}\mapsto (-1)^{(m+e)e}\,e_{m,2e}~.
\ee 
This is actually an algebra isomorphism, with the charges in the subscripts of $e_\gamma$ and $\tilde e_{\tilde \gamma}$ being related as $\gamma=F_\ast(\tilde \gamma)$. 
For $m$ odd, this gives us a map:
\be 
\frak{\underline f}_*:\qquad \tilde \CK_{m,e}^2\to \CK_{m,2e}~,
\ee 
thus mapping such terms on the right hand side of~\eqref{KSNf2} to those of~\eqref{KSNf0}. The map for even $m$ is more intricate due to the signs, but one can show that:
\be\label{2coverHom1}
\frak{\underline f}_*:\qquad \tilde \CK_{0,1}^4 \tilde\CK_{0,2}^{-2}\to \CK_{0,2}^{-2}~.
\ee 
This proves the first equation in~\eqref{rels between M and M SU2expl}, and therefore establishes~\eqref{CoverRelInv} as explained in section~\ref{subsec:hom and BPS rels}. This equation was first established in~\cite[section~2.2]{Gaiotto:2010okc}  in terms of symplectomorphisms.

\medskip
\noindent
{\bf Homomorphism induced by the pull-up functor.}
Consider the homomorphism~\eqref{eq:frakfuplam} given by
\be 
\label{Hom2}
\frak{\underline f}^*\;:\; \frak{g}_{Q}\to \underline{\frak{g}}_{\tilde Q} \; :\;  e_{m,2e}\mapsto \frac{1}{2}\, \tilde e_{2(m,e)}~,
\ee 
where subscripts of $e_\gamma$ and $\tilde e_{\tilde \gamma}$ are here related through the pull-up functor, $F^\ast(\gamma)=2\tilde \gamma$. This map is injective on the $\Gamma$-graded KS algebra $\frak{g}_{Q}$ restricted to the physical charges, which are those with even electric charge --- {\it i.e.} setting $e\in \Z$ in~\eqref{Hom2}; see also footnote~\ref{footnote:Z21form}.
 The homomorphism~\eqref{Hom2} is not surjective. Instead, its image is the subalgebra $\frak{g}^{(2)}_{\tilde Q}\subset \frak{g}_{\tilde Q}$ with $\tilde \gamma$ divisible by $2$. Note also that the composition of the two homomorphisms doubles the electric-magnetic charge in the $N_f=2$ theory:
\be 
\frak{\underline f}^*\circ\frak{\underline f}_*:\qquad \tilde e_{m,e}\to (-1)^{(m+e)e}\,\tilde e_{2(m,e)}~.
\ee  
The projection $\mathfrak{h}^{(2)}$ of $\frak{g}_{\tilde Q}$ to $\frak{g}_{\tilde Q}^{(2)}$ projects $\tilde \CK_{m,e}$ to elements with charge $(m,e)$ divisible by $2$,
 \be
\mathfrak{h}^{(2)}(\tilde \CK_{m,e})=\left\{\begin{array}{rr} \tilde \CK_{m,e},& \quad 2\vert (m,e)~, \\ 
\tilde \CK_{2(m,e)}^{1/4}, & 2\! \not |\, (m,e)~.
\end{array}\right.
 \ee
Considering $\tilde \CV_{m,e}$ as defined in~\eqref{eq:Vgamma}, the projection $\mathfrak{h}^{(2)}$ maps $\tilde \CV_{m,e}\to 1$ for $(m,e)$ not divisible by 2. Thus, $\mathfrak{h}^{(2)}$ maps the monodromy $\mathbb{\underline M}_{\tilde Q}$ (\ref{KSNf2}) to
\be \label{KSNf2v2}
\mathbb{\underline M}^{(2)}_{\tilde Q}=\left\{\begin{array}{rr}
\tilde \CK^{1/2}_{-2,2} \tilde \CK_{2,0}^{1/2},& \quad \zeta_{1,\alpha} <\zeta_{2,\alpha}~, \\
(\tilde \CK_{2,0}^{1/2}\tilde \CK_{2,2}^{1/2}\tilde \CK_{2,4}^{1/2}\cdots)\tilde \CK_{0,2}^{-1}(\cdots \tilde \CK_{-2,6}^{1/2}\tilde \CK_{-2,4}^{1/2}\tilde \CK_{-2,2}^{1/2}),& \quad \zeta_{1,\alpha} >\zeta_{2,\alpha}~. 
\end{array}
\right. 
\ee 
The injective homomorphism~\eqref{Hom2} therefore maps $\mathbb{ M}_Q$ to $\mathbb{\underline M}^{(2)}_{\tilde Q}$, with corresponding identification~\eqref{RatRelInv2} between the BPS invariants.   

It is important to note that the relation~\eqref{RatRelInv2} cannot be understood as a direct consequence of the algebra homomorphism since the latter is not surjective. Instead, the equivalence of $\mathbb{\underline M}_{\tilde Q}^{(2)}$ and $\mathbb{M}_{Q}$ is a consequence of the fact that the support of the BPS invariants in tables \ref{BPSInvK2} and \ref{BPSInvQK2cover} is only on primitive dimension vectors, in fact on a one-dimensional sublattice of the two-dimensional charge lattice. For more general (non-tame) quivers, crossing walls of marginal stability will lead to non-vanishing $\Omega^{\tilde Q}(\tilde \gamma)$ for charges divisible by  $|\G|$. In other words, wall-crossing and the action of $\mathfrak{h}^{(|\G|)}$ do not commute in general,  because changing the order in $\mathbb{M}_{\t Q}$ of those $\tilde \CV_{\tilde \gamma}$ which are in the kernel of $\mathfrak{h}^{(|\G|)}$  generates new BPS states missed by first acting with $\mathfrak{h}^{(|\G|)}$. We will see this at play in the example of the 3-cover of $K_3$ discussed in subsection~\ref{sec:Sym3Cover}. 

\medskip\noindent
{\bf Covering formulas for refined BPS invariants.} 
 It is natural to seek to uplift the covering relations~\eqref{CoverRelInv} and~\eqref{RatRelInv2} to relations between refined invariants by using the refined KS product $\bM_Q(y)$ defined in~\eqref{eq:MQy}.

 Let us first consider the homomorphism induced by the push-down functor. As discussed in section~\ref{subsec:symgalcov}, we assign non-trivial single-centred invariants $\Omega_S^{\tilde Q}$ to the nodes of $\t Q$, which gives rise to the relation~\eqref{eq:OmQOmtQy2}. 
The assignment~\eqref{eq:assignOmS} for this $\Z_2$ cover means that we assign $\tilde y^{1/2}$ to one of the $\alpha=0$ node in each block $y_{j,\alpha}$ ($\alpha=0,1$),  and $\tilde y^{-1/2}$ to the $\alpha=1$ node, with $\Omega_S^{\tilde Q}(\tilde \gamma)=0$ for all other $\tilde \gamma \in \tilde \Gamma$. Then,  the relation~\eqref{eq:OmQOmtQy2} between the refined invariants of $Q$ and $\tilde Q$ with $\gamma=(m,2e)$ becomes:
\be 
\label{CoverRefInv}
\sum_{\tilde \gamma | F_\ast(\tilde \gamma)=\gamma} \bar \Omega^{\tilde Q}(\tilde \gamma,\tilde y)=(-1)^{(m+1)e}\,(y+y^{-1})\, \bar \Omega^{K_2}(\gamma,y)~,\qquad \tilde y^{1/2}=y~.
\ee 
Consider for instance $\gamma=(0,2)$ (that is, $N=(1,1)$) with $\zeta_{1}> \zeta_2$. The sum over BPS invariants of $\tilde Q$ consists of the 4 permutations of $\tilde N=(1,0;1,0)$. The MPS Coulomb-branch formula gives us
\be
\label{eq:CF02}
\sum_{\tilde \gamma | F_\ast(\tilde \gamma)=\gamma} \Omega^{\tilde Q}(\tilde \gamma,\tilde y)=\sum_{\alpha,\beta\in \{0,1\}} \Omega_S^{\tilde Q}( \gamma_{1,\alpha},\tilde y)\,\Omega_S^{\tilde Q}( \gamma_{2,\beta},\tilde y)
\ee
Given our assignment of single-centred invariants, this reduces to $\tilde y+2+\tilde y^{-1}$, which indeed agrees with~\eqref{CoverRefInv}. As another example, consider $\gamma=(-1,4)$. The non-vanishing contributions to the sum in~\eqref{CoverRefInv} are from $\tilde N=(1,1;1,0)$, $(2,0;0,1)$ and permutations within the blocks. Then, the MPS formula gives us
\be 
\begin{split}
\sum_{\tilde \gamma | F_\ast(\tilde \gamma)=\gamma}\Omega^{\tilde Q}(\tilde \gamma,\tilde y)&=\Omega_S^{\tilde Q}( \gamma_{1,0},\tilde y)\Omega_S^{\tilde Q}( \gamma_{1,1},\tilde y) \sum_{\alpha=0,1}\Omega_S^{\tilde Q}( \gamma_{2,\alpha},\tilde y)\\
& +\frac{1}{2} \sum_{\alpha,\beta\in \{0,1\}}\left( \Omega_S^{\tilde Q}( \gamma_{1,\alpha},\tilde y)^2-\Omega_S^{\tilde Q}( \gamma_{1,\alpha},\tilde y^2) \right) \Omega_S^{\tilde Q}( \gamma_{2,\beta},\tilde y)~,
\end{split}
\ee
which reduces to
$\tilde y^{1/2}+\tilde y^{-1/2}$,  
in agreement with~\eqref{CoverRefInv} since $\Omega^{K_2}(\gamma,y)=1$ for $\gamma=(-1,4)$. As a last example, consider the non-primitive charge $\gamma=(0,4)$ (that is, $N=(2,2)$), where we can find:
\be
\sum_{\tilde \gamma | F_\ast(\tilde \gamma)=\gamma} \bar \Omega(\tilde \gamma,\tilde y)=-\tilde y-\tilde y^{-1}+\frac{1}{2}\frac{\tilde y-\tilde y^{-1}}{\tilde y^2-\tilde y^{-2}}\,(\tilde y^2+2+\tilde y^{-2})=-\frac{1}{2}(\tilde y+\tilde y^{-1})
\ee 
for the rational invariants with non-zero contribution, to be compared with 
\be
\bar \Omega(\gamma,y)=\frac{1}{2} \frac{y-y^{-1}}{y^2-y^{-2}}\,(-y^2-y^{-2})
 = -\frac{1}{2}\frac{y^2+y^{-2}}{y+y^{-1}}~,
\ee 
which again agrees with~\eqref{CoverRefInv}.

The second Galois covering relation~\eqref{RatRelInv2} also generalises to the refined setting. Here, it is not necessary to introduce bespoke single-centred invariants as above, as the simple  $\Omega_S^{\tilde Q}(\gamma_{j,\alpha})=1$ for each node (and zero otherwise) does the job. We propose that the general relation for refined invariants is:
\be
\label{eq:OmRel2}
\sum_{\tilde \gamma | F_\ast(\tilde \gamma)=\gamma}\bar \Omega(2\tilde \gamma,\tilde y)=   \frac{\bar \Omega(\gamma,y)}{y^{1/2}+y^{-1/2}}~, \qquad \tilde y^2=y~.
\ee 
Considering $\gamma=(0,2)$ for instance, the rational refined invariants reorganise as:
\be 
\sum_{\tilde \gamma | F_\ast(\tilde \gamma)=\gamma}\bar \Omega(2\tilde \gamma,\tilde y) =-\tilde y-\tilde y^{-1}+\frac{4}{2}\frac{\tilde y-\tilde y^{-1}}{\tilde y^2-\tilde y^{-2}}
=\frac{-\tilde y^2-\tilde y^{-2}}{\tilde y+\tilde y^{-1}}~,
\ee 
in agreement with \eqref{eq:OmRel2}. Other examples can be checked similarly.

\medskip
\noindent
{\bf Homomorphisms of quantum tori.}  Analogously to our discussion of the homomorphisms between KS algebras, we can understand the relations between refined invariants from homomorphisms between the quantum tori associated to $\mathfrak{g}_Q$ and $\mathfrak{\underline g}_{\tilde Q}$. Following \eqref{eq:deftU}, we introduce $\t \CU_{m,e}$ as the product over all $\t \gamma$ with $F_*(\t\gamma)=(m,2e)$,
\be
\label{eq:deftCU}
\tilde \CU_{m,e}=\exp\left( \sum_{\substack{ f_1,f_2\\\tilde \gamma=(m,e,f_1,f_2)}} \sum_{n=1}^\infty \frac{\Omega(\tilde \gamma,\tilde y^n)}{n}\frac{\hat {\underline e}_{n(m,e)}}{\tilde y^n-\tilde y^{-n} }\right).
\ee
The refinement of the pull-down homomorphism reads:
\be 
\mathfrak{\underline f}_*:\qquad \hat {\underline e}_{m,e}\to (-1)^{(m+e)e}\, \hat {e}_{m,2e}~,
\ee 
as anticipated in~\eqref{eq:frakQT}. 
Using the single-centred invariants \eqref{eq:assignOmS} discussed above, one finds that, for the charges $\gamma=(1,0)$ and $\gamma=(-1,2)$ of the two nodes of $Q$, 
\be
\mathfrak{\underline f}_*:\qquad\tilde \CU_{m,e} \to \CU_{m,2e},
\ee 
with $\tilde \CU_{m,e}$ as in~\eqref{eq:deftCU} and $\CU_{m,2e}$ as in~\eqref{eq:DefU}. More generally, for $m$ odd and $\zeta_1>\zeta_2$, this follows  from~\eqref{CoverRefInv}. For $m$ even, the product for the charges $(0,1)$ and $(0,2)$ is mapped to $\CU_{0,2}$,
\be
\mathfrak{\underline f}_*:\qquad\tilde \CU_{0,1} \tilde \CU_{0,2} \to \CU_{0,2}~.
\ee 
We therefore realise that the motivic KS products are indeed related as follows:
\be 
\mathfrak{\underline f}_*:\qquad\mathbb{\underline M}_{\tilde Q}(\tilde y) \to \mathbb{M}_Q(\tilde y^{1/2})=\mathbb{M}_Q(y),\qquad \tilde y^{1/2}=y~.
\ee 
Finally, for the alternative relation~\eqref{eq:OmRel2} between refined invariants, the refined analogue of the action of the injective homomorphism $\frak{\underline f}^*$ on the generators of the quantum tori reads:
\be 
\frak{\underline f}^*:\qquad  \hat {e}_{m,2e}\to  \hat {\underline e}_{2(m,e)}~,\qquad  y\to \tilde y^2~.
\ee 
Then, the refined monodromy operators are related as:
\be 
\mathbb{M}_{Q}(y) \to \mathbb{\underline M}^{(2)}_{\tilde Q}(\tilde y)~,\qquad \tilde y^2=y~.
\ee 
Note that this relation first appeared in~\cite[Eq. (5.19)]{Cecotti:2015qha}.

\subsection{Symmetric 3-cover of the Kronecker quiver \texorpdfstring{$K_3$}{K3}}
\label{sec:Sym3Cover}

\begin{figure}[ht]
\centering
\begin{tikzpicture}[baseline=1mm, every node/.style={circle, draw, minimum size=1cm}, thick]
\node (L0) at (0,0)   {$\t\gamma_{1,0}$};
\node (L1) at (0,-1.5){$\t\gamma_{1,1}$};
\node (L2) at (0,-3)  {$\t\gamma_{1,2}$};
\node (R0) at (3,0)   {$\t\gamma_{2,0}$};
\node (R1) at (3,-1.5){$\t\gamma_{2,1}$};
\node (R2) at (3,-3)  {$\t\gamma_{2,2}$};
\foreach \i in {0,1,2} {
  \foreach \j in {0,1,2} {
    \draw[->-=0.4] (L\i) -- (R\j);
  }
}
\end{tikzpicture}
\qquad\qquad
\begin{tikzpicture}[baseline=1mm, every node/.style={circle, draw, minimum size=1cm}, thick]
\node (1) at (0,-1.5) {$\gamma_{1}$};
\node (2) at (2.4,-1.5) {$\gamma_{2}$};
\draw[->-=0.5] (1.25)  to[bend left=15]  (2.155);
\draw[->-=0.5] (1.0)   to                (2.180);
\draw[->-=0.5] (1.335) to[bend right=15] (2.205);
\end{tikzpicture}
\caption{The $\Z_3$ symmetric cover $\t K_3$ (\textsc{Left}) of the $K_3$ quiver (\textsc{Right}).}
\label{fig:k3_z3_cover}
\end{figure}
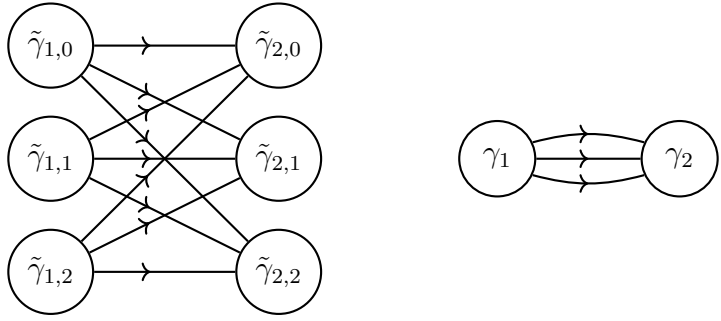

Our next example is the Kronecker quiver with three arrows, $Q=K_3$. Consider the $\G=\Z_3$ grading $d_a=(0,1,2)$, and the corresponding symmetric cover quiver $\t Q=\t K_3$ shown in figure~\ref{fig:k3_z3_cover}. These quivers are not BPS quivers of known 4d $\CN=2$ SQFTs, but it is still interesting to explore their BPS invariants to further test our covering formulas. Such quivers can also appear as sub-quivers of larger CY$_3$ quivers --- see for instance the $E_0$ quiver and its $E_6$ cover in figure~\ref{quiver E6 orbi}, which is obviously an extension of the symmetric cover $F:\t K_3\rightarrow K_3$. Another special feature of this $K_3$ example is that $\xi(F_\ast(\t\gamma), \t\gamma)=1$ for any $\t\gamma$. For Galois covers of this type, our general formula~\eqref{eq:CoverBPSInv} simplifies to a relation for integer invariants: 
\be
\label{eq:trivialcocycle}
\Omega^Q(\gamma; \zeta) ={1\ov |\G|} \sum_{\t\gamma\, |\,
  F_\ast(\t\gamma)=\gamma}  
\Omega^{\t Q}(\t\gamma; F^\ast(\zeta))~.
\ee

We may take the quiver basis $\gamma_{j}$ and $\t\gamma_{j, \alpha}$ ($j=1,2$ and $\alpha=0,1,2$) for the charge lattices $\Gamma\cong \Z^2$ and $\t\Gamma\cong \Z^6$, respectively, with the pull down-down functor acting as $F_\ast(\t\gamma_{j,\alpha})=\gamma_j$. 
For $\zeta_1<\zeta_2$ and $\tilde \zeta_{1,\alpha}<\tilde \zeta_{2,\beta}$,  the quivers $K_3$ and $\t K_3$ admit a finite number of BPS states corresponding to the quiver nodes. For
$\zeta_1>\zeta_2$ and $\tilde \zeta_1>\tilde \zeta_2$, we list some quiver invariants $\Omega^Q(N,\zeta)$ and $\Omega^{\t Q}(\t N,\t \zeta)$ of various dimension vectors in table~\ref{BPSInvQK3cover}. Using these results, we can easily check that~\eqref{eq:trivialcocycle} indeed holds --- that is, we find:
\be 
\label{eq:DTK3}
\sum_{\tilde \gamma | F_\ast(\tilde\gamma)=\gamma} \Omega^{\tilde Q}(\tilde \gamma,\tilde \zeta)=3\,\Omega^{K_3}(\gamma, \zeta)~,\qquad \quad \gamma=F_\ast(\tilde \gamma)~.
\ee

\begin{table}[ht]
\centering
\renewcommand{\arraystretch}{1.4}
\begin{tabular}{|c|r|}
\hline
$N$ & $\Omega^Q(N,\zeta)$\\
\hline
$(1,1)$ & $3$\\
$(2,2)$ & $-6$\\
$(3,3)$ & $18$\\
$(4,4)$ & $-84$\\
$(2,1)$ & $3$\\
$(4,2)$ & $-6$\\
$(6,3)$ & $18$\\
\hline
\end{tabular}
\hspace{1cm}
\begin{tabular}{|c|r|r|}
\hline
$\tilde N$ & $\Omega^{\t Q}(\tilde N, \tilde \zeta)$ & \# permutations\\
\hline\hline
$(1,0,0;1,0,0)$ & $1$ & 9 \\
$(1,1,0;1,1,0)$ & $-2$ & 9 \\
$(2,0,0;1,1,0)$ & $0$ & 18 \\
$(2,0,0;2,0,0)$ & $0$ & 9 \\
$(1,1,1;1,1,1)$ & $18$ & 1 \\
$(2,1,0;2,1,0)$ & $0$ & 36 \\
$(2,1,0;1,1,1)$ & $3$ & 12 \\
$(2,2,0;2,2,0)$ & $0$ & 9 \\
$(2,1,1;2,1,1)$ & $-20$ & 9 \\
$(2,2,0;2,1,1)$ & $-4$ & 18 \\
$(3,1,0;2,1,1)$ & $0$ & 36 \\\hline
$(1,1,0;1,0,0)$ & $1$ & 9 \\
$(2,0,0;1,0,0)$ & $0$ & 9 \\
$(2,1,1;1,1,0)$ & $-2$ & 9 \\
$(2,2,0;1,1,0)$ & $0$ & 9 \\
\hline
\end{tabular}
\caption{\textsc{Left:} a sample of non-vanishing BPS invariants $\Omega(N,\zeta)$ for $Q=K_3$ with $\zeta_1>\zeta_2$. \textsc{Right:} BPS invariants $\Omega(\tilde N,\tilde\zeta)$ for the $3$-cover of $K_3$ for $\tilde\zeta=F^*(\zeta)$. The first set of entries corresponds to charges of the form $F_*(\tilde\gamma)=n(\gamma_1+\gamma_2)$, the last four to $F_*(\tilde\gamma)=n(2\gamma_1+\gamma_2)$.}
\label{BPSInvQK3cover}
\end{table}

\medskip
\noindent
{\bf KS monodromy operator and homomorphisms.} Let us understand this covering relation using the surjective KS algebra homomorphism. In the finite chamber $\zeta_1<\zeta_2$, we simply have:
\be
\label{K3finch}
\mathbb{M}_Q=\CK_{0,1}\CK_{1,0},
\ee
where the subscript in $\CK_{N_1,N_2}$ corresponds to the charge $\gamma= N_1\gamma_1+N_2\gamma_2$. As before, we consider  the $\Gamma$-graded Lie algebra $\mathfrak{\underline g}_{\tilde Q}$ for $\tilde Q$, and we directly see that
\be
\label{K3Cfinch}
\mathbb{\underline M}_{\tilde Q}=\tilde \CK^3_{0,1}\tilde \CK^3_{1,0}
\ee 
in the finite chamber $\t\zeta=F^\ast(\zeta)$ with $\zeta_1<\zeta_2$. In the other chamber, $\zeta_1>\zeta_2$, the spectrum becomes much wilder --- indeed, technically speaking, $K_3$ is of wild representation type. For simplicity, let us just consider the product for charges with a fixed BPS ray --- here we pick the ray corresponding to the
central charge of $\gamma_1+\gamma_2$ and $\tilde \gamma_{1,\alpha}+\tilde
\gamma_{2,\beta}$, respectively. Using the MPS formula to determine the BPS invariants, we obtain:
\be 
\label{K3infch}
\mathbb{M}_Q=\CK^3_{1,1} \CK^{-6}_{2,2}\CK^{18}_{3,3}\CK^{-84}_{4,4}\cdots~,
\ee 
and
\be 
\label{K3Cinfch}
\mathbb{\underline M}_{\tilde Q}=\tilde \CK^9_{1,1}\tilde \CK^{-18}_{2,2}\tilde\CK^{54}_{3,3}\tilde\CK^{-252}_{4,4}\cdots~.
\ee
Similar computations can be performed for any possible ray on the central-charge plane. Now, as per the general formalism of section~\ref{sec:AlgHoms}, we have the surjective homomorphism:
\be 
\frak{\underline f}_*\;:\; \mathfrak{\underline g}_{\tilde Q}\to \mathfrak{g}_{Q}\; :\;  3\,\tilde e_{\gamma} \mapsto e_{\gamma},
\ee 
which acts on $\t \CK_\gamma$ very simply as
\be
\tilde \CK^3_{\gamma}\to \CK_\gamma~.
\ee 
This homomorphism relates the KS products and also the full non-commutative monodromy operators for these quivers --- indeed, we directly see that $\frak{\underline f}_*(\mathbb{\underline M}_{\tilde Q})=\mathbb{M}_{ Q}$ in the finite chamber, thereby proving the covering relation~\eqref{eq:DTK3} for the rational BPS invariants for all BPS states in all chambers.

\medskip
\noindent
{\bf A comment on the injective homomorphism.}
We may also consider the homomorphism 
\be
\mathfrak{\underline f}^*\;:\; \mathfrak{g}_{Q} \to \mathfrak{\underline g}_{\tilde Q} \; : \; e_{\gamma}\mapsto \frac{1}{3}\,\tilde e_{3\gamma}~,
\ee 
which is injective but not surjective, since it only maps to elements of the subalgebra $\mathfrak{\underline g}_{\tilde Q}^{(3)}\subset \mathfrak{\underline g}_{\tilde Q}$ consisting of elements $\tilde e_{\gamma}$ with $\gamma$ divisible by 3. Hence $\mathfrak{\underline f}^*$ acts on $\CK_\gamma$ as:
\be
\begin{split}
&3\! \not |\, \gamma:\qquad  \CK_\gamma\to \tilde \CK_{\gamma}^3~,\\
&3\,|\,  \gamma:\qquad \CK_{\gamma/3}^3\to \tilde \CK_{\gamma}~.
\end{split}
\ee 
While this relates the finite-chamber monodromy operators~\eqref{K3finch} and~\eqref{K3Cfinch}, this does not relate them in the other chamber, as is clear already from~\eqref{K3infch} and \eqref{K3Cinfch}. This is due to the fact that the homomorphism is not surjective: the generic support of the BPS invariants of the $K_3$ quiver does not allow us to repeat the argument we could make for the symmetric cover of $K_2$, for instance.

\medskip
\noindent
{\bf Refined BPS invariants.} A relation between refined BPS invariants follows as above by introducing non-trivial single-centred invariants for the nodes of $\tilde Q$. Following~\eqref{eq:assignOmS}, we set $\tilde y^{2/3}$, $1$ and $\tilde y^{-2/3}$ for the three nodes $\gamma_{j, \alpha}$ in each block $j=1,2$. We then find, for all the cases we checked, that
\be
\label{RefInvGC3}
\sum_{\tilde \gamma | F_\ast(\tilde\gamma)=\gamma} \Omega^{\tilde Q}(\tilde \gamma,\tilde y)=(y^2+1+y^{-2})\,\Omega_{Q}(\gamma, y)~,\qquad \tilde y^{1/3}=y~.
\ee 
This agrees with the general proposal~\eqref{eq:OmQOmtQy2}.

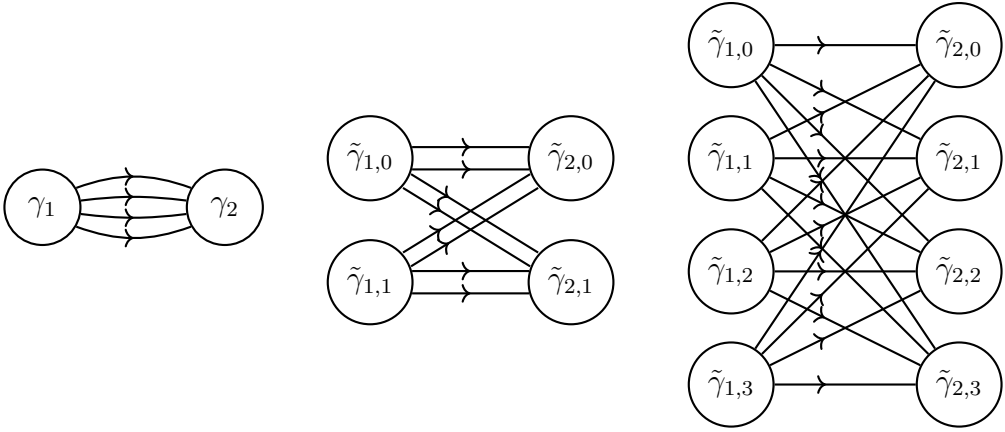
\begin{figure}[t]
\centering
\begin{tikzpicture}[baseline=1mm, every node/.style={circle, draw, minimum size=1cm}, thick]
\node[] (1) at (0,-0.65){$\gamma_{1}$};
\node[] (2) at (2.4,-0.65){$\gamma_{2}$};
\draw[->-=0.5] (1.30)  to[bend left=20]  (2.150);
\draw[->-=0.5] (1.10)  to[bend left=7]   (2.170);
\draw[->-=0.5] (1.350) to[bend right=7]  (2.190);
\draw[->-=0.5] (1.330) to[bend right=20] (2.210);
\end{tikzpicture}%
\qquad
\begin{tikzpicture}[baseline=1mm, every node/.style={circle, draw, minimum size=0.9cm}, thick]
\node[] (1) []{$\t\gamma_{1,0}$};
\node[] (2) [right =1.5 of 1]{$\t\gamma_{2,0}$};
\node[] (3) [below =.5 of 1]{$\t\gamma_{1,1}$};
\node[] (4) [right =1.5 of 3]{$\t\gamma_{2,1}$};
\draw[->-=0.5] (1.15)  to (2.165);
\draw[->-=0.5] (1.345) to (2.195);
\begin{scope}[transform canvas={xshift=1.5pt, yshift=2.6pt}]
  \draw[->-=0.3] (1) to (4);
\end{scope}
\begin{scope}[transform canvas={xshift=-1.5pt, yshift=-2.6pt}]
  \draw[->-=0.3] (1) to (4);
\end{scope}
\draw[->-=0.5] (3.15)  to (4.165);
\draw[->-=0.5] (3.345) to (4.195);
\begin{scope}[transform canvas={xshift=-1.5pt, yshift=2.6pt}]
  \draw[->-=0.3] (3) to (2);
\end{scope}
\begin{scope}[transform canvas={xshift=1.5pt, yshift=-2.6pt}]
  \draw[->-=0.3] (3) to (2);
\end{scope}
\end{tikzpicture}
\qquad
\begin{tikzpicture}[baseline=1mm, every node/.style={circle, draw, minimum size=1cm}, thick]
\node (L0) at (0,1.5)   {$\t\gamma_{1,0}$};
\node (L1) at (0,0){$\t\gamma_{1,1}$};
\node (L2) at (0,-1.5)  {$\t\gamma_{1,2}$};
\node (L3) at (0,-3) {$\t\gamma_{1,3}$};
\node (R0) at (3,1.5)   {$\t\gamma_{2,0}$};
\node (R1) at (3,0){$\t\gamma_{2,1}$};
\node (R2) at (3,-1.5)  {$\t\gamma_{2,2}$};
\node (R3) at (3,-3) {$\t\gamma_{2,3}$};
\foreach \i in {0,1,2,3} {
  \foreach \j in {0,1,2,3} {
    \draw[->-=0.365] (L\i) -- (R\j);
  }
}
\end{tikzpicture}
\caption{The $K_4$ quiver (\textsc{Left}) together with its $\Z_2$ (\textsc{Middle}) and $\Z_4$ (\textsc{Right}) symmetric covers.}
\label{fig:k4_z2_z4_cover}
\end{figure} 

\subsection{Symmetric covers of the Kronecker quiver \texorpdfstring{$K_4$}{K4}}
Consider next the Kronecker quiver $K_4$. It admits two symmetric covers, a 2-cover and a 4-cover, as shown in figure~\ref{fig:k4_z2_z4_cover}. We list a sample of BPS invariants for $K_4$ in table~\ref{BPSInvtildeQ_K4} (\textsc{Left}) together with BPS invariants for the 2-cover of $K_4$ (\textsc{Right}). The invariants for the
4-cover are listed in table~\ref{BPSInvQK44cover}. One can again check the general relation \eqref{eq:CoverBPSInv} in many explicit examples. Because the number of arrows between nodes is even for both $Q=K_4$ and its symmetric 2-cover, \eqref{eq:CoverBPSInv} simplifies in this case to~\eqref{eq:trivialcocycle}, namely:
\be
2\,\Omega^{K_4}(\gamma,\zeta)=\sum_{\tilde \gamma| F_\ast(\tilde
  \gamma)=\gamma} \Omega^{\tilde Q}(\tilde \gamma,\tilde \zeta).
\ee 
This relation also directly follows from the action on the KS monodromies from the homomorphism $\mathfrak{\underline f}_*$ between the Lie algebras,
\be 
\mathfrak{\underline f}_*\,:\,\mathfrak{\underline g}_{\tilde Q}\to \mathfrak{g}_{Q}\,:\,\ 2\tilde e_{\gamma}\mapsto e_{\gamma}.
\ee

\begin{table} 
\center
  \renewcommand{\arraystretch}{1.4}
\begin{tabular}{|c|r|}
\hline 
$N$ & $\Omega^{Q}(N,\zeta)$\\
\hline 
$(1,1)$ & $-4$\\
$(2,2)$ & $-16$\\
$(3,3)$ & $-144$\\
$(4,4)$ & $-1632$\\
\hline
\end{tabular}
\hspace{1cm}
\begin{tabular}{|c|r|r|}
\hline 
$\t N$ & $\Omega^{\t Q}(\t N,\tilde \zeta)$ & \# permutations\\
\hline \hline
$(1,0;1,0)$ & $-2$ & 4 \\
$(2,0;1,1)$ & $-4$ & 4 \\
$(2,0;2,0)$ & $0$ & 4 \\
$(1,1;1,1)$ & $-16$ & 1 \\
$(3,0;3,0)$ & $0$ & 4 \\
$(2,1;3,0)$ & $-6$ & 8 \\
$(2,1;2,1)$ & $-60$ & 4 \\
$(4,0;4,0)$ & $0$ & 4 \\
$(3,1;4,0)$ & $-8$ & 8 \\
$(2,2;4,0)$ & $-24$ & 4 \\
$(3,1;3,1)$ & $-160$ & 4 \\
$(2,2;3,1)$ & $-392$ & 4 \\
$(2,2;2,2)$ & $-896$ & 1 \\
\hline 
\end{tabular}
\caption{\textsc{Left:} a sample of BPS invariants $\Omega(N,\zeta)$ for $Q=K_4$ with $\zeta_1>\zeta_2$. \textsc{Right:} BPS invariants $\Omega(\t N,\tilde \zeta)$ for the
  $2$-cover of $K_4$ with $\tilde \zeta=F^\ast(\zeta)$. }
\label{BPSInvtildeQ_K4}
\label{BPSInvQK4dcover}
\end{table}

\begin{table} 
  \center
\renewcommand{\arraystretch}{1.4}
\begin{tabular}{|c|r|r|}
\hline 
$\t N$ & $\Omega^{\t Q}(\t N,\tilde \zeta)$ & \# permutations\\
\hline \hline
$(1,0,0,0;1,0,0,0)$ & $1$ & 16 \\
$(1,1,0,0;1,1,0,0)$ & $-2$ & 36 \\
$(1,1,1,0;2,1,0,0)$ & $3$ & 96 \\
$(1,1,1,0;1,1,1,0)$ & $18$ & 16 \\
$(1,1,1,1;3,1,0,0)$ & $-4$ & 24 \\
$(2,1,1,0;2,2,0,0)$ & $-4$ & 144 \\
$(1,1,1,1;2,2,0,0)$ & $-24$ & 24 \\
$(2,1,1,0;2,1,1,0)$ & $-20$ & 144 \\
$(1,1,1,1;2,1,1,0)$ & $-96$ & 24 \\
$(1,1,1,1;1,1,1,1)$ & $-384$ & 1 \\
\hline 
\end{tabular}
\caption{BPS invariants $\Omega(\t N,\tilde \zeta)$
  for the $4$-cover of $K_4$ with $\sum_{\alpha} N_{j,\alpha}\leq 4$ and $\tilde \zeta=F^\ast(\zeta)$. }
\label{BPSInvQK44cover}
\end{table}


\noindent
For the 4-cover, the covering relation \eqref{eq:CoverBPSInv} specialises to
\be
4\,\bar \Omega(\gamma,\zeta)=\sum_{\tilde \gamma| F_\ast(\tilde
  \gamma)=\gamma} (-1)^{N_1N_2} \bar \Omega(\tilde \gamma,\tilde \zeta)~,
\ee
where $\gamma=N_1\gamma_1+N_2\gamma_2$. Using the data displayed in table~\ref{BPSInvQK44cover}, one finds that the first terms of the KS
monodromy for charges of the form $F^*(\t\gamma)=n(\gamma_1+\gamma_2)$
are given by,
\be 
\tilde \CK^{16}_{1,1}\tilde \CK^{-72}_{2,2}\tilde\CK^{576}_{3,3}\tilde\CK^{-6528}_{4,4}\cdots
\ee
For this $\Z_4$ cover, we have the homomorphism~\eqref{eq:frakfdownlam} which reads:
\be
\mathfrak{\underline f}_*\,:\,\mathfrak{\underline g}_{\tilde Q}\to \mathfrak{g}_{Q}\,:\, 4\tilde e_{N_1,N_2}\mapsto (-1)^{N_1N_2} e_{N_1,N_2},
\ee 
and which gives the transformation
\be 
\tilde \CK^{16}_{1,1}\tilde \CK^{-8}_{2,2}\mapsto  \CK^{4}_{1,1}
\ee 
such that we arrive at
\be 
\CK^{-4}_{1,1} \CK^{-16}_{2,2} \CK^{-144}_{3,3} \CK^{-1632}_{4,4}\cdots,
\ee 
in agreement with table~\ref{BPSInvtildeQ_K4}.

\subsection{Symmetric covers with loops: \texorpdfstring{$E_0-E_6$}{E0-E6} pair and \texorpdfstring{$E_1-E_5$}{E1-E5} pair}
\label{sec:E0E6}

{\bf A $\Z_3$ cover with superpotential.}  As an example of a quiver with closed loops and a superpotential, let us consider the $E_0-E_6$ Galois pair whose quivers were displayed in figure~\ref{quiver E6 orbi}. 
To test our formula, we tabulate a number of BPS invariants for
both quivers in table~\ref{BPSInvtildeQ_E0} for
$\zeta=(1,0,-1)$ and $\t\zeta=F^\ast(\zeta)$. For $\KK E_0$ (that is, the local $\mathbb{P}^2$ geometry), the Coulomb branch formula determines these numbers with single-centred invariants $\Omega_S(\gamma_j)=1$ for $\gamma_j\in Q_0$, $\Omega_S(n \delta)=-1$ with $\delta=\gamma_1+\gamma_2+\gamma_3$ and $0$ otherwise \cite{Beaujard:2020sgs, Mozgovoy:2020has}. In \cite{Beaujard:2021fsk}, such $\Omega_S$ are connected to the BPS invariant for the trivial stability condition, $\zeta_j=0$.

The 5d BPS quiver of the $\KK E_6$ theory corresponds to the local (pseudo-)$dP_6$ geometry. It has many distinct charges in the kernel of $\langle -, \t\gamma\rangle$, corresponding to the D0-branes and to six other `flavour' D2-brane charges. The charges of the form $\t\gamma=n \sum_{j,\alpha} \gamma_{j,\alpha}$ correspond to $n$ D0-branes, in which case we have $F_*(\t \gamma)=3n\delta$. The corresponding $\Omega_S$ can be deduced from the toric fan, and it is such that $\Omega(\t \gamma)=-9$~\cite{Mozgovoy:2020has}.  The other pure flavour charges $\t \gamma$ are such that $F_*(\t \gamma)=n\delta$ for $n$ not divisible by 3. Their $\Omega_S$'s are not fully determined, but it is known that they are equal to either 0 or $y$. The Galois-cover perspective leads us to the following conjecture for $n=1$. First we note that the 27 permutations of dimension vectors $(1,0,0;1,0,0;1,0,0)$ all correspond to charges $\t \gamma$ with $F_*(\t \gamma)=\delta$. Their support is on a cyclic subquiver of $\KK E_6$, namely the affine Dynkin quiver  $\hat A_2$, which consists of three nodes and three directed arrows forming a 3-cycle. However, only 18 of these 3-cycles appear in the superpotential~\eqref{eq:WE6}, which means that 18 out of the 27 dimension vectors correspond to a subquiver in the mutation class of the $A_3$ quiver~\cite{Alim:2011kw}. For these charges, the expected dimension is negative, $-1$, such that the associated $\Omega_S(\t \gamma)$ vanish. On the other hand, for 9 out of 27 permutations,\footnote{These correspond to charges $\t\gamma_{0,0}+\t\gamma_{1,0}+\t\gamma_{2,0}$, $\t\gamma_{0,0}+\t\gamma_{1,1}+\t\gamma_{2,2}$, $\t \gamma_{0,0}+\t \gamma_{1,2}+\t \gamma_{2,1}$ and their $\mathbb{Z}_3$ images. Thus the $\G$-labels ($\alpha$ of $\gamma_{j,\alpha}$) add up to $0\mod 3$ for these linear combinations.} the superpotential does not give a constraint and the expected dimension equals $1$. Since the moduli space is actually non-compact, $\CM_{\t\gamma}\cong \mathbb{C}$, we naturally assign to these 9 charges:
\be\label{OmegaS E6 A2}
\Omega_S(\t \gamma, \t y)=-\t y~, \qquad \text{hence}\quad \Omega_S(\t \gamma)=-1~.
\ee
We conjecture similarly that the single-centred invariant also equals $-1$ for 9 of the 27 permutations of $(1,1,0;1,1,0;1,1,0)$.

Comparison of the tables demonstrates that for primitive dimension vectors $\gamma\in \Gamma$, the covering relation~\eqref{eq:CoverBPSInv} is satisfied for $\G=\Z_3$. Moreover, with the conjectured assignment for $\Omega_S^{\t Q}$, the relation continues to hold for non-primitive dimension vectors of the form $n\delta$.

To extend the relation to refined invariants, we introduce single-centred indices $\tilde y^{\pm 2/3}$ and 1 for the three vertices in each block of the $E_6$ quiver of figure~\ref{quiver E6 orbi},  as discussed in subsection~\ref{subsec:symgalcov}. The computation is similar to the examples discussed in subsections~\ref{sec:Symm2Cover}-\ref{sec:Sym3Cover}. We checked explicitly for the non-trivial examples $N=(1,2,1)$ and $(3,3,1)$  that the refined invariants indeed satisfy the covering relation~\eqref{eq:OmQOmtQy2}.

\begin{table} 
  \center
  \renewcommand{\arraystretch}{1.4}
\begin{tabular}{|c|r|}
\hline 
$N$ & $\Omega^Q(N, \zeta)$\\
\hline 
$(n,n,n)$ & $-3$ \\
$(1,2,1)$ & $9$ \\
$(3,3,1)$ & $-38$ \\
\hline
\end{tabular}
\hspace{1cm}
\begin{tabular}{|c|r|r|}
\hline 
$\tilde N$ & $\Omega^{\t Q}(\tilde N,\tilde \zeta)$ & \# permutations\\
\hline \hline
$(1,0,0;1,0,0;1,0,0)$ & $-1$ & 9 out of 27\\
$(1,0,0;1,0,0;0,1,0)$ & $0$ & 18 out of 27 \\
$(n,n,n;n,n,n;n,n,n)$ & $-9$ & 1 \\
$(1,0,0;2,0,0;1,0,0)$ & $0$ & 27 \\
$(1,0,0;1,1,0;1,0,0)$ & $1$ & 27 \\
$(2,1,0;1,1,1;1,0,0)$ & $-2$ & 18 \\
$(1,1,1;2,1,0;1,0,0)$ & $-2$ & 18 \\
$(1,1,1;1,1,1;1,0,0)$ & $-14$ & 3 \\
\hline 
\end{tabular}
\caption{\textsc{Left:} Non-vanishing BPS invariants $\Omega(N,\zeta)$ for the $E_0$
  quiver; the quiver on the right-hand side of figure~\ref{quiver E6
    orbi}, with $\zeta=(1,0,-1)$. \textsc{Right: } BPS invariants $\Omega(\tilde N,\tilde \zeta)$ for the
  $E_6$ quiver; the quiver on the left-hand side of figure~\ref{quiver
    E6 orbi} and $\tilde \zeta=F^\ast(\zeta)$. }
\label{BPSInvtildeQ_E0}
\end{table}

\medskip\noindent
{\bf A $\Z_2$ cover with superpotential.}
As a further example of a Galois cover of a quiver with superpotential, consider the $E_1-E_5$ Galois pair for Phase (b) of the local $\mathbb{F}_0$ geometry, as discussed in section~\ref{sec:Galois_covers_En}. The two quivers are displayed in figure~\ref{fig:E1E5cover}. We order the central charges as
\be 
Z_1=Z_3, \qquad Z_2=Z_4,\qquad {\rm arg}(Z_1) > {\rm arg}(Z_2),
\ee
or $\zeta_1=\zeta_3=1$ and $\zeta_2=\zeta_4=-1$. Note that this choice
does not preserve the block structure of the $E_1$ quiver. We tabulate BPS invariants for the $E_1$ quiver $Q$ and its cover $\t Q$, the $E_5$ quiver, in table~\ref{BPSInvtildeQ_E1}. Comparing the invariants again confirms the general relation \eqref{eq:CoverBPSInv} for primitive charges $\gamma$. 

For non-primitive charges $\gamma$, the $\KK E_5$ quiver gives rise to a higher-dimensional flavour lattice in the kernel of $\langle-,\t\gamma \rangle$ similarly to the previous example, and the comparison of BPS invariants for such charges proceeds analogously. For $\gamma=4n\delta$ with $\delta=\sum_{j=1}^4 \gamma_j$, the BPS invariant follows from the index for D0-branes. Moreover for $\gamma=\delta$, the details of the superpotential $W_{dP_5^{(d)}}$ \eqref{eq:WdP5} work out neatly. There are 16 permutations of $\t N=(1,0;1,0;1,0;1,0)$, which correspond to $F_*(\t \gamma)=\delta$ and which are supported on a subquiver of $\t Q$ with four nodes. For the two arrows from $\t \gamma_{4,\sigma}$ to $\t \gamma_{1,\alpha}$, the restriction of $W_{dP_5^{(d)}}$ to the subquiver gives rise to one relation for 8 permutations\footnote{These correspond to the charges $\t\gamma_{1,\alpha}+\t\gamma_{2,\beta}+\t\gamma_{3,\rho}+\t\gamma_{4,\sigma}$, with $\beta\neq \rho$.}, and two relations for the other 8 permutations. Hence, $\Omega^{\t Q}$ equals $-1$ for the first 8 and $0$ for the others. 
We conjecture that the BPS invariants for other dimension vectors work out such that the general relation is satisfied.

\begin{table} 
  \center
    \renewcommand{\arraystretch}{1.4}
\begin{tabular}{|c|r|}
\hline 
$N$ & $\Omega^{Q}(N,\zeta)$\\
\hline 
$(n,n,n,n)$ & $-4$ \\
$(n,n,n-1,n-1)$ & $-2$ \\
$(n-1,n-1,n,n)$ & $-2$ \\
\hline
\end{tabular}
\begin{tabular}{|c|r|r|}
\hline 
$\tilde N$ & $\Omega^{\t Q}(\tilde N,\tilde \zeta)$ & \# perm's\\
\hline \hline
$(1,0;0,1;1,0;1,0)$ & $-1$ & 8 out of 16 \\
$(1,0;1,0;1,0;1,0)$ & $0$ & 8 out of 16 \\
$(n,n;n,n;n,n;n,n)$ & $-8$ & 1 \\
$(1,1;1,1;1,0;1,0)$ & $1$ & 4 \\
$(1,0;1,0;1,1;1,1)$ & $1$ & 4 \\
\hline
\end{tabular}
\caption{\textsc{Left:} Non-vanishing BPS invariants $\Omega(N,\zeta)$ for the
  $E_1$ quiver; the quiver on the right hand side of figure~\ref{fig:E1E5cover}, with $\zeta=(1,-1,1,-1)$ \cite{Longhi:2021qvz}. \textsc{Right:} BPS invariants $\Omega(\tilde N,\tilde \zeta)$ for the
  $E_5$ quiver; the quiver on the left hand side of figure~\ref{fig:E1E5cover}, with $\tilde \zeta=F^\ast(\zeta)$.}
\label{BPSInvtildeQ_E1}
\end{table}

\subsection{Non-symmetric covers of \texorpdfstring{$K_2$}{K2}, \texorpdfstring{$K_3$}{K3}, \texorpdfstring{$\mathbb{C}^3$}{C3} and the conifold}\label{sec:nonsymGC}
 
In this final subsection, we consider four examples of non-symmetric Galois covers ---that is, Galois covers for which the arrow gradings are not symmetrically assigned. We consider quivers without loop in subsection~\ref{sec:Z3K2}~and~\ref{sec:Z2K3}, and CY$_3$ quivers with superpotentials in subsections~\ref{sec:C3C3Z3} and~\ref{sec:ConP1P1}. 

\subsubsection{\texorpdfstring{$\mathbb{Z}_3$}{Z3} cover of \texorpdfstring{$K_2$}{K2}}
\label{sec:Z3K2}

A simple example of a non-symmetric Galois cover is the $\Z_n$ cover of the $K_2$ quiver with gradings 0 and 1 assigned to the two arrows of $K_2$. The resulting quiver $\tilde Q$ has $2n$ nodes with charges $\{\tilde \gamma_{j,\alpha}\}$, $j=1,2$, $\alpha=1,\dots, n$, and arrows satisfying
\begin{equation}
    \langle \tilde \gamma_{1,\alpha},\tilde\gamma_{2,\alpha}\rangle=\langle\tilde\gamma_{1,\alpha},\tilde\gamma_{2,\alpha+1}\rangle=1,
  \end{equation}
where the second index is considered modulo $n$, {\it i.e.} such that $\langle\tilde\gamma_{1,n-1},\tilde\gamma_{2,0}\rangle=1$. We consider $n=3$ in detail. The quiver $\tilde Q$ for this case is displayed in figure~\ref{GQuiver3}. Since $\tilde \zeta=F^\ast(\zeta)$ is on a wall of
marginal stability, we introduce a small
perturbation $\tilde \zeta'$ away from $\tilde \zeta$. To be specific, the FI
parameters are chosen as
\be
\label{eq:tzeta'3cover}
\begin{split}
\tilde \zeta' &=\left(1,1-\frac{1}{\sqrt{2000}},1-\frac{1}{\sqrt{3000}},-1,-1+\frac{1}{\sqrt{4000}},-1+\frac{1}{\sqrt{5000}}\right),\\
\tilde \zeta&=F^\ast(\zeta)=(1,1,1,-1,-1,-1),\\
\zeta&=(1,-1).
\end{split}
\ee
We first evaluate BPS invariants for $\t\zeta'$, and subsequently use the KS wall-crossing formula to determine the invariants on the wall $\t\zeta$. A sample of BPS invariants for both $\t\zeta'$ and $\t\zeta$ are given in
table~\ref{BPSInvQK23cover}.

\begin{table} 
  \center
\renewcommand{\arraystretch}{1.4}
\begin{tabular}{|c|r|r|r|}
\hline 
$\tilde N$ & $\Omega^{\t Q}(\tilde N,\tilde \zeta')$ & $\Omega^{\t Q}(\tilde N,\tilde \zeta)$ & \# permutations\\
\hline \hline
$(1,0,0;0,0,0)$ & 1 & 1 & 3 \\
$(0,0,0;1,0,0)$ & 1 & 1 & 3 \\
$(1,0,0;1,0,0)$ & 1 & 1 & 3 \\
$(1,0,0;0,1,0)$ & 1 & 1 & 3\\
$(1,0,0;0,0,1)$ & 0 & 0 & 3 \\
$(1,1,0;1,0,0)$ & 0 & 0 & 6\\
$(1,1,0;0,1,0)$ & 1 & 1 & 3\\
$(2,0,0;1,0,0)$& 0 &0 & 18\\
$(2,0,0;2,0,0)$& 0 & 0 & 9\\
$(2,0,0;1,1,0)$& 0 & 0 & 18\\
$(1,1,0;1,1,0)$& 1 & 1/2 & 1\\
$(1,0,1;1,0,1)$& 0 & 1/2 & 1\\
$(0,1,1;0,1,1)$& 0 & 1/2 & 1\\  
$(1,0,1;1,1,0)$& 1 & 1/2 & 1\\
$(1,1,0;0,1,1)$& 0 & 1/2 & 1\\
$(0,1,1;1,1,0)$& 0 & 1/2 & 1\\
$(1,0,1;0,1,1)$& 0 & 0 & 1 \\
$(1,1,0;1,0,1)$& 0 & 0 & 1 \\
$(0,1,1;1,1,0)$& 0 & 0 & 1 \\
$(1,1,1;1,1,1)$&  $-2$ & $-2$ & 1\\
\hline 
\end{tabular}
\caption{BPS invariants $\Omega(\t N,\tilde \zeta)$ for the non-symmetric $3$-cover of $K_2$ with $\zeta'$ as in~\protect\eqref{eq:tzeta'3cover} and $\tilde \zeta=F^\ast((1,-1))$.}
\label{BPSInvQK23cover}
\end{table}

The corresponding BPS invariants for $Q=K_2$ in this chamber can be read off from table~\eqref{BPSInvK2} or~\eqref{KSNf0}. Comparing this with table~\ref{BPSInvQK23cover}
demonstrates that the invariants satisfy the specialisation of \eqref{eq:CoverBPSInv},
\be\label{rel example Z3 cover of K2}
\begin{split}
  &3\,\bar \Omega^{K_2}(N,\zeta)= \sum_{\tilde N\vert F_\ast(\tilde
  N)=N}
(-1)^{\sum_{\alpha=0}^2 \tilde N_{1,\alpha} \tilde  N_{2,\alpha}+ \tilde  N_{1,\alpha} \tilde  N_{2,\alpha+1}}
\bar \Omega^{\tilde Q}(\tilde N,\tilde \zeta).
\end{split}
\ee 
For example, for $N=(2,2)$ the right-hand side reads
\be
6\times \frac{1}{4}+6\times 1/2\times (-1)=-3/2,
\ee
which matches with the left-hand side, $3\times 1/4 \times (-2)=-3/2$.
Moreover, we observe from table~\ref{BPSInvQK23cover} that the BPS invariants $\Omega^{\tilde Q}(\tilde
N,\tilde \zeta)$ corresponding to a
$\mathbb{Z}_3$ orbit are identical, as expected, while this is not the
case for the invariants at $\tilde \zeta\neq F^\ast(\zeta)$. 

We can establish~\eqref{rel example Z3 cover of K2} using the KS algebra homomorphism introduced in section~\ref{sec:GenGalHom}. Given a charge
$\tilde \gamma=\sum_{j,\alpha} N_{j,\alpha} \tilde\gamma_{j,\alpha}$,  we define the $\Z_3$-invariant algebra elements ${\tilde e}^\G_{\tilde \gamma}\in
\mathfrak{g}^{\G}_{\tilde Q}$ as:
\be 
\label{eq:flnonsym}
 {\tilde e}^\G_{\tilde \gamma}=\sum_{\beta\in \mathbb{Z}_3}
\tilde e_{\sum_{j} \tilde N_{j,\alpha} \tilde\gamma_{j,(\alpha+\beta)}}~. 
\ee 
We then have the surjective Lie algebra homomorphism~\eqref{eq:fGlam}, namely:
\be
\mathfrak{f}^\G_*:\quad  {\tilde e}^\G_{\tilde \gamma} \mapsto
(-1)^{\sum_{\alpha=0}^2 \tilde N_{1,\alpha} \tilde  N_{2,\alpha}+ \tilde  N_{1,\alpha} \tilde  N_{2,\alpha+1}}\, 
e_{F_\ast(\tilde \gamma)}~.
\ee 
For example, we have three pre-images of  $e_{\gamma_1+\gamma_2}$,
\begin{equation}
    \begin{aligned}
        &\tilde e_{\tilde\gamma_{1,0}+\tilde\gamma_{2,0}}+\tilde e_{\tilde\gamma_{1,1}+\tilde\gamma_{2,1}}+\tilde e_{\tilde\gamma_{1,2}+\tilde\gamma_{2,2}}\to e_{\gamma_1+\gamma_2}~,\\
        &\tilde
        e_{\tilde\gamma_{1,0}+\tilde\gamma_{2,1}}+\tilde e_{\tilde\gamma_{1,1}+\tilde\gamma_{2,2}}+\tilde
        e_{\tilde\gamma_{1,2}+\tilde\gamma_{2,0}}\to
        e_{\gamma_1+\gamma_2}~,\\
            &\tilde e_{\tilde\gamma_{1,0}+\tilde\gamma_{2,2}}+\tilde e_{\tilde\gamma_{1,1}+\tilde\gamma_{2,0}}+\tilde e_{\tilde\gamma_{1,2}+\tilde\gamma_{2,1}}\to e_{\gamma_1+\gamma_2}~.
    \end{aligned}
\end{equation}
It is straightforward to explore other examples. A few BPS invariants are given in table~\ref{BPSInvQK24cover} for the $n=4$ cover of $K_2$ with $d_a=(0,1)$. For simplicity, we have restricted to dimension vectors which are independent of a small perturbation away from $F^\ast(\zeta)$. One may again verify the agreement with the general formula~\eqref{eq:CoverBPSInv}.
\begin{table} 
\center
\renewcommand{\arraystretch}{1.4}
\begin{tabular}{|c|r|r|}
\hline 
$\tilde N$ & $\Omega(\tilde N,\tilde \zeta)$ & \# permutations\\
\hline \hline
$(1,0,0,0;0,0,0,0)$ & 1 & 4 \\
$(0,0,0,0;1,0,0,0)$ & 1 & 4 \\
$(1,0,0,0;1,0,0,0)$ & 1 & 4 \\
$(1,0,0,0;0,1,0,0)$ & 1 & 4\\
$(1,1,0,0;0,1,0,0)$ & 1 & 4\\
$(1,1,1,1;1,1,1,1)$ & $-2$ & 1\\
\hline 
\end{tabular}
\caption{A few non-vanishing BPS invariants $\Omega(\tilde N, \tilde \zeta)$ for the $4$-cover of $K_2$ with gradings $d_a=(0,1)$. }
\label{BPSInvQK24cover}
\end{table}

\subsubsection{$\mathbb{Z}_2$ cover of $K_3$}
\label{sec:Z2K3}

\begin{figure}
\begin{center}
\begin{tikzpicture}[baseline=1mm,baseline=1mm,every node/.style={circle,draw},thick]
\node[] (1) []{$\tilde \gamma_{1,0}$};
\node[] (2) [right =1.5 of 1]{$\tilde \gamma_{2,0}$};
\node[] (3) [below =1.3 of 1]{$\tilde \gamma_{1,1}$};
\node[] (4) [right =1.5 of 3]{$\tilde \gamma_{2,1}$};
\draw[->-=0.5] (1.20) to   (2.160);
\draw[->-=0.5] (1.340) to   (2.200);
\draw[->-=0.3] (1) to   (4);
\draw[->-=0.5] (3.20) to   (4.160); 
\draw[->-=0.5] (3.340) to   (4.200);
\draw[->-=0.3] (3) to   (2);
\end{tikzpicture}
\caption{Quiver $\tilde Q$ corresponding to the $\mathbb{Z}_2$ Galois cover of $K_3$ with grading $d_a=(0,0,1)$.}
\label{fig:Z2K3}
\end{center}
\end{figure}
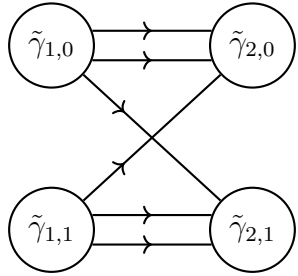

As a second example of a non-symmetric Galois cover, consider the $\mathbb{Z}_2$-cover $\tilde Q$ of the $K_3$ quiver, with gradings $d_1=d_2=0$ and $d_3=1$, which is displayed in figure~\ref{fig:Z2K3}. This Galois covering quiver $\tilde Q$ has four nodes with charges $\{\tilde \gamma_{j,\alpha}\}$, $j=1,2$ and $\alpha=0,1$, and with the arrows determined by
\be 
\left<\tilde \gamma_{1,0},\tilde \gamma_{2,0}\right>=\left<\tilde\gamma_{1,1},\tilde\gamma_{2,1}\right>=2~,\text{\,\,and\,\,} \left<\tilde\gamma_{1,0},\tilde\gamma_{2,1}\right>=\left<\tilde\gamma_{1,1},\tilde\gamma_{2,0}\right>=1~.
\ee
We choose two FI parameters $\tilde \zeta'$, $\tilde \zeta$ as well as $\zeta$ as:
\be 
\begin{split}
&\tilde \zeta'=(1,1-1/\sqrt{2000},-1,-1+1/{3000}),\\
&\tilde \zeta=(1,1,-1,-1),\\
&\zeta=(1,-1),
\end{split}
\ee 
with $\tilde \zeta'$ a small perturbation of $\tilde
\zeta=F^\ast(\zeta)$. Note that $\t\zeta$ sits on a wall of marginal stability for many choices of $\tilde N$.

\begin{table} 
\center
\renewcommand{\arraystretch}{1.2}
\begin{tabular}{|c|r|r|r|r|}
\hline 
$\tilde N$ & $\Omega^{\t Q}(\tilde N;\tilde\zeta')$ & $\mu(N,\zeta')$ & $\Omega^{\t Q}(\tilde N;\tilde \zeta)$ & \# permutations\\
\hline \hline
$(1,0;1,0)$ & $-2$ & 0 & $-2$ & 1 \\
$(0,1;0,1)$ & $-2$ & $-0.002$ & $-2$ & 1 \\
$(1,0;0,1)$ & 1 & 0.009 & 1 & 1 \\
$(0,1;1,0)$ & 1 & $-0.01$ & 1 & 1 \\
$(2,0;1,0)$ & 1 & & 1 & 2 \\
$(2,0;0,1)$ & 0 & & 0 & 2 \\
$(1,1;1,0)$ & $-2$ & & $-2$ & 2 \\
$(1,0;1,1)$ & $-2$ & & $-2$ & 2 \\
$(1,0;2,0)$ & 1 & & 1 & 2 \\
$(1,0;0,2)$ & 0 & & 0 & 2 \\
$(1,0;1,1)$ & $-2$ & & $-2$ & 2 \\
$(2,0;2,0)$ & 0 & & $0$ & 4 \\
$(2,0;1,1)$ & $-2$ & 0.005 & $-1$ & 1 \\
$(0,2;1,1)$ & 0 & $-0.007$ & $-1$ & 1 \\
$(1,1;2,0)$ & $-2$ & $-0.006$ & $-1$ & 1 \\
$(1,1;0,2)$ & 0 & 0.004 & $-1$ & 1 \\
$(1,1;1,1)$ & $-6$ & $-0.001$ & $-6$ & 1\\
$(3,0;3,0)$ & $0$ &  & 0 & 4\\
$(3,0;2,1)$ & $3$ &  & $5/6$ & 1\\
$(3,0;1,2)$ & $0$ &  & $1/3$ & 1\\
$(0,3;2,1)$ & $0$ &  & $1/3$ & 1\\
$(0,3;1,2)$ & $0$ &  & $5/6$  & 1\\
$(2,1;3,0)$ & $3$ &  & $5/6$ & 1\\
$(2,1;0,3)$ & $0$ &  & $1/3$ & 1\\
$(1,2;3,0)$ & $0$ &  & $1/3$ & 1\\
$(1,2;0,3)$ & $0$ &  & $5/6$ & 1\\
$(2,1;2,1)$ & $-12$ & & $-32/3$ & 1\\
$(2,1;1,2)$ & $5$ & & $19/3$ & 1\\
$(1,2;2,1)$ & $5$ & & $19/3$ & 1\\
$(1,2;1,2)$ & $-10$ & & $-32/3$ & 1\\
\hline 
\end{tabular}
\caption{BPS invariants $\Omega(\tilde N,\t \zeta)$ for the non-symmetric $2$-cover $\tilde K_3$ of $K_3$ with grading $(0,0,1)$. }
\label{BPSInvQK3cover001}
\end{table}

We list some BPS invariants $\Omega^{\t Q}(\tilde N; \tilde \zeta')$ and
$\Omega^{\t Q}(\tilde N; \tilde \zeta)$  in table~\ref{BPSInvQK3cover001}. The integers $\Omega^{\t Q}(\tilde N; \tilde \zeta)$ are
determined using the KS wall-crossing formula as described around~\eqref{eq:sameslope}.%
Comparison with the left-hand side of table~\ref{BPSInvQK3cover} for $K_3$ allows us to check the expected covering relation
\be 
\label{eq:BPSInvId}
2\,\bar \Omega^{K_3}(N,\zeta)=\sum_{\tilde \gamma\vert
  F_\ast(\tilde \gamma)=\gamma} (-1)^{\tilde N_{1,0} \tilde N_{2,0}+\tilde N_{1,1}\tilde N_{2,1}}\, \bar \Omega^{\tilde Q}(\tilde N,\tilde \zeta),
\ee 
for $N=(1,1), (2,1), (1,2), (2,2), (3,3)$. The case of $N=(3,3)$ is
quite subtle; in particular, the covering formula does not hold for the stability $\t\zeta'$, but it does for $\t\zeta$. Note also that the entries in table~\ref{BPSInvQK3cover} are
dependent on the specific choice of perturbation away from $\t\zeta$ --- for example, for $\tilde
\zeta''=(1.01,0.98,-0.99,-1.03)$, the last four entries of the second
column are $-10,5,9,-10$, and we would not satisfy the covering relation either. At the degenerate stability condition $\tilde \zeta$, the Galois covering relation does hold, remarkably. To see this, we need to use the Baker--Campbell--Hausdorff formula~\eqref{eq:BCH} to write the monodromy operators such that contributions with the same slope enter as a single exponent, as prescribed in~\eqref{eq:sameslope}. 

As expected, the numerical values in the fourth column for $\tilde \zeta$ are invariant under the $\mathbb{Z}_2$ action, while the invariants in the second column for $\tilde \zeta'$ are not. 
We can again define a surjective homomorphism from $\mathfrak{g}^{\mathbb{Z}_2}_{\tilde Q}$ to $\mathfrak{g}_{Q}$, namely:
\be 
\begin{split}
&\mathfrak{f}_*^{\G}(\tilde e_{(\tilde N_{1,0} , \tilde N_{1,1};\tilde N_{2,0},\tilde N_{2,1})}+ \tilde e_{(\tilde N_{1,1},\tilde N_{1,0};\tilde N_{2,1},\tilde N_{2,0})})\\
&\qquad \qquad = (-1)^{\tilde N_{1,0}\tilde N_{2,0}+\tilde N_{1,1}\tilde N_{2,1}}\,e_{(\tilde N_{1,0}+\tilde N_{1,1},\tilde N_{2,0}+\tilde N_{2,1})}
\end{split}
\ee
This establishes~\eqref{subsec:hom and BPS rels} as explained in section~\ref{subsec:hom and BPS rels}. 
Since we are not aware of an extension of this map to the non-commutative
torus~$\hat e_\gamma$, it is unclear what is the correct generalisation of~\eqref{eq:BPSInvId} to refined invariants.

\subsubsection{$\mathbb{C}^3$ and $\mathbb{C}^3/\mathbb{Z}_3$}
\label{sec:C3C3Z3}

\begin{figure}[t]
\begin{center}
 \begin{tikzpicture}[baseline=1mm, every node/.style={circle,draw},thick]
\node[] (1) []{$\gamma_1$};
\draw[->-=0.5] (1.30) to[looseness=20]   (1.150); 
\draw[->-=0.5] (1.40) to[looseness=15]   (1.140); 
\draw[->-=0.5] (1.50) to[looseness=10]   (1.130); 
\end{tikzpicture} 
\caption{BPS quiver for local $\mathbb{C}^3$.}
\label{QuiverC3}
\end{center}
\end{figure}
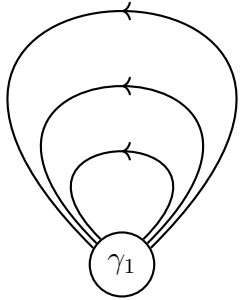

As a further example of a non-symmetric Galois pair, we consider the quiver $Q$ for the  local Calabi--Yau $\mathbb{C}^3$ given in figure~\ref{QuiverC3} and with superpotential 
\be\label{W C3}
W={\rm Tr}\left(a_1a_2a_3-a_1a_3a_2\right). 
\ee
The quiver contains three loops, which are the three arrows connecting the node to itself. The refined and unrefined BPS invariants equal for all $N>0$ \cite{Behrend:2009dc, Mozgovoy:2011ps},
\be
\Omega(N,\zeta;y)=-y^3~,\qquad \Omega(N,\zeta)=-1~.
\ee
We consider the $\mathbb{Z}_3$ cover of the $\C^3$ quiver with gradings $d_1=d_2=1$ and $d_3=-2$ for the three arrows in $Q$. The resulting Galois cover is the 5d BPS quiver of the $\KK E_0$ theory displayed on the right in figure~\ref{quiver E6 orbi}. Its superpotential~\eqref{eq:WE0} is the Galois uplift of~\eqref{W C3} obtained according to~\eqref{tilde W}. Of course, this Galois pair is also a simple example of the general discussion of section~\ref{subsec:Galois and orbifold}, namely of the fact that Galois covers are abelian orbifolds --- here we have that $Z(\CA_Q)\cong \C[a_1, a_2, a_3]$, and the Galois cover gives us the CY$_3$ quiver for the orbifold~$\C^3/\Z_3$~\cite{Kachru:1998ys}, which is also the quiver for the local $\mathbb{P}^2$ geometry obtained upon crepant resolution.

\begin{table} 
  \center
  \renewcommand{\arraystretch}{1.4}
\begin{tabular}{|c|r|r|}
\hline 
$\tilde N$ & $\Omega(\tilde N, \tilde \zeta)$ & \# permutations \\
\hline 
$(1,0,0)$ & $1$ & 3 \\
$(1,1,0)$ & $3/2$ & 3 \\
$(0,1,2)$ & $0$ & 6 \\
$(n,n,n)$ & $-3$ & 1 \\
\hline
\end{tabular}
\caption{Non-vanishing BPS invariants $\Omega(\tilde N, \tilde \zeta)$ for the $E_0$
  quiver with $F_\ast(\tilde N)\leq 3$; the quiver on the right hand side of figure~\ref{quiver E6
    orbi}, with $\t \zeta=(0,0,0)$.}
\label{BPSInvtildeQ_E0_000}
\end{table}

Since the gradings are not $\G$-symmetric, this is a non-symmetric Galois cover. As in the previous sections, we evaluate the BPS invariants for $\t\zeta=F^\ast(\zeta)$ using the KS wall-crossing formula. We list a few invariants in table~\ref{BPSInvtildeQ_E0_000} for $\tilde \zeta=(0,0,0)$. With the cocycle  
\be
\xi(\tilde \gamma,\gamma)=(-1)^{\tilde N_0 \tilde N_1+\tilde N_1 \tilde N_2+\tilde N_2 \tilde N_0-\tilde N_0^2 -\tilde N_1^2 - \tilde N_2^2}~,
\ee
one verifies that the numerical BPS invariants again satisfy the general relation~\eqref{eq:CoverBPSInv}.

\subsubsection{Conifold and local $\mathbb{F}_0$}
\label{sec:ConP1P1}
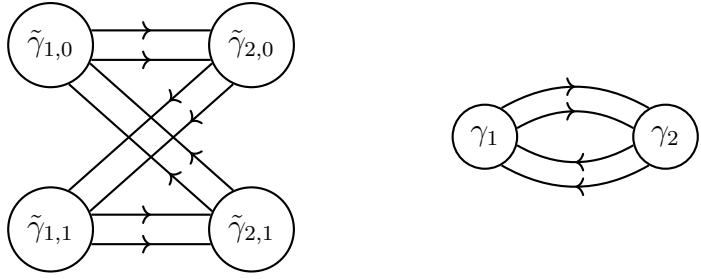
\begin{figure}
\begin{center}
\begin{tikzpicture}[baseline=1mm,baseline=1mm,every node/.style={circle,draw},thick]
\node[] (1) []{$\tilde \gamma_{1,0}$};
\node[] (2) [right =1.5 of 1]{$\tilde \gamma_{2,0}$};
\node[] (3) [below =1.3 of 1]{$\tilde \gamma_{1,1}$};
\node[] (4) [right =1.5 of 3]{$\tilde \gamma_{2,1}$};
\draw[->-=0.5] (1.20) to   (2.160);
\draw[->-=0.5] (1.340) to   (2.200);
\draw[->-=0.3] (4.115) to   (1.335);
\draw[->-=0.3] (4.155) to   (1.295);
\draw[->-=0.5] (3.20) to   (4.160); 
\draw[->-=0.5] (3.340) to   (4.200);
\draw[->-=0.3] (2.205) to   (3.65);
\draw[->-=0.3] (2.245) to   (3.25);
\node[] (5)  [below right=0.5 and 5.0 of 1]{$\gamma_1$};
\node[] (6) [right =1.5 of 5]{$\gamma_2$};
\draw[->-=0.5] (5.15) to[bend left=30]  (6.165);
\draw[->-=0.5] (5.60) to[bend left=30]  (6.120);
\draw[->-=0.5] (6.195) to[bend left=30] (5.345);
\draw[->-=0.5] (6.240) to[bend left=30] (5.300);
\end{tikzpicture}
\caption{\textsc{Right:} Quiver $Q$ corresponding to the conifold. \textsc{Left:} Quiver $\tilde Q$ corresponding to the $\mathbb{Z}_2$ Galois cover of $Q$ with grading $0$ for the two arrows from $\gamma_1$ to $\gamma_2$, and grading $\pm 1$ for the reverse arrows.}
\label{QuiverConifold}
\end{center}
\end{figure}

Yet another example of a Galois pair between CY$_3$ quivers is provided by the conifold and the local $\mathbb{F}_0$ geometry. The conifold quiver is displayed on the right of figure~\ref{QuiverConifold}. Its superpotential reads:
\be
W_{\rm Conifold}=\sum_{a,b\in \{1,2\}}\epsilon_{ab} {\rm Tr}\left( a_{12}^1\, a_{21}^a\, a_{12}^2\, a_{21}^b \right)~.
\ee
For a generic choice of stability, say $\zeta=(1,-1)$, the non-vanishing BPS invariants read, for $n\geq 1$ \cite{Szendroi:2007nu, Gholampour:2009, Morrison:2012, Banerjee:2019apt},
\be
\begin{split}
&\Omega(n(\gamma_1+\gamma_2),\zeta;y)=-y^3-y,\\ &\Omega(n\gamma_1+(n-1)\gamma_2,\zeta;y)=\Omega((n-1)\gamma_1+n\gamma_2,\zeta;y)=1~.
\end{split}
\ee

\noindent
The $\mathbb{F}_0$ quiver (Phase (a)) is obtained as a $\Z_2$ cover $\t Q$ of the conifold quiver by choosing the $\Z_2$ gradings $d_{a_{12}^a}=0$ for the two arrows from $\gamma_1$ to $\gamma_2$ and $d_{a_{21}^a}= 1$ for the two arrows from $\gamma_2$ to $\gamma_1$. The Galois cover is then the BPS quiver for local $\mathbb{F}_0$ displayed on the left in figure~\ref{QuiverConifold} --- recall that it also previously appeared on the right-hand side of figure~\ref{Z4 symmetric quivers E5 E1}, with the superpotential $\t W=W_{\mathbb{F}_0}$ given in~\eqref{eq:WF0}. While this Galois pair does not correspond to a symmetric Galois cover, the BPS invariants $\Omega^{\tilde Q}(\tilde \gamma,\tilde \zeta)$ are independent of a small perturbation away from $\tilde \zeta =F^\ast(\zeta)$ with $\zeta=(1,-1)$. The BPS spectrum is known for this choice of $\tilde \zeta$ \cite[Eq. (1.2)]{Longhi:2021qvz}, and listed in table~\ref{QuiverP1P1}. Comparison of the unrefined BPS invariants of the pair again confirms the general formula (\ref{eq:CoverBPSInv}). Since $\xi(\gamma,\t\gamma)=1$ for all $\t \gamma$, the covering relation simplifies to~\eqref{eq:trivialcocycle}.

\begin{table} 
  \center
  \renewcommand{\arraystretch}{1.4}
\begin{tabular}{|c|r|r|}
\hline 
$\tilde N$ & $\Omega(\tilde N, \tilde \zeta;y)$ & \# permutations \\
\hline 
$(n,0;n-1,0)$ & 1 & 2 \\
$(n-1,0;n,0)$ & 1 & 2 \\
$(n-1,n;n-1,n)$ & $-y^{-1}-y$ & 2 \\
$(n,n;n,n)$ & $-y^{-1}(1+y^2)^2$ & 1 \\
\hline
\end{tabular}
\caption{Non-vanishing BPS invariants $\Omega^{\t Q}(\tilde N, \tilde \zeta)$ for the $\mathbb{F}_0$
  quiver; the quiver on the right-hand side of figure~\ref{quiver E6
    orbi}, with $\tilde \zeta=(1,1,-1,-1)$.}
\label{QuiverP1P1}
\end{table}

\renewcommand{\baselinestretch}{1}
\footnotesize
\bibliography{references} 
\bibliographystyle{JHEP} 
\end{document}